\documentclass[usenatbib,usegraphicx]{mn2e}
\input epsf
\usepackage{graphics}
\usepackage{epsfig}
\usepackage{url}
\usepackage{amsmath}
\usepackage{longtable}
\usepackage{lscape}

\newcommand{\kms}{~$km~s^{-1}$}

\def\degr{\hbox{$^\circ$}}
\def\deg{\hbox{$^\circ$}}

\newcommand{\pvm}{position-velocity diagram}
\newcommand{\SNR}{signal-to-noise ratio}
\newcommand{\snr}{signal-to-noise ratio}
\newcommand{\FOV}{field-of-view}
\newcommand{\fov}{field-of-view}
\newcommand{\FP}{Fabry-Perot}

\newcommand{\PSF}{PSF}
\newcommand{\vdm}{velocity dispersion map}
\newcommand{\vdms}{velocity dispersion maps}

\newcommand{\Vdms}{Velocity dispersion maps}
\newcommand{\vfs}{velocity fields}
\newcommand{\vf}{velocity field}
\newcommand{\Vf}{Velocity field}
\newcommand{\Vfs}{Velocity fields}
\newcommand{\rc}{rotation curve}
\newcommand{\rcs}{rotation curves}
\newcommand{\Rc}{Rotation curve}

\newcommand{\pa}{position angle}
\newcommand{\pas}{position angles}

\newcommand{\TF}{Tully-Fisher}
\newcommand{\Ha} {H$\alpha$}
\newcommand{\ha} {H$\alpha$}
\newcommand{\hz} {high redshift}

\newcommand{\lz} {low redshift}

\def\jnl@style{\it}
\def\aaref@jnl#1{{\jnl@style#1}}

\def\aaref@jnl#1{{\jnl@style#1}}

\def\aj{\aaref@jnl{AJ}}                   
\def\araa{\aaref@jnl{ARA\&A}}             
\def\apj{\aaref@jnl{ApJ}}                 
\def\apjl{\aaref@jnl{ApJ}}                
\def\apjs{\aaref@jnl{ApJS}}               
\def\ao{\aaref@jnl{Appl.~Opt.}}           
\def\apss{\aaref@jnl{Ap\&SS}}             
\def\aap{\aaref@jnl{A\&A}}                
\def\aapr{\aaref@jnl{A\&A~Rev.}}          
\def\aaps{\aaref@jnl{A\&AS}}              
\def\azh{\aaref@jnl{AZh}}                 
\def\baas{\aaref@jnl{BAAS}}               
\def\jrasc{\aaref@jnl{JRASC}}             
\def\memras{\aaref@jnl{MmRAS}}            
\def\mnras{\aaref@jnl{MNRAS}}             
\def\pra{\aaref@jnl{Phys.~Rev.~A}}        
\def\prb{\aaref@jnl{Phys.~Rev.~B}}        
\def\prc{\aaref@jnl{Phys.~Rev.~C}}        
\def\prd{\aaref@jnl{Phys.~Rev.~D}}        
\def\pre{\aaref@jnl{Phys.~Rev.~E}}        
\def\prl{\aaref@jnl{Phys.~Rev.~Lett.}}    
\def\pasp{\aaref@jnl{PASP}}               
\def\pasj{\aaref@jnl{PASJ}}               
\def\qjras{\aaref@jnl{QJRAS}}             
\def\skytel{\aaref@jnl{S\&T}}             
\def\solphys{\aaref@jnl{Sol.~Phys.}}      
\def\sovast{\aaref@jnl{Soviet~Ast.}}      
\def\ssr{\aaref@jnl{Space~Sci.~Rev.}}     
\def\zap{\aaref@jnl{ZAp}}                 
\def\nat{\aaref@jnl{Nature}}              
\def\iaucirc{\aaref@jnl{IAU~Circ.}}       
\def\aplett{\aaref@jnl{Astrophys.~Lett.}} 
\def\apspr{\aaref@jnl{Astrophys.~Space~Phys.~Res.}}
\def\bain{\aaref@jnl{Bull.~Astron.~Inst.~Netherlands}} 
\def\fcp{\aaref@jnl{Fund.~Cosmic~Phys.}}  
\def\gca{\aaref@jnl{Geochim.~Cosmochim.~Acta}}   
\def\grl{\aaref@jnl{Geophys.~Res.~Lett.}} 
\def\jcp{\aaref@jnl{J.~Chem.~Phys.}}      
\def\jgr{\aaref@jnl{J.~Geophys.~Res.}}    
\def\jqsrt{\aaref@jnl{J.~Quant.~Spec.~Radiat.~Transf.}}
\def\memsai{\aaref@jnl{Mem.~Soc.~Astron.~Italiana}}
\def\nphysa{\aaref@jnl{Nucl.~Phys.~A}}   
\def\physrep{\aaref@jnl{Phys.~Rep.}}   
\def\physscr{\aaref@jnl{Phys.~Scr}}   
\def\planss{\aaref@jnl{Planet.~Space~Sci.}}   
\def\procspie{\aaref@jnl{Proc.~SPIE}}   

\title[GHASP VIII. Dynamical evolution in high-z disks] {Evidence for strong dynamical evolution in disk galaxies through the last 11 Gyr.
\emph{GHASP VIII: A local reference sample of rotating disk
galaxies for high redshift studies}}

\author[B. Epinat et al.]
{Epinat B.$^{1,2}$\thanks{E-mail: benoit.epinat@ast.obs-mip.fr}, Amram P.$^{1}$, Balkowski C.$^{3}$, Marcelin M.$^{1}$\\
$^{1}$Laboratoire d'Astrophysique de Marseille, Universit\'e de Provence, CNRS, 38 rue Fr\'ed\'eric Joliot-Curie,\\ F-13388 Marseille Cedex 13, France\\
$^{2}$Laboratoire d'Astrophysique de Toulouse-Tarbes, Universit\'e de Toulouse, CNRS, 14 Avenue \'Edouard Belin,\\ F-31400 Toulouse, France\\
$^{3}$Galaxies Etoiles Physique et Instrumentation, Observatoire de Paris-Meudon, Universit\'e Paris VII, 5 Place Jules Janssen,\\ F-92195 Meudon, France.\\
}

\date{Accepted 2009 September 09. Received 2009 September 08; in original form 2009 April 24}

\pubyear{2009}

\begin{document}

\maketitle

\begin{abstract}

Due to their large distances, high redshift galaxies are observed
at a very low spatial resolution. In order to disentangle the evolution of galaxy kinematics from low
resolution effects, we have used Fabry-Perot 3D \Ha\ data-cubes of 153 nearby isolated
galaxies selected from the Gassendi \Ha\ survey of SPirals (GHASP) to simulate data-cubes of galaxies at redshift $z=1.7$ using
a pixel size of $0.125''$ and a $0.5''$ seeing. We have
derived \Ha\ flux, velocity and velocity dispersion maps. From
these data, we show that the inner velocity gradient is
lowered and is responsible for a peak in the velocity dispersion
map. This signature in the \vdm\ can be used to make a kinematical classification, but misses 30\%\ of the regular rotating disks in our sample. Toy-models of rotating disks have been built to recover the
kinematical parameters and the \rcs\ from low resolution data. The
poor resolution makes the kinematical inclination uncertain and
the position of galaxy center difficult to recover. The position angle of the major axis is retrieved with an accuracy higher than 5\degr\ for 70\% of the sample. Toy-models
also enable to retrieve statistically the maximum velocity and
the mean velocity dispersion of galaxies with a satisfying accuracy. This validates the use
of the Tully-Fisher relation for high redshift galaxies but the loss of resolution induces a lower slope of the relation despite the beam smearing corrections.
We conclude that the main kinematic parameters are
better constrained for galaxies with an optical radius at least as large as
three times the seeing. The simulated data have been compared to
actual high redshift galaxies data observed with VLT/SINFONI,
Keck/OSIRIS and VLT/GIRAFFE in the redshift range $3>z>0.4$, allowing to follow galaxy evolution from eleven to four Gyr.
For rotation-dominated galaxies, we find that the use of the velocity dispersion central peak as a
signature of rotating disks may misclassify slow and solid body rotators. This is the case for $\sim30$\%\ of our sample.
%
%
%
We show that the projected local data cannot reproduce the high velocity dispersion observed in high redshift galaxies except when no beam smearing correction is applied.
%
%
This unambiguously means that, unlike local evolved galaxies, there exists
at high redshift at least a population of disk galaxies for which
a large fraction of the dynamical support is due to random
motions. We should nevertheless insure that these features are not
due to important selection biases before concluding that the
formation of an unstable and transient gaseous disk is a general
galaxy formation process.

\end{abstract}


\section{Introduction}

Formation and evolution of galactic disks is one of the most
important unsolved questions of extragalactic astronomy and is
probably a key clue to merge cosmological models and galaxy
building-up mechanisms. The understanding of the rate and the
processes followed by galaxies of different masses to assemble,
the relative importance of mergers versus continuous gas accretion
infall onto the disk, the connection between bulge and disk
formation and more widely the dynamical evolution, the rate of
metal enrichment, the evolution of ratio between the baryonic and
dark matter masses and mass distribution, the angular momentum
transfers during these processes are among some of the fundamental and open questions.

Since the mid 1990s, large ground-based telescopes combined with
space observatory multiwavelength observations allow to tackle observationally the
question of galaxy formation. The challenges for the future are
also to extend the study of galaxy formation to the earliest phases,
at $z>6$, and to chart the progress of galaxy formation in detail
down to lower redshifts. Morphological and photometric studies
point out that high redshift galaxies do not show well-defined
shapes and their colors indicate a rapid star formation.  Galaxies
undergo strong evolution from irregular clumps of star formation
into the Hubble sequence valid in the local universe
\citep{Papovich:2005}. Global properties such as stellar mass,
population age, star formation rate, large-scale gaseous outflows,
active galactic nucleus fraction have been extensively studied by
numerous authors \citep{Dickinson:2003,Steidel:2004,Reddy:2006}.
The epoch of galaxy formation may span over a broad period
probably over 5 Gyrs. At redshifts $z\sim2$, galaxies are thought
to be accumulating the majority of their stellar mass and a wide
variety of evolutionary states from young and active star-forming
to massive and passively evolving galaxies are observed. At
redshifts $z\sim1$, the pattern of spiral and elliptical galaxies
observed in the nearby universe has settled into place 
even if the fraction of peculiar galaxies is higher \citep{Glazebrook:1995,Abraham:1996,Lotz:2008}. However,
it is still unknown whether the majority of star formation occurs
in flattened disk-like or alternatively in non-equilibrium
systems. More widely, it is clear that we do not yet understand
the dynamical state of galaxies during this period in which they
are forming the bulk of their stars \citep{Law:2007}.

More than one thousand high redshift galaxies (mainly Lyman-break
galaxies up to $z\sim3$) have a spectroscopic redshift (e.g.
\citealp{Steidel:2003}). Samples of galaxies have been observed
with long slit spectrographs to study their kinematics and
dynamics. Pioneer observations of $z\sim1$ disk galaxies have been
obtained by \citet{Vogt:1996,Vogt:1997}. Observations
of galaxies at higher redshift were more recently obtained
\citep{Erb:2003,Erb:2004,Erb:2006,Weiner:2006,Kassin:2007}.
Studies using NIR
slit or integral field unit (IFU) spectroscopy of \ha~emission
under seeing limited conditions have suggested that at least a
subset of high redshift galaxies have a disk-like morphology and
show large organized rotation, which may indicate the formation of an
early galactic disk \citep{Erb:2003}.
\cite{Kassin:2007} showed from long slit spectroscopy kinematical data and HST restframe B-band morphology that a correlation between peculiar kinematics and peculiar or merger-like morphology exists at $z\sim1$.

However, at high
redshift, the small angular size of the galaxies
($\sim0.5-1.5$\arcsec), comparable to the size of the seeing halo
which imposes to set-up a large width for the slit, is a serious
observational difficulty. The difficulty is even enhanced by the
fact that irregular galaxy morphology may induce possible strong
misalignment of the slit with respect to the kinematic major axis.
Moreover, with slit spectroscopy, it is not possible to study
internal kinematics features like spiral arms or bars. For these
reasons, the use of integral field spectroscopy has been overcome
using seeing-limited and adaptive optics (AO) assisted integral field
unit spectroscopy to obtain two-dimensional maps of these
galaxies. Due to obvious observational difficulties, the advent of
large telescopes and specialized focal instrumentations were
necessary to map in 3D some of these galaxies. Nowadays,
kinematics and dynamics of intermediate to high redshift
($0.4<z<3$) galaxies are being increasingly studied with integral
field instruments on 8/10-meters class telescopes. IMAGES survey
\citep{Flores:2006,Puech:2006,Yangetal:2008,Neicheletal:2008,Puechetal:2008,Rodrigues:2008}
contains 63 \vfs~and \vdms~of intermediate galaxies ($0.4<z<0.75$)
observed with the integral-field spectrograph FLAMES/GIRAFFE at
the VLT in the optical, in order to probe the dynamical evolution, in
particular in the Tully-Fisher relation.
The SINS survey has been carried out with the integral-field
spectrograph SINFONI at the VLT (\citealp{Forster-Schreiberetal:2009}
and references therein).  They have analyzed the 2D \Ha\
kinematics for 63 high redshift galaxies ($1.3<z<2.6$) in the near
infra-red (among 80 galaxies observed). They realized sub-kpc
resolution AO assisted observations using SINFONI
for eight galaxies plus four $z\sim3$ Lyman Break Galaxies.
%
%
Similar programs are under progress also using SINFONI at redshift
$\sim1.5$ (\citealp{Epinat:2009c}, \citealp{Queyrel:2009}, \citeauthor{Contini:2008} in preparation)
and using OSIRIS at Keck Observatory at redshift $\sim 1.5$
\citep{Wright:2007,Wright:2009} and $z\sim 3$
\citep{Law:2007,Law:2009}.

The question of the assembly of galaxies via major dissipative
mergers or internal secular processes has been recently intensely debated in the literature.  Based on the analysis of \Ha\
\vfs, \vdms\ and flux distributions, all the different teams
advocated that disk candidates are distinguishable from merger
candidates.
\citet{Forster-Schreiberetal:2009} classified the whole SINS sample and concluded that a third of galaxies has
rotation-dominated kinematics, another third is composed of interacting or
merging systems and the last third has
dispersion-dominated kinematics.
\citet{Epinat:2009c} reached the same conclusions from the MASSIV pilot run.
\citet{Wright:2009} and \citet{Law:2009} gave conclusions
compatible with this classification.
However, the large picture that emerges in terms of galaxy
formation is still a bit confused.
\citet{Genzeletal:2008,Genzeletal:2006} and \citet{Forster-Schreiberetal:2006}
claimed that a secular process of assembly forms bulges and
disks in massive galaxies at $z\sim2$. \citet{Robertson:2008} nevertheless suggested that the observation of \hz\ disk galaxies like the one presented in \citet{Genzeletal:2006} is consistent with the hypothesis that gas-rich mergers play an important role in disk formation at high redshift. \citet{Law:2007, Law:2009} and \citet{
Nesvadba:2008} argued that galaxies display irregular kinematics
more related to merging or gas cooling systems than rotating disks
and concluded that the high velocity dispersions observed in most
of the galaxies at $z\sim2$ may be due neither to a `merger' nor to a `disk', but
to the result of instabilities related to cold gas accretion
becoming dynamically dominant. \citet{Epinat:2009c} advocated that
several processes are acting at these epochs. Among them, merging
seems to play a key role. Close pairs of galaxies expected to
merge in less than 1~Gyr, indicate that the hierarchical build up
of galaxies at the peak of star formation is fully in progress.
The dominant `perturbed rotators' may include a significant
fraction of galaxies with minor mergers in progress or cold gas
accretion along streams of the cosmic web, producing a high
velocity dispersion.

The unusual kinematics, the high gas fraction
and star formation rates in high redshift galaxies
have been observed quite recently and attempts to explain them have been done.
One explanation is that these young galaxies may have experienced gas-rich major or minor mergers (e.g. \citealp{Semelin:2002,Robertson:2008}). An
alternative or complementary scenario may be that early-stage
galactic disks accrete large amounts of low angular momentum gas
from the cosmic web and thus content huge quantities of cold gas
which fragments and collapses to form violent starbursts (e.g. \citealp{Immeli:2004b,Bournaud:2007,Elmegreen:2007}).

In that
last scenario, large star formation may have happened in
dispersion-dominated transitory disks rather than in rotationally
supported gaseous disks as predicted in current galaxy formation
theories.  Through secular evolution processes, these unstable
disks may lead to the formation of the nowadays bulges and thick
disks. Filamentary gas accretion mechanisms should be no more
observable nowadays since large amounts of low angular momentum
cold gas do not exist anymore. As a consequence, merging is the
only mechanism able to fuel galaxies with large amounts of fresh
gas in the local universe while at higher redshifts alternative
mechanisms may have been in strong concurrence.

Selection effects in the different observations are induced by the
relatively low number of galaxies studied and cosmic variance
effects. For obvious observational reasons, preferentially
extended and bright emission lines galaxies were selected.  The
prevalence of large velocity shears (large galaxies) or large
velocity dispersions (mergers, etc.) in these sources may thus be a
product of the selection criteria.

At high redshift, the best seeing-limited observations cover
$\sim5~kpc$ and provide only 2 or 3 spatial resolution elements
across the major axis of a typical galaxy. Seeing limited studies
may miss velocity structures on spatial scales smaller than that
of the seeing halo, thus these kinematical measurements are
insufficient to claim rotation without using a model to deconvolve
the beam smearing effect. The use of IFU instead of long slit
spectrograph minimizes the problem but does not solve it
completely. Current integral field surveys at redshift $z>0.5$ lack
of a reference that would be affected by the same observation and
methodological biases. This is for instance necessary to probe a
possible evolution in the Tully-Fisher relation or to probe a
possible evolution in the dynamical support (rotation or
dispersion). A solution is the use of N-body/hydrodynamical
simulations of galaxies projected at high redshift as done by
\citet{Kronberger:2007}. A complementary approach, tackled in this
work, is to use real data and project them at high redshift, with
the same observing conditions as the real \hz~observations.

In section \ref{previous} we describe previous simulations of high
redshift data from nearby kinematical data. In section
\ref{observations_simulations}, we describe the GHASP subsample
selection and the simulation of redshifted galaxies. We test the validity of a galaxy classification based on the kinematical maps in section \ref{classificationflores}. We present
the velocity maps analysis method in section \ref{fitting_method},
we comment the results in section \ref{analysis}, then discuss
them in section \ref{discussion}. A conclusion is provided in
section \ref{conclusion}. The model used to recover the high
resolution \vfs\ and \rcs\ from the projected local data set of
galaxies is more widely detailed in Appendix \ref{model}. The
fit parameters and the beam smearing parameter for each
galaxy are given in Appendix \ref{table}. The maps of local sample
projected at high redshift are displayed in Appendix \ref{maps}
and the \rcs~corresponding to actual data and different models are
given in Appendix \ref{rcz}. Appendixes \ref{table}, \ref{maps} and \ref{rcz} are provided online only.

Throughrout this paper we use a standard cosmology with
$H_0=71$\kms $~Mpc^{-1}$, $\Omega_m=0.27$, and
$\Omega_\Lambda=0.73$. We have chosen to project our sample at the critical cosmological scale of redshift 1.7 which is in addition representative of the scale of galaxies from four to eleven Gyr ($0.4<z<3$).
In such a cosmology, at redshift $z=1.7$, $1''$ corresponds to $8.56~kpc$.

\section{Local galaxies to simulate distant galaxies}
\label{previous}

To learn about galaxy evolution, a method is to compare
primordial galaxies to nowadays ones. Because of their large
distances, \hz~galaxies are obviously not observable with the same
spatial sampling as low redshift galaxies. To compare nearby and distant
galaxies, it is thus necessary to disentangle distance effects
from evolution ones.

Due to the lost of spatial resolution, \emph{(i)} it is difficult
to disentangle rotators from mergers; \emph{(ii)} the
determination of the kinematical parameters (\pa~of the major
axis, center, inclination, systemic velocities) is more difficult; \emph{(iii)} the structures within the
galaxies (bars, rings, spiral arms, bubbles, etc.) as well as the
disk/bulbe/halo mass distributions in the inner regions are
smoothed when not erased.

The comparison between nearby galaxies projected at high redshift
and observed distant galaxies can help identifying signatures of
mergers, kinematical parameters and internal galaxy features and
shapes.

Even at low redshift, while the spatial resolution is high enough
to allow detailed analysis, controversy may exist on the nature
and on the history of peculiar galaxies such as interacting,
mergers or starforming galaxies. This is the case for instance for
the nearby gas rich Hickson compact group HCG 31 which displays a
low velocity dispersion ($\sim60$\kms) and an intense star
formation rate. Three scenarios have been put forward to explain
the nature of this object: \emph{(i)} these are two systems that
are in a pre-merger phase
\citep{Amram:2004,Verdes-Montenegro:2005,Amram:2007}, \emph{(ii)}
the system is a late-stage merger \citep{Williams:1991} or
\emph{(iii)} it is a single interacting galaxy
\citep{Richer:2003}. At $z=0.013$, the actual redshift of the
group, high spatial and spectral Fabry-Perot observations allow to
observe that the broader \ha~profiles (larger than 30\kms) are
located in the overlapping regions between the two main galaxies (HCG
31 A and C). This clearly maps the shock between the two galaxies
and the subsequent starburst regions
\citep{Amram:2004,Amram:2007}. What would tell us the observations of a
compact group like HCG 31 ($z=0.013$) when observed at
higher redshift ? To illustrate the answer to this question for
this specific compact group, beam smearing effects have been
tested by \citet{Amram:2008}. At $z=0.15$, it becomes
already difficult to count how many galaxies are involved in the
system and the broadening of the \ha~profiles would be interpreted
as an indicator of rotating disk. This system could thus be
catalogued as a rotator instead of a merger \citep{Flores:2006}.
At $z=0.60$, disentangling the system is a real challenge. This
illustrates the difficulty to retrieve the true nature and the
history of \hz~galaxies from observations affected by a too small spatial resolution.

As illustrated by the previous example, spatial resampling of
nearby galaxies has already been used to simulate
distant galaxies in order to interpret integral field data as well as
long slit observations
\citep{Rix:1997,Weiner:2006,Flores:2006,Puechetal:2008,Shapiroetal:2008,Amram:2008},
but a systematic comparison has never been done for a large local
reference sample.

The systematics induced by the beam smearing effects have been
studied in \citet{Amram:2008} who have projected the data cube of
the galaxies used to study the local Tully-Fisher (TF) relation
for CGs \citep{Mendes-de-Oliveira:2003} at different redshifts.
They pointed out several features: \emph{(i)} \hz~galaxies have
smoother rotation curves than local galaxies, a
``solid-bodyfication'' of the rotation curve is observed;
\emph{(ii)} nothing indicates that the maximum velocity of the
rotation curve is reached, leading to uncertainties in the
Tully-Fisher relation determination.

In order to analyze the kinematics of \hz~galaxies, control
samples of nearby galaxies, with well studied kinematics, are
necessary. Compact groups are probably extreme cases difficult to
describe even if they have probably been more frequent in the past
than nowadays. Close-by interacting galaxies may also lead to
inextricable confusion if the separation between the galaxies is
not large enough to disentangle the individual galaxies.  Star
forming galaxies dominated by bright HII regions producing strong
winds may also lead to misinterpretation when the spatial
resolution is not high enough to access the main mass
component.  Before studying these difficult kinds of galaxies
which will be considered in further works, in the present paper we
have considered more quiescent galaxies. We study the smoothing of
these signatures by using the GHASP sample in order to simulate
\hz~galaxies. The aim of this work is to know whether atmospheric
seeing may mask more complex structures than simple flattened
disk-like configuration and to test different models enabling to
recover the structures and the kinematic parameters.

\section{The sample} \label{observations_simulations}

\subsection{GHASP: the local dataset}

\FP\ observations from the GHASP survey
\citep{Epinat:2008b,Epinat:2008a} have been used for this work.
The GHASP sample contains 203 local galaxies, mainly isolated
spirals and irregulars, observed through their \ha\ line. These
data consist of high spectral resolution ($\sim5-10$\kms) and
seeing-limited data cubes. Nearby galaxies present a broad range
of luminosities/masses and morphological types and provide a wide
range of kinematical signatures (shape of the \vfs\ and of the
\rcs\ as well as presence of non circular structures like bars,
spiral arms, etc.).
%
%
This sample is thus particularly well adapted to be compared with
what is thought to be the ancestors of the actual rotating disks.

We have corrected some local distances computed from the Hubble law using the systemic
velocities, since the Hubble constant in the GHASP paper
($H_0=75$\kms $~Mpc^{-1}$) differs from the one used in the
present paper ($H_0=71$\kms $~Mpc^{-1}$).

\subsection{The redshifted dataset}

153 galaxies belonging to the GHASP sample
have been projected to redshift $z=1.7$ and constitute the
so-called ``redshifted dataset''.  We describe in this section the
selection criteria and the techniques applied to project the data
cubes taking into account several constraints (distance,
foreground contaminations, seeing, resampling, etc.) and to compute
the moment maps.

\subsubsection{Physical length scale}

\begin{figure}
\begin{center}
\includegraphics[width=8.5cm]{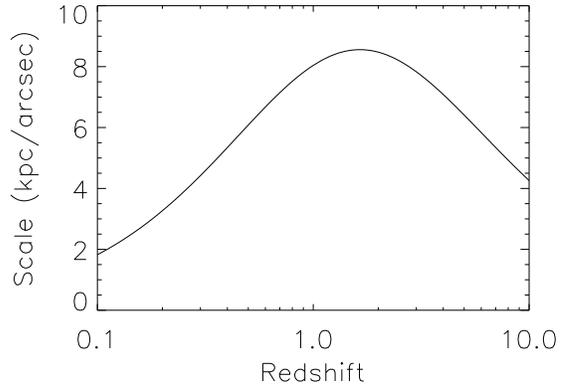}
\end{center}
\caption{Evolution of the physical length scale with the redshift
using the canonical cosmological parameters
$H_0=71$\kms$~Mpc^{-1}$, $\Omega_m=0.27$, and
$\Omega_\Lambda=0.73$.} \label{scale}
\end{figure}

Considering the standard cosmology chosen in this paper and
ignoring evolutionary effects, the angular size of galaxies
decreases with the distance from redshifts $z\sim0$ to $z\sim1.7$
and thus increases for redhsifts $z>1.7$ (see Figure \ref{scale}).
We have chosen to set the galaxies at their lower
angular size, i.e. at the redshift $z=1.7$  leading to a physical
scale of $8.56~kpc/arcsec$. This physical scale is representative of \hz\ galaxies
for which 3D observations are available today. Indeed, the
physical scale of $8.56~kpc/arcsec$ computed at $z=1.7$ decreases
only by $20$\% in the range of redshifts $z\sim0.63-4.34$. Thus,
this physical scale correctly matches actual observations of
\hz\ galaxies done with integral field spectroscopy instruments
such as SINFONI
\citep{Forster-Schreiberetal:2009},
OSIRIS \citep{Law:2009,Wright:2007,Wright:2009} and FLAMES/GIRAFFE
\citep{Flores:2006,Yangetal:2008}.

\subsubsection{Flux re-scaling}

Direct comparison between low and high redshift galaxy fluxes is not straightforward since high redshift galaxies do have higher star formation rates and higher luminosities than at low redshift. Nevertheless, instead of giving arbitrary units for \Ha\ fluxes, we
have computed the expected flux $F_{l}$ at redshift $z=1.7$ for
each galaxy, using the flux $F_0$ computed from the calibration in
\citet{Epinat:2008b}, the distance $d$ of the galaxy and the
luminous distance at redshift 1.7 ($d_l=12.865~Gpc$) using
equation \ref{fluxz}:

\begin{equation}
F_{l}=F_0\times\frac{d^{2}}{d_l^{2}} \label{fluxz}
\end{equation}
\Ha\ monochromatic maps presented in Appendix \ref{maps} have been
calibrated using equation \ref{fluxz}.

\subsubsection{Cleaning from background contaminations}

In order to exclude most of the foreground stars from the Milky
Way as well as to reduce residual night sky lines contribution,
regions where no ionized gas was detected in the local data cubes
have been masked on each channel. Indeed, sky contribution is large since it is integrated over a large angular size
(around 10\arcmin\ square).


\subsubsection{Blurring, resampling and noise addition}
\label{blurr}


The wavelength range of the data cube has been extended from 24 to 72
channels in order to remove interfringe effects due to the
spectrum periodicity of \FP\ interferometers \citep{Epinat:thesis}. Each channel of the
cube has been blurred by a two dimensional gaussian simulating the
seeing. The width of this gaussian has been computed taking into account
the seeing measured on the $z=0$ data so that the seeing halo for
redsfhited galaxies is simulated by a two dimensional gaussian
function of 0.5\arcsec\ FWHM. This halo of 0.5\arcsec matches the
best average spatial resolution that can be reached without
AO. This operation is computed in the Fourier space.

The spatial sampling has been set to 0.125\arcsec, to mimic the
SINFONI pixel size.
To avoid any interpolation, the binning is
the merging of an integer number of real pixels that corresponds
to the closest simulated size obtained for a redshift $z=1.7$. The
ratio seeing/pixel size has been set to be identical for each galaxy.
Thus the mean scale for our sample is $8.5~kpc/arcsec$ with a
standard deviation of $0.3~kpc/arcsec$.

In the present study, no spectral binning or smoothing has been
applied in order to dissociate these two resolution effects on 3D
data (this test will be done in a forthcoming work).

No noise has been added in the datacubes. Our goal is to study the beam smearing effects in the data to test the ability to recover the kinematical parameters. Indeed, if noise is added on the spectra simultaneously to blurring, it will not be straightforward to unambiguously disentangle the lack of spatial resolution from the low \snr.
Adding noise reduces the
detectability at low intensity levels, does not strongly bias
velocity distribution but affects velocity dispersion measurements. The \snr\ of the simulated
data (ranging from $\sim 3$ to $\sim 50$) is higher than real \hz\ observations (ranging from $\sim 2$ to $\sim 10$, e.g. \citealp{Epinat:2009c}).
The \snr\ slightly varies
from one galaxy to the other since the binning is not the same.

\subsubsection{Cleaning procedures on redshifted data}

A cleaning procedure has been applied on redshifted data to remove spurious measurements outside of the galaxies. We used the following criteria that ensure to avoid discontinuities on the edges of the velocity fields: the
velocity dispersion must be larger than 5\kms, which is lower than the spectral resolution of our data, and the
\snr\ (defined as the ratio of the \Ha\ monochromatic flux over
the RMS among spectral elements in the continuum at
that pixel
times the full width of the line) must
be larger than 2.7. This cleaning corresponds to the maps presented in Appendix \ref{maps}.

\subsubsection{Computating the moment maps for redshifted data}

The different maps have been computed using the barycenter method
described in \citet{Daigle:2006b} and already used to compute the
local maps in \citet{Epinat:2008b,Epinat:2008a}. \Vdms\ heve been
corrected from the spectral \PSF\ considered to be described by a
gaussian function using the following classical relation:
\begin{equation}
\sigma^2_{corr}=\sigma^2_{obs}-\sigma^2_{PSF}
\end{equation}

\subsubsection{Selection of a sub-sample}
\label{subsample}

\begin{figure}
\begin{center}
\includegraphics[width=7.0cm]{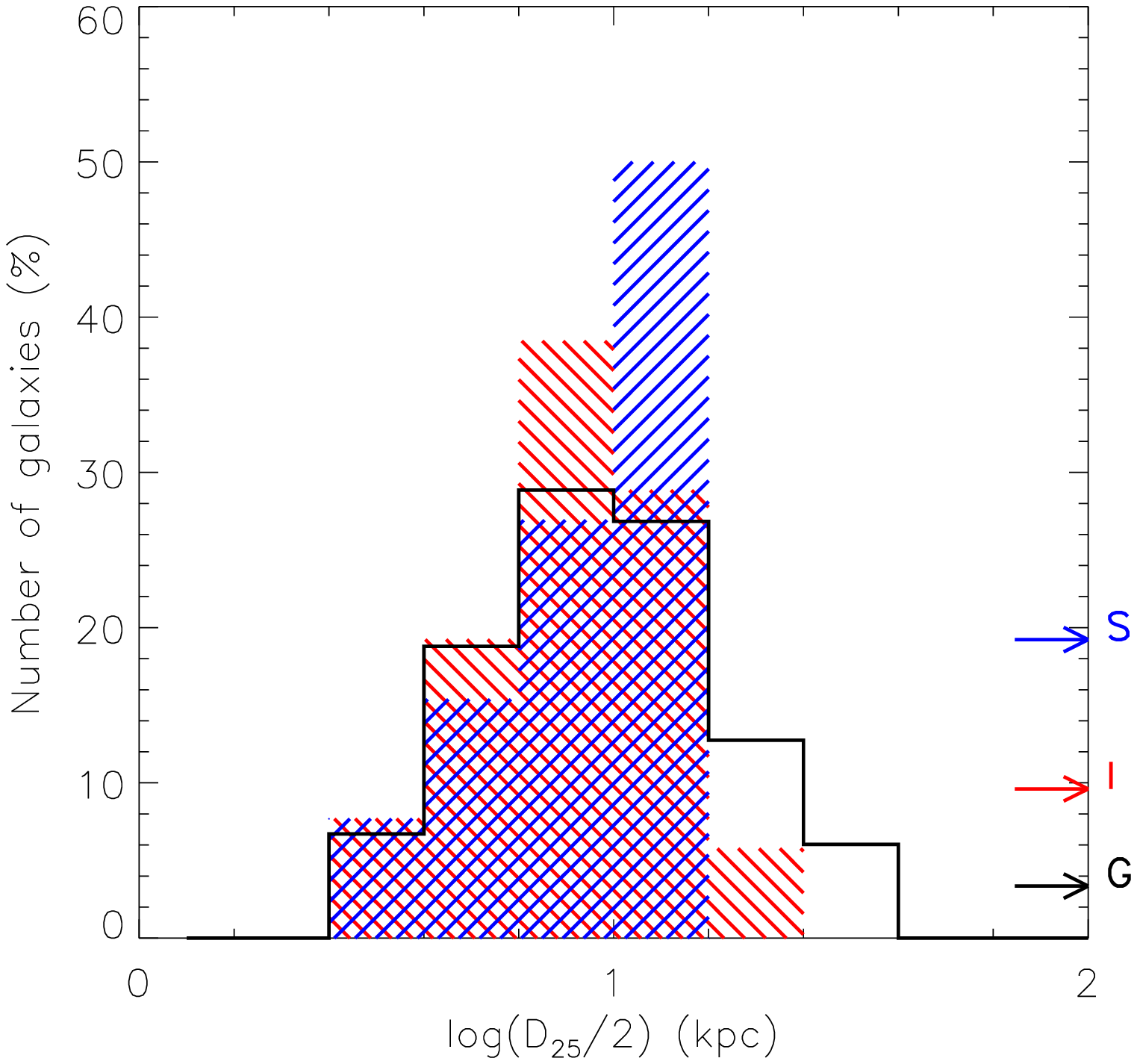}
\includegraphics[width=7.0cm]{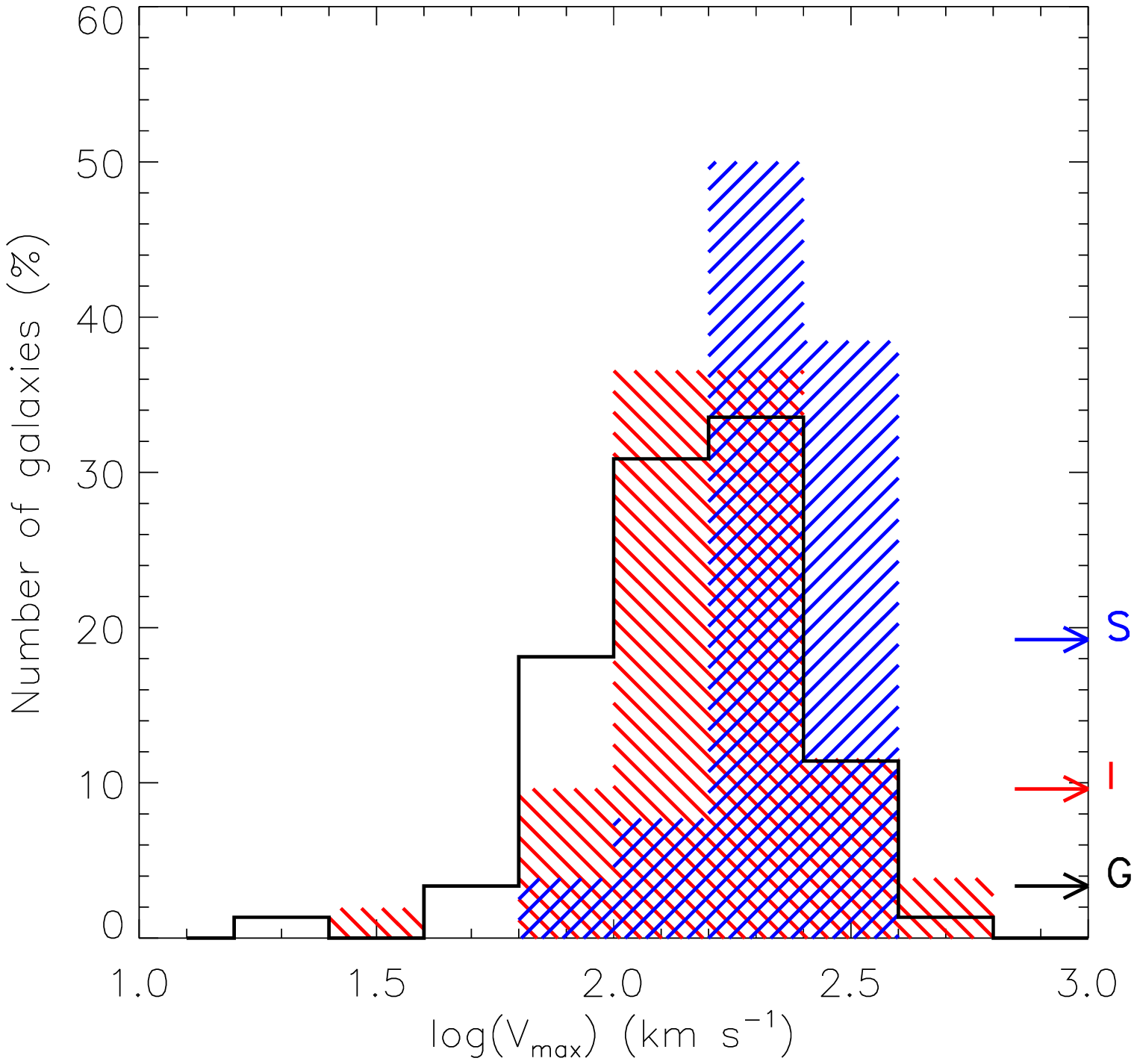}
\includegraphics[width=7.0cm]{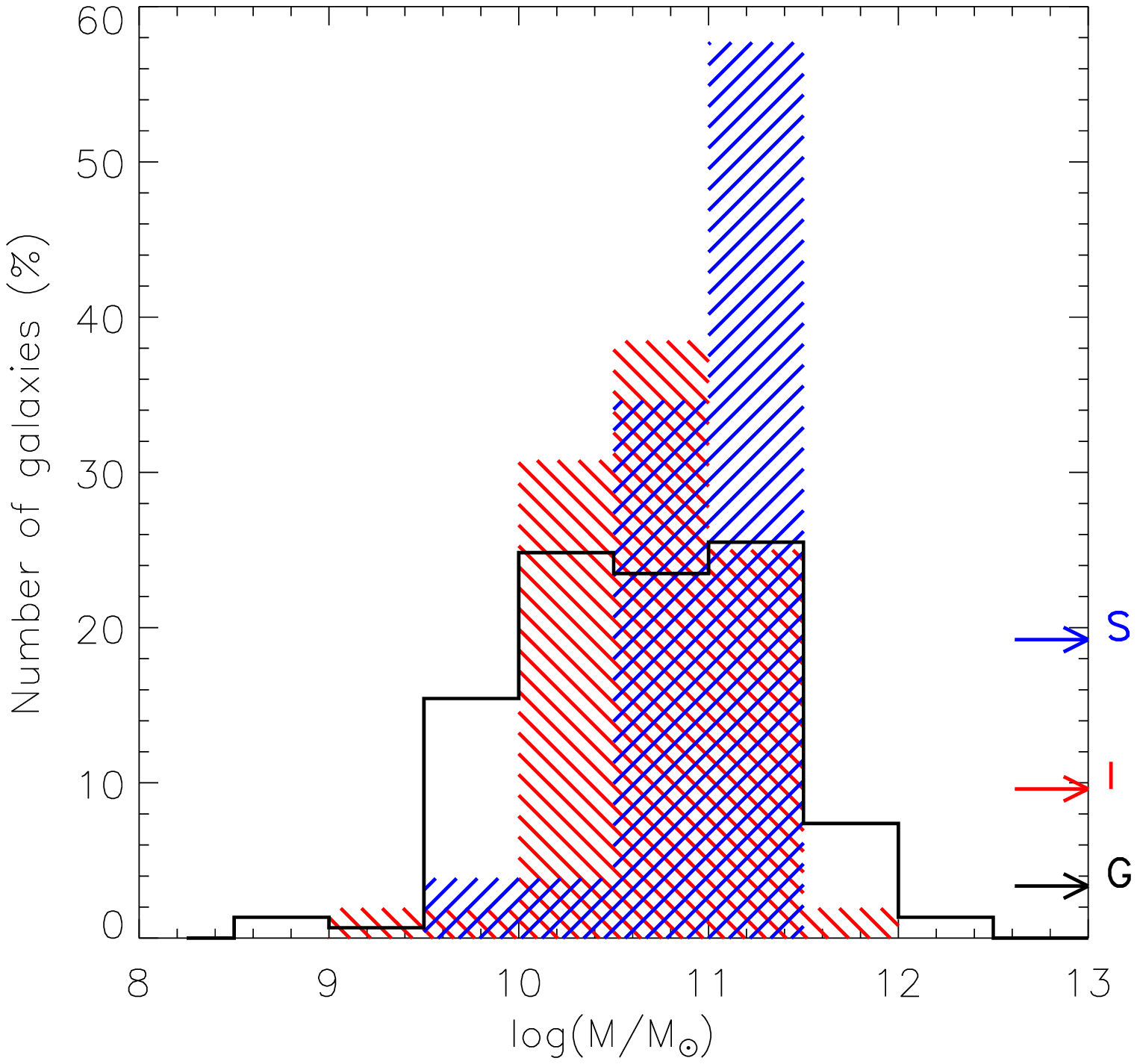}
\end{center}
\caption{Relative distribution of galaxy properties. Top: optical
radius; Middle: maximum rotation velocity; Bottom: masses.  The
black stairs indicates the GHASP local sub-sample, the red
hatchings the IMAGES sample and the blue hatchings the SINS
sample. In order to show the respective size of the samples (153,
63 and 26 galaxies respectively for GHASP, IMAGES and SINS),
arrows and letters with the same colors indicate five galaxies for
each sample (G for GHASP, I for IMAGES and S for SINS).}
\label{histo_masses}
\end{figure}

To avoid artifacts and to produce a realistic sample, only a
sub-sample of GHASP galaxies has been used to simulate galaxies at
high redshift.  Some galaxies coming from the GHASP sample have
been rejected. The selection criteria are described hereafter.

\emph{I.} Actual observations of \hz\ galaxies are limited in flux
and in size.  We have discarded too small and too faint galaxies
and ``uncomplete'' observations:

\emph{(i)} galaxies with an optical radius smaller than $3.2~kpc$
($3/4$ of the seeing at $z=1.7$);

\emph{(ii)} galaxies having less than 20 pixels after cleaning;

\emph{(iii)} galaxies with an integrated \Ha\ emission fainter than
$10^{-22}~W~m^{-2}$;

(iv) ``uncomplete'' observations, i.e. galaxies larger than the
\fov\ (see \citealp{Epinat:2008b}) (for which \Ha\ emission is
missed, the comparison has been made with \Ha\ images when
available from NED database) as well as galaxies showing a non
uniform \Ha\ emission due to filter transmission problems.

\emph{II.} Pair galaxies are analyzed separately when their
angular separation at high redshit is large enough (typically
0.5\arcsec) to clearly disentangle them. Only the two pairs UGC
5931/5935 and UGC 8709/NGC 5296 are presented on the same maps in
Appendix \ref{maps}.  Only UGC 8709 is analyzed since it is clear
on projected maps that NGC 5296 is only a small satellite. The
pair UGC 5931/5935 is the only one that could be interpreted as a
single galaxy at redshift $z\sim1.7$ thus the couple is analyzed as a
single galaxy, UGC 5931.

In summary, the sub-sample resulting from these criteria contains
153 galaxies (or close pair galaxies) among the 203 GHASP
galaxies. Thus, our sample contains 153 simulated high \SNR, high
spectral resolution ($\sim10$\kms), sky subtracted data cubes of
galaxies observed at a redshift $z\sim1.7$ under good seeing
conditions ($0.5''$ seeing) with a $0.125''$ spatial sampling.

\begin{figure*}
\begin{minipage}{180mm}
\begin{center}
\begin{tabular}{cc}
UGC 07901 (unambiguous case) & UGC 05414 \emph{(i)} \\
\includegraphics[width=8.75cm]{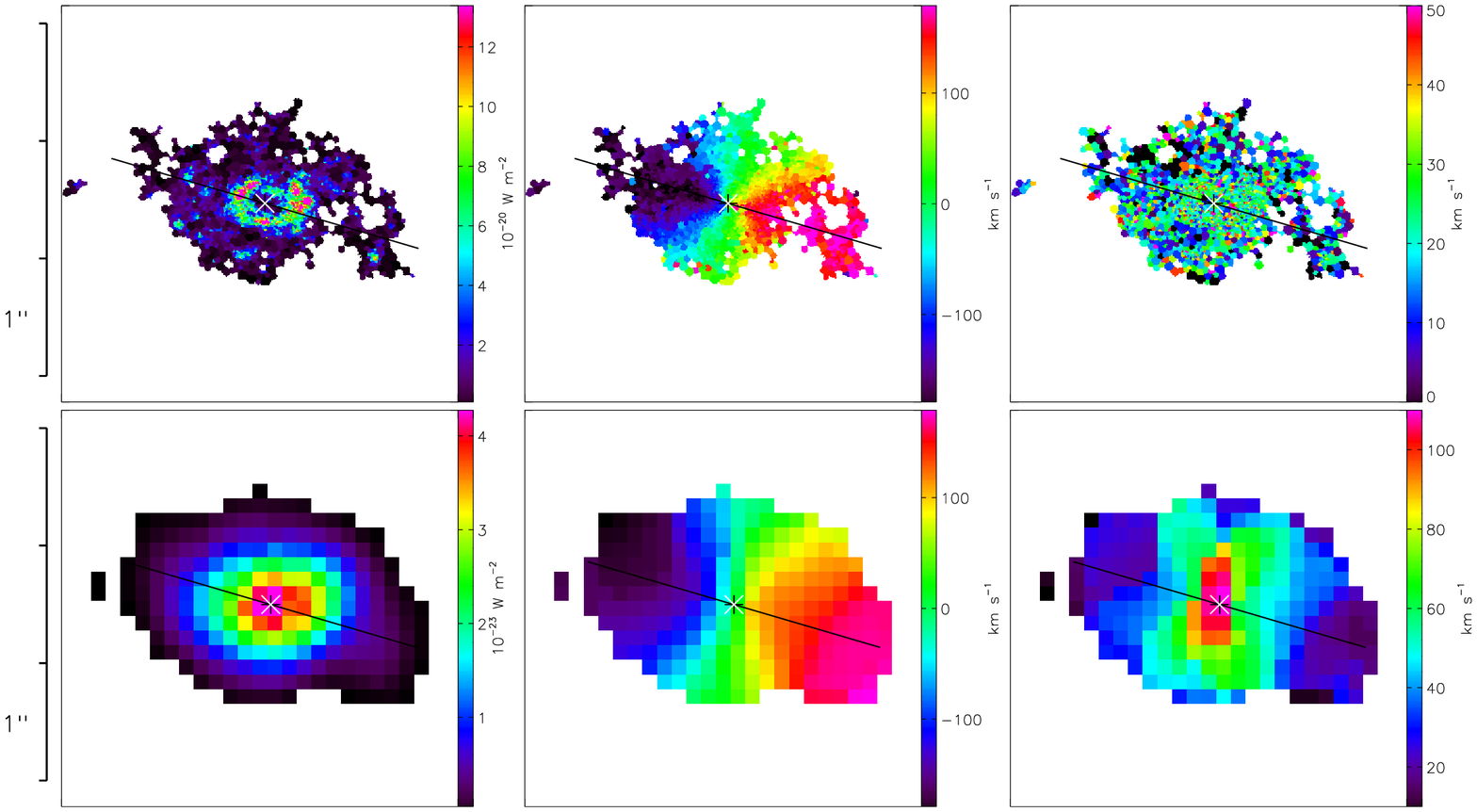} &
\includegraphics[width=8.75cm]{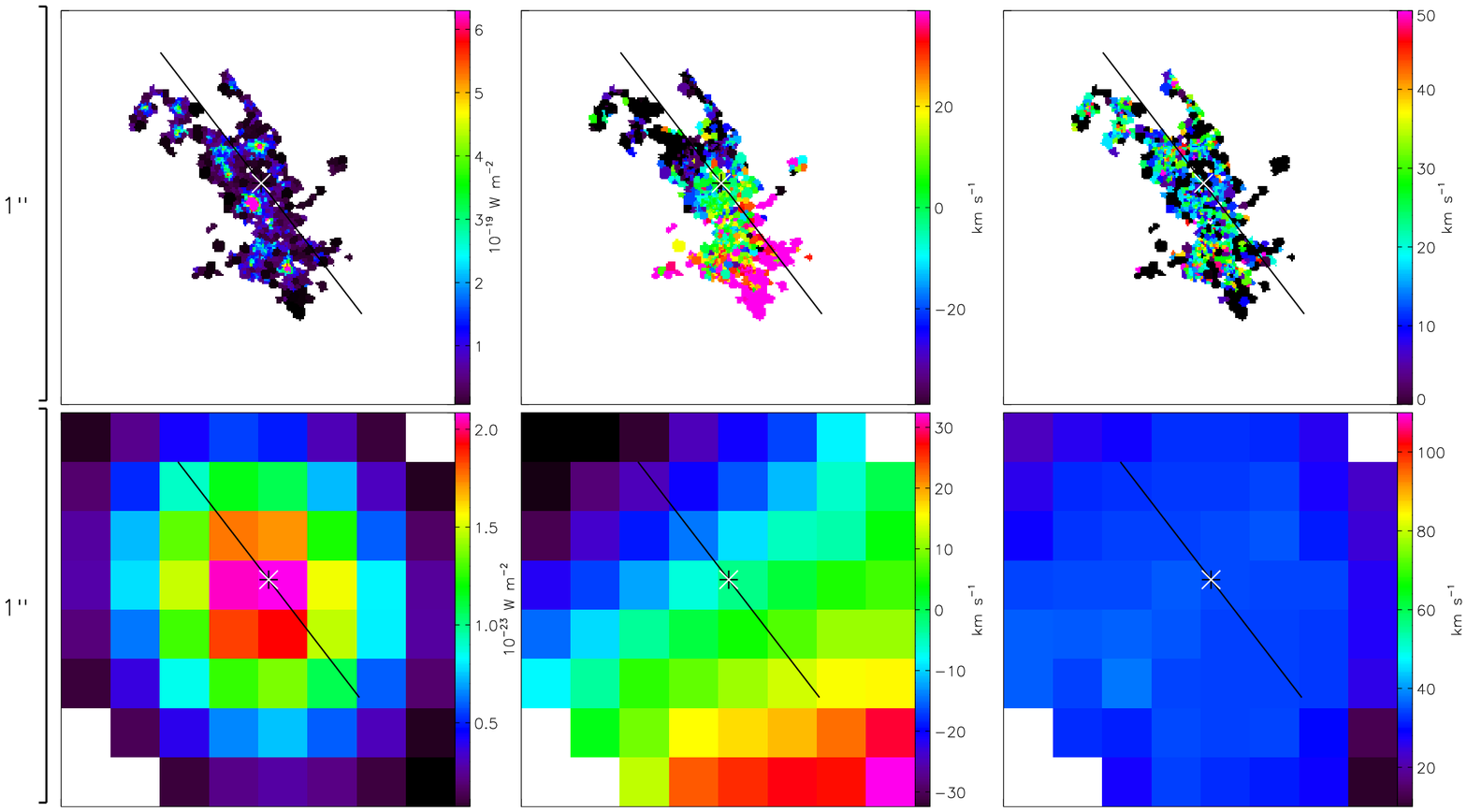}\\
UGC 07853 \emph{(ii)} & UGC 05789 \emph{(iii)} \\
\includegraphics[width=8.75cm]{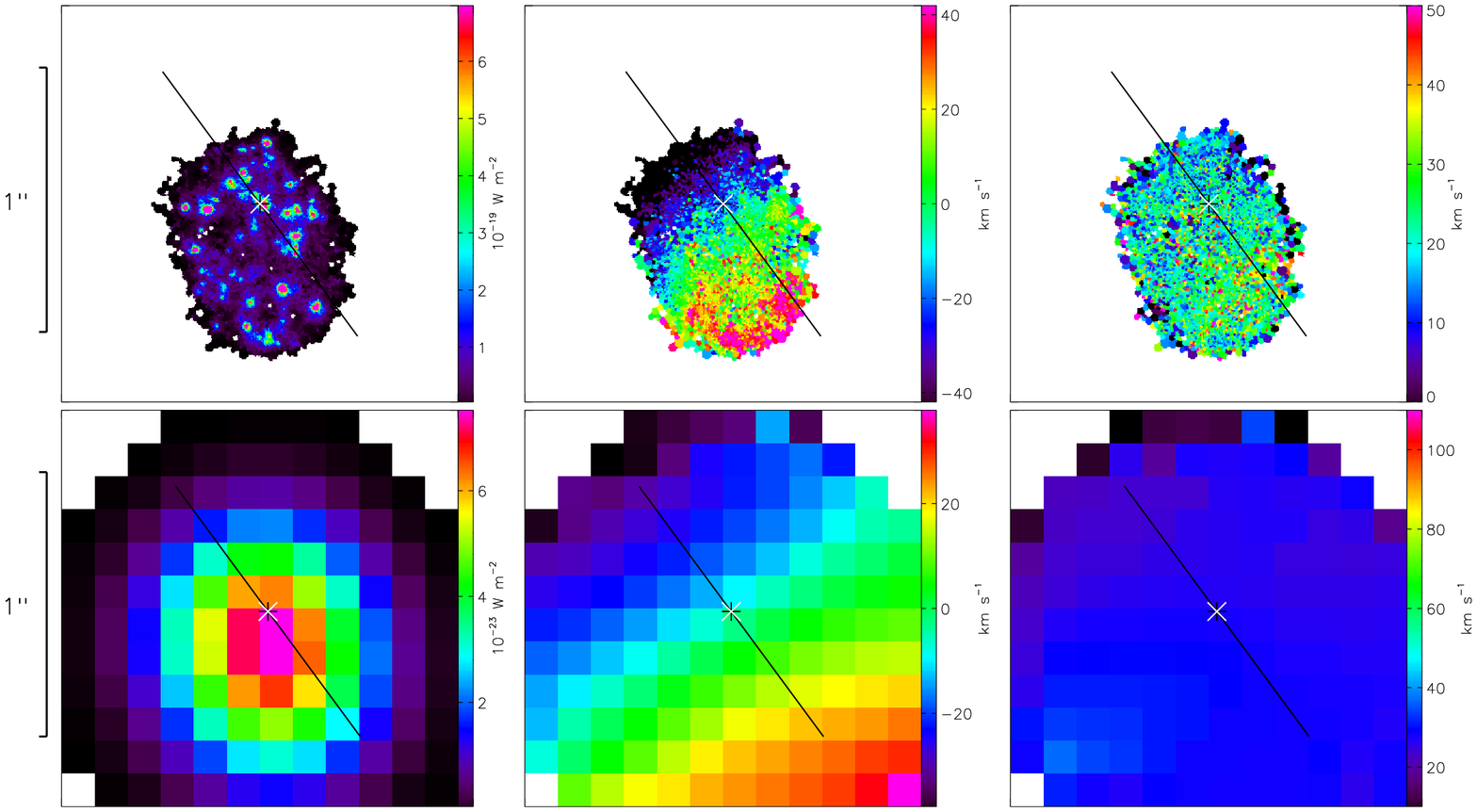} &
\includegraphics[width=8.75cm]{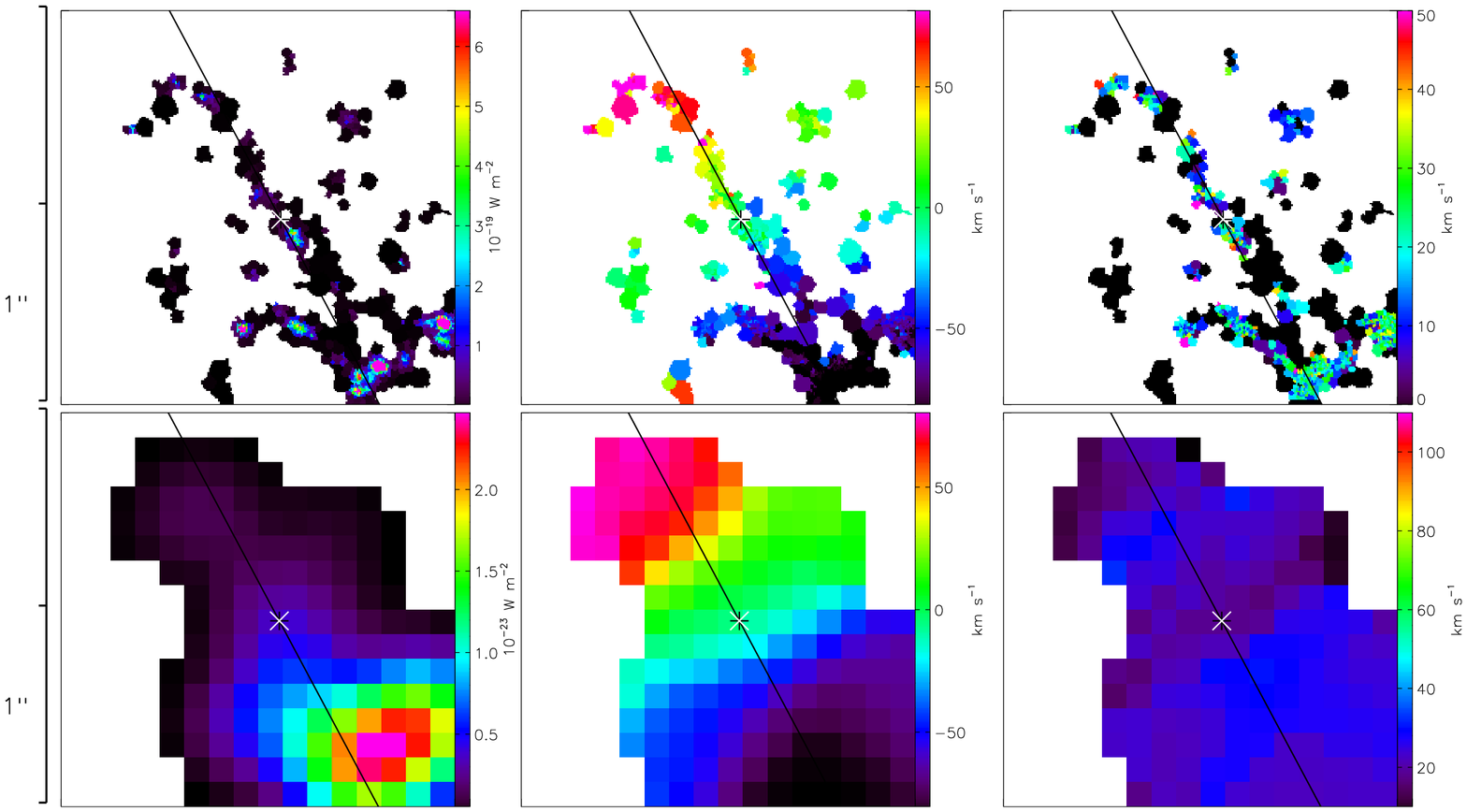}\\
UGC 10310 \emph{(iv)} & UGC 04820 \emph{(v)} \\
\includegraphics[width=8.75cm]{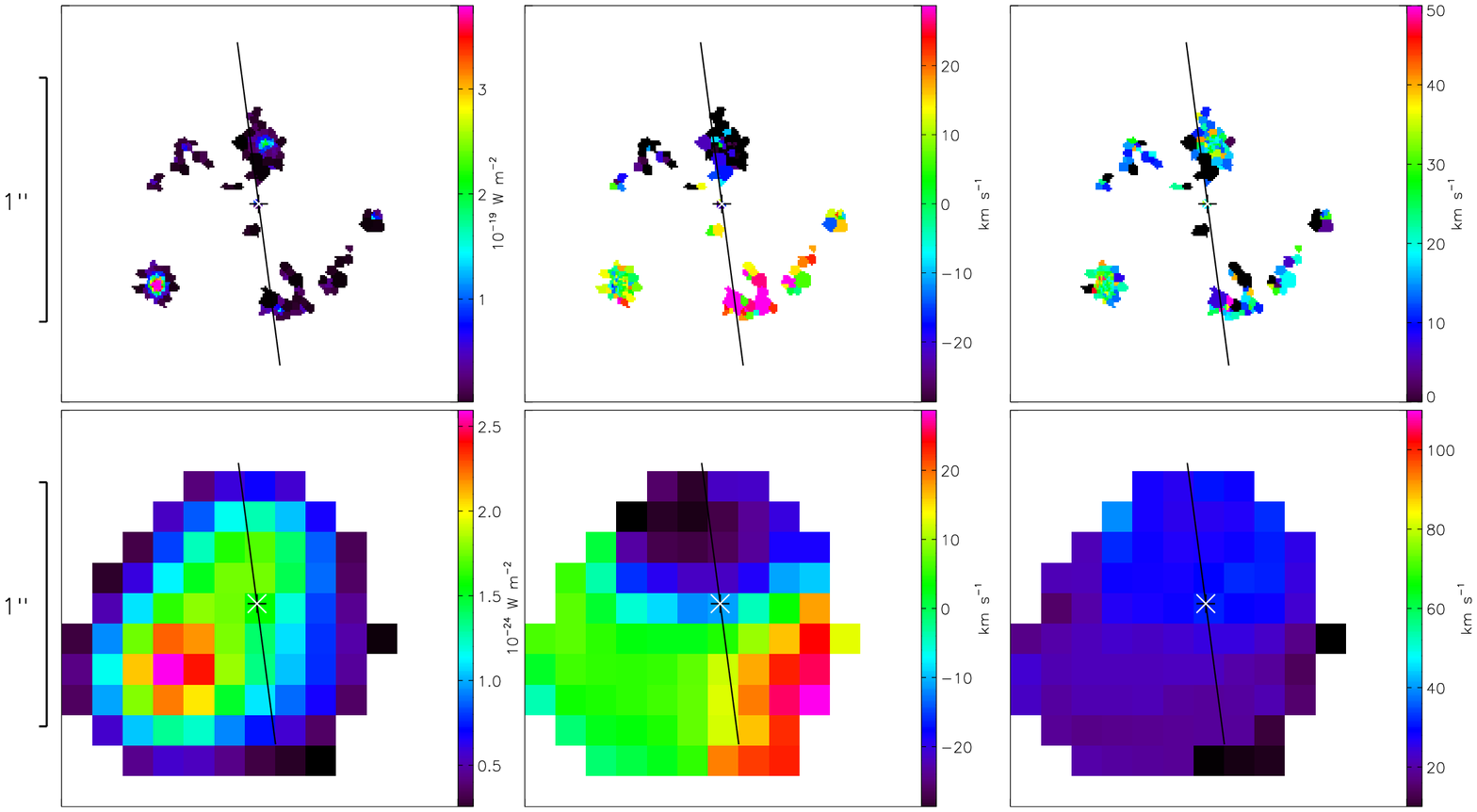} &
\includegraphics[width=8.75cm]{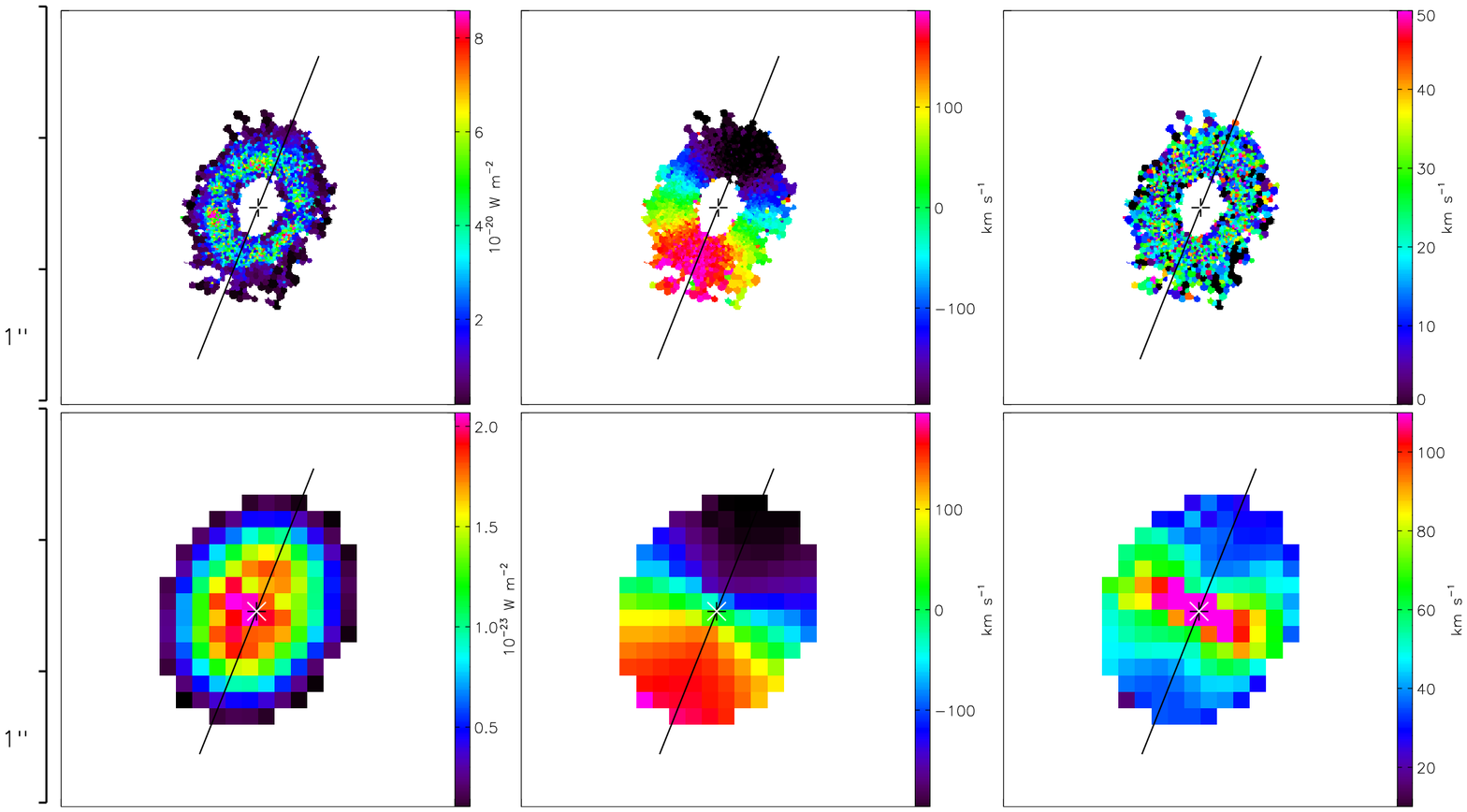}\\
UGC 05556 \emph{(vi)}  & UGC 05931 \emph{(vii)} \\
\includegraphics[width=8.75cm]{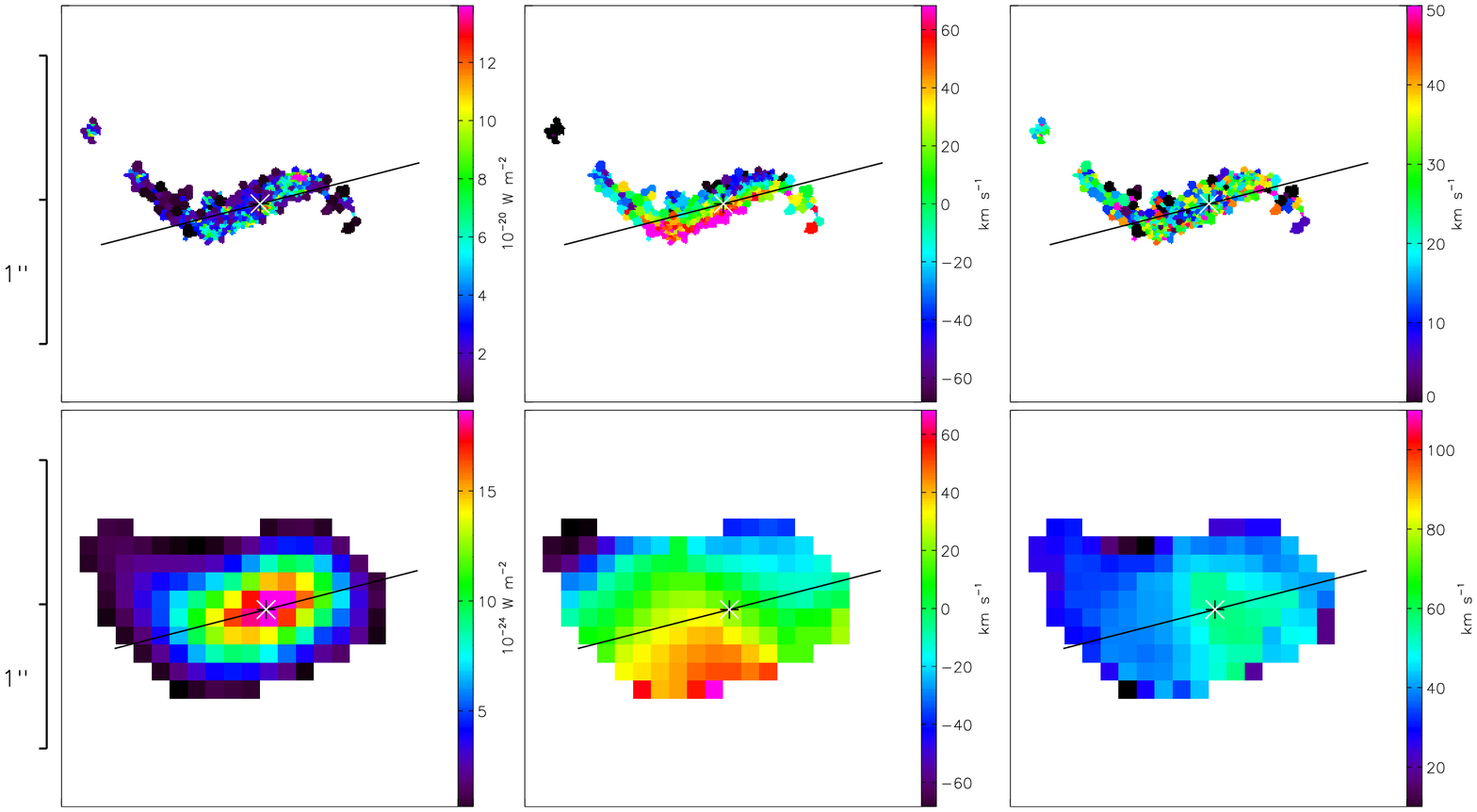} &
\includegraphics[width=8.75cm]{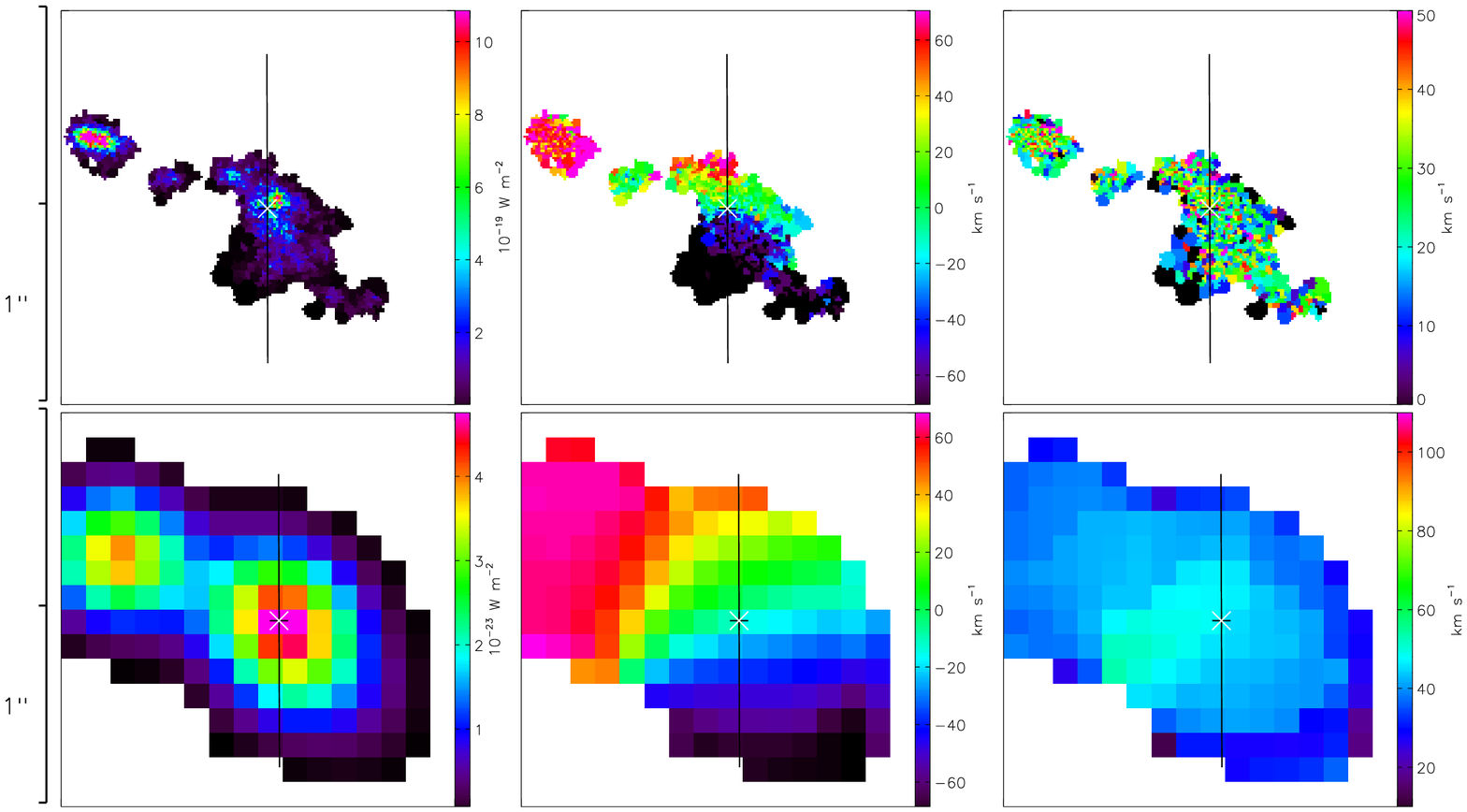}\\
\end{tabular}
\end{center}
\caption{Spatial resolution effects illustrated on eight galaxies illustrating an unambiguous case and the cases described from \emph{(i)} to \emph{(vii)} in section \ref{classificationflores}.
The following comments concern each galaxy. Top line: actual high resolution data at $z=0$. Bottom line: data
projected at $z=1.7$. The spatial scale is labelled in arcsecond
on the left side of both lines. From left to right:
\Ha\ monochromatic maps, \vfs\ and \vdms. The rainbow scale on the
right side of each image represents the flux for the first column
and the line-of-sight velocities corrected from instrumental
function for the two next columns. The black and white double
crosses mark the kinematical center at low redshift, while the
black line represents the major axis and ends at the optical
radius. More projected galaxies are presented in Appendix
\ref{maps}.} \label{mapsu7901}
\end{minipage}
\end{figure*}

Some examples of the original and blurred maps are given in Figure
\ref{mapsu7901}:
for each galaxy, the top line presents the actual maps already presented in \citet{Epinat:2008b,Epinat:2008a} whereas the bottom one corresponds to the blurred maps for the same galaxy projected at redshift $1.7$.
The whole set is presented in Appendix \ref{maps}:
the original XDSS image, as well as the blurred \ha\ flux,
\vf\ and the \vdms\ are given for each galaxy of the sub-sample. On
each map of Figure \ref{mapsu7901} and Appendix \ref{maps}, the white and black double crosses mark the center used
for the analysis while the black line represents the major axis
used or derived from the analysis. This line ends at the optical
radius taken from the RC3 catalog (see Table \ref{table_modz0}).

In Appendix \ref{rcz}, we present the \rcs\ of redshifted galaxies.
The black dots correspond to the \rc\ along the major axis (determined from high resolution data, see Table \ref{table_modz0}). The velocities are measured on the \vf\ for the pixels intercepted by the major axis and are deprojected from inclination.
The colored lines are the high resolution \rcs\ obtained from the models fit on the \vfs\ (see section \ref{fitting_method}).
The red-open triangles correspond to the high resolution
\rcs\ from \citet{Epinat:2008b,Epinat:2008a}. These authors have computed the \rcs\ from \Ha\ data cubes obtained from adaptive binning techniques based on Voronoi tessellations. Original improvements, based on the whole 2D \vf\ and on the power spectrum of the residual \vf\ rather than the classical method using fit in annuli or tilted ring model has been used to compute the \rcs.
The kinematical parameters (inclination, position angle, systemic velocity and center) were not allowed to vary with the radius.



\subsubsection{Distribution of the sub-sample}

Figure \ref{histo_masses} presents the relative distribution for
the three following galaxy parameters: optical radius
($D_{25}/2$), maximum rotation velocity ($V_{max}$) and total mass
($M$) computed within the optical radius for three different
samples.
\begin{equation}\label{masse}
    M=\frac{V_{max}^2\times~D_{25}/2}{G}
\end{equation}

These samples are: \emph{(i)} the GHASP sub-sample previously
defined (black stairs); \emph{(ii)} the IMAGES sample (red
hatchings) observed with FLAMES/GIRAFFE
\citep{Flores:2006,Puech:2006,Yangetal:2008,Neicheletal:2008,Puechetal:2008}
and \emph{(iii)} the 26 galaxies from SINS sample (blue hatchings) for which these measurements are available so far, and that are mainly classified as rotating disks \citep{Forster-Schreiberetal:2009,Cresci:2009}. For comparison, the total
amount of galaxies of each sample being different (153 for GHASP,
63 for IMAGES and 26 for SINS), we have marked on
the histograms of Figure \ref{histo_masses} a reference level of
five galaxies for each sample with arrows of the same color as the
histograms (G for GHASP, I for IMAGES and S for SINS). The
GHASP local sample contains galaxies over a broader mass range
resulting from larger galaxies and slowest rotators than the two
other samples. The lack of very large galaxies at high redshift
can be explained by both evolution effect and observational biases
due to a poorer \SNR, inducing underestimated radii.
We may also notice that GHASP barred galaxies are on average
smaller than unbarred galaxies. This biases the comparison that
can be done between barred and unbarred galaxies since we expect
the parameters determination accuracy to be correlated with the
size of redshifted galaxies. The bias induced between barred and
unbarred GHASP galaxies does not affect the global comparison with
\hz\ galaxies. Moreover, even if high redshift and local
distributions are different, the simulated maps are suited for
studying biases in the kinematical parameters determination since
the GHASP sub-sample covers the whole mass, extent and velocity
ranges observed at high redshift.
It is however interesting to notice that almost no high redshift galaxies {from both IMAGES and SINS samples} are slow rotators even if they are on average smaller objects. This is probably due to magnitude selection effects and could indicate that no \ha\ is detected in the outer regions (nevertheless, \citealp{Cresci:2009} found a good agreement between the radii measured in $K$-band and in \Ha). Moreover, high redshift samples are not selected in a statistically complete way since they aim at observing galaxies with resolved kinematics.

\subsubsection{\Vf\ extent}
\label{zorglub}

As already underlined in paragraph \ref{blurr},
our redshifted sample benefits from a high \snr, thus, our \vfs\ are
probably more extended than what observation facilities would
enable for real high redshift observations. The extent of local
\vfs\ is close to the optical radius value as underlined by
\citet{Garrido:2005}. The mean value of optical radius for our
GHASP sub-sample is $11~kpc$ (median value is $9~kpc$), with a
dispersion of $7~kpc$. The lowest value is $3~kpc$ and the highest
value is $35~kpc$. 
For comparison, we have converted half light radii ($r_{1/2}$) taken from the literature for high redshift objects into optical radii ($r_{opt}$) assuming an exponential distribution of light: $r_{opt}=1.9r_{1/2}$.
\par
IMAGES galaxies observed with FLAMES/GIRAFFE in the
redshift range $0.4<z<0.75$ by \citet{Neicheletal:2008}
have a mean optical radius of $9~kpc$ with a scatter of
$\pm4~kpc$, which is comparable to our sample. Their smallest galaxy
is $2.9~kpc$ and the largest is $19.5~kpc$.
On average, the 16 galaxies observed with OSIRIS by
\citet{Law:2009} with redshifts from 2 to 3 extend up to $1.1\pm0.3~kpc$. These values are very low. This could be partially attributed to the different estimators. Indeed, the disk
dimensions are deduced from the ionized gas flux map, which is
not completely suitable for comparison.
The four $z\sim1.5$ redshift galaxies observed by
\citet{Wright:2007} with OSIRIS using AO extend up to $4.9\pm1.1~kpc$ in optical radius.
\citet{Forster-Schreiberetal:2006,Forster-Schreiberetal:2009} and \citet{Cresci:2009} have provided half light radius measurements for 26 galaxies (mainly for rotating disks) out of the 63 SINS galaxies. The mean optical radius is $9.1\pm3.3~kpc$, the smallest galaxy radius is $3.2~kpc$ and the largest one is $14.5~kpc$, which is still slightly smaller than for the GHASP sample.
The nine galaxies with redshift ranging between 1 and 1.5 presented by \citet{Epinat:2009c} have optical radii of $10.3\pm4.1~kpc$. The sizes are ranging from $5.3~kpc$ to $17.3~kpc$.
Except for \hz\ galaxies observed with the OSIRIS
instrument that uses AO facility \citep{Law:2009,Wright:2007,Wright:2009}, the extent of high redshift galaxy
\vfs\ is rather similar to the ones of our sub-sample. However, there is no case for galaxies larger than $20~kpc$ as already
noticed from the histogram in Figure \ref{histo_masses}.
The smaller extent of observations with AO facility could be explained by the use of a very small pixel scale (50 mas for both OSIRIS and SINFONI in AO mode) that induces a loss in flux detection. Indeed, for constant surface brightness objects, it is necessary to use longer exposures when using a smaller pixel scale to reach a given \snr, even with a negligible read-out noise.

On the other hand, due to selection criteria effects on high
redshift sample, we would expect to observe large galaxies but
evolution processes have the opposite effect. In conclusion, since
local data have a better \snr\ and on average a larger spatial
extent, in section \ref{analysis}, we have truncated the images of
all the galaxies at the optical radius to mimic \hz\ galaxies. However, the maps presented in Appendix \ref{maps} are not truncated.

\subsection{Biases induced by spatial resolution effects}
\label{biasesproj}

At redshift $z=0$, the use of optical spectroscopy is the best way
to probe the inner shape of \rcs\ since the inner regions are
usually not well resolved with HI radio observation (for GHASP
data already observed in HI in the WHISP survey by
\citealp{Noordermeer:2005}, the typical resolution is
$\sim5~kpc$). Optical \rcs\ are not always extended enough to
determinate reliable maximum velocities \citep{Garrido:2005}.
Complementarily, HI data are used to trace the outer regions of
\rcs\ since HI generally extends further away. At high redshift,
the situation regarding the spatial resolution in optical or in
infrared becomes comparable to HI at local redshift, but still with
a smaller extent. Thus, the biases due to spatial resolution effects
for our sample are somewhat similar to HI beam smearing effects
for local galaxies (see section \ref{BeamSmearingParameter} for a discussion on the
beam smearing parameter). Our projected sample gives a good
opportunity to revisit these biases, and to point out specific
biases in the optical or in the infrared since we exactly know how
the high resolution kinematical maps look like.


\begin{figure}
\begin{center}
\includegraphics[width=8.5cm]{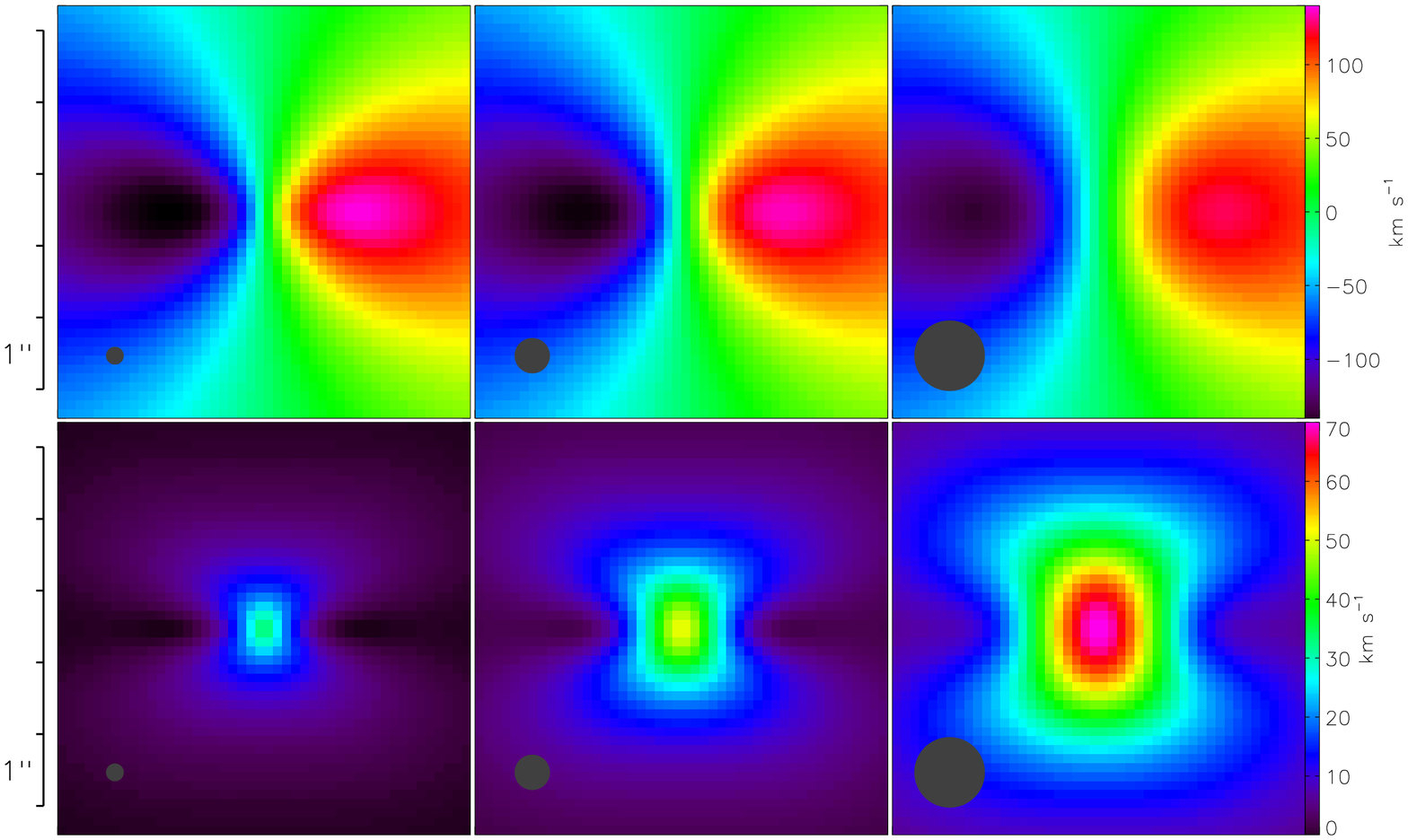}
\end{center}
\caption{Beam smearing effects on a simulation (\vfs\ on the top
line and \vdms\ on the bottom line) depending on increasing
blurring parameter. From left to right, the seeing (represented by
a dark disk on the six images) increases from 0.25\arcsec\ to
1\arcsec. The pixel size is 0.125\arcsec. The disk scale length is
set to 5 kpc (observed at $z=1.7$), the inclination is 45\deg\ and
the maximum velocity in the plane of the disk is 200\kms.}
\label{6VF}
\end{figure}

From the comparison between the original and redshifted maps given
in Figure \ref{mapsu7901} in the case of UGC 07901 (top-left), we note that:

\emph{(i)}  The apparent size of the galaxy seems to be enlarged
while in fact, flux limits reduce it. Indeed, the emitting regions
in the blurred images are artificially extended toward outer
regions of the galaxy where there is in fact no emission. This is
due to beam smearing that spreads out the flux over the PSF. In
actual observations, depending on the \snr, these faint outer
regions should not exist.

\emph{(ii)} Concerning the \Ha\ monochromatic map, we totally lose
the details of the inner ring distribution and the emission is
only present in the central peak of the blurred images.

\emph{(iii)} The velocity gradient is lowered along the major axis
while the velocity gradient is increased across the minor axis.
Indeed, both \vfs\ nevertheless present the usual ``spider'' shape.
However, the isovelocity lines are more open for the \hz\ galaxy than
for the $z=0$ galaxy.
If one does not take into account the beam smearing, this could be interpreted
as a lower inclination for the redshifted galaxy (the same
conclusion would be reached by looking at the morphology due to
the fact that the relative enlargement is higher for the minor axis than for the
major axis).


\emph{(iv)} The \vdms\ are quite different. The bottom-right map shows the velocity dispersion affected by beam smearing, the top-right map displays the velocity dispersion for each point of the galaxy, referred hereafter as the local velocity dispersion and noted $\sigma$. We aim to measure this quantity in order to estimate the pressure support for both nearby and distant galaxies.
The local velocity dispersion does not display any strong feature whereas the \vdm\ at
high redshift clearly shows a central peak elongated along the
minor axis.
As already discussed by other authors (e.g. \citealp{Weiner:2006, Flores:2006}), this peak is only due to beam smearing effects: for
each pixel, the resulting line is the combination of lines at
various wavelengths (velocities) weighted by the real flux and is
thus enlarged. The enlargement is maximum where the projected
velocity gradient is the highest (see Appendix \ref{model} for details). 

\begin{figure}
\begin{center}
\includegraphics[width=8.5cm]{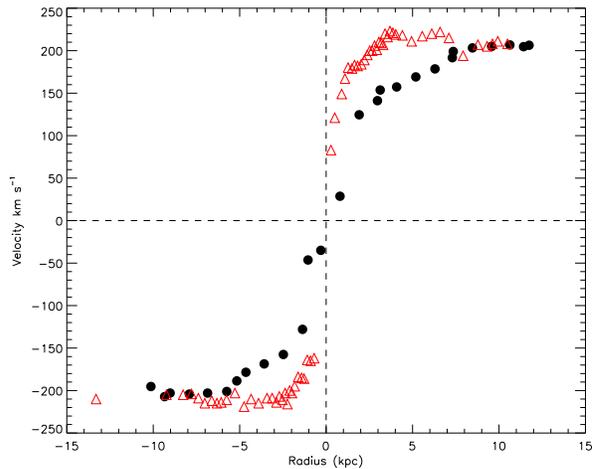}
\end{center}
\caption{Example of a \rc\ obtained for a redshifted galaxy. Both \rcs\ have been computed from both local and projected UGC
7901 \vfs\ presented in
Figure \ref{mapsu7901}: red-open triangles
correspond to local full resolution data while black dots
come from the data projected at $z=1.7$.} \label{rcu7901}
\end{figure}

Both redshifted \vf\ and \vdm\ contain information on the true \vf\ itself. 
In Figure \ref{6VF}, in order to illustrate this effect that is responsible for both points \emph{(iii)} and \emph{(iv)},
an exponential disk model has been drawn in order to
compute \vfs\ and \vdms\ with increasing seeing ranging from
0.25\arcsec\ to 1\arcsec. The disk scale length has been set to
$5~kpc$ and the maximum velocity of the \rc\ to 200\kms. The
inclination has been fixed to 45\degr. The flux contribution follows an
exponential disk, and the local velocity dispersion $\sigma$ is null
everywhere. We observe that
the velocity shear vanishes whereas the velocity dispersion peak
increases. The behavior would be the same with an increasing pixel
size or with a decreasing disk scale length. If the local velocity dispersion has a constant value $\sigma$ in the field, the resulting \vdm\ is the quadratic sum of $\sigma$ with the previously computed blurred \vdm. It results that the peak is more attenuated for galaxies with a high local velocity dispersion.

In addition to these effects on the maps, the beam smearing will modify the shape of the \rc, which
will eventually look like a solid body \rc. This is illustrated in
Figure \ref{rcu7901} where the \rc\ at low redshift (red-open
triangles) is over-plotted on the \rc\ derived from the major axis
of the redshifted \vf\ (black dots). At high redshift, the inner velocity gradient is
lowered whereas the outer gradient becomes higher. It can be
noticed that the maximum velocity seems to be reached at larger
radii (around $8~kpc$ instead of $3~kpc$) on both the \vf\ and the
\rc\ of the projected galaxy. This is due to the fact that, for this specific ring galaxy, the blurred
\Ha\ distribution is dominated by the contribution of
the ring. Since it is close to the center, the velocity of the
ring has a strong weight and is reached rapidly. At larger radii,
the contribution of the \Ha\ ring remains important and tends to lower
the plateau. For most of the nearby galaxies observed at high spatial resolution with a flatter
\Ha\ distribution, the inner slope is also shallowed (see Appendix
\ref{rcz}, e.g. UGC 11872). For galaxies with a lower extent, the
maximum velocity is not reached on the \rc\ (see Appendix
\ref{rcz}, e.g. UGC 528). With HI data at low redshift, this would
not be true since, the extent being larger, the external plateau
could be reached. However, mostly for massive spiral
galaxies, we see that the maximum velocity is reached close to the
center in \Ha, the \rc\ may even be decreasing afterward.
\citet{Epinat:2008a} suggested that this could be a possible
explanation for the difference observed in \TF\ relations obtained
from \Ha\ and HI data for the most massive galaxies.


\section{Kinematical signatures of high redshift rotating disks}
\label{classificationflores}

\subsection{Kinematical classification}
\label{classificationkin}

The \vdm\ feature discussed in the previous section is typical of a rotating disk with a rather
uniform flux distribution and with a projected velocity gradient
larger than 100\kms, due to the strong inner velocity gradient. \citet{Flores:2006} use this signature to
provide a dynamical classification for \hz\ galaxies (for
$z\sim0.6$): ``rotating disks'' present a central velocity
dispersion peak, ``perturbed rotators'' show a
peak (slightly) offset from the center and objects having ``complex
kinematics'' (e.g. mergers) display  featureless \vdms.
%
%
The GHASP sample contains mainly rotating disks, thus we can use it to probe
this classification. We find that around 70\% of the sample would be correctly
classified (i.e. entering in the category ``rotating disks'').
Nevertheless, the remaining fraction of the sample would be
misclassified for the following reasons (see Figure \ref{mapsu7901} for illustrations of each case):

\emph{(i)} disks in rotation with a low velocity gradient
(face-on, low mass galaxies, high velocity dispersion in the
\vf\ with respect to the rotation velocity amplitude) show a very
faint or no central velocity dispersion peak (see Appendix
\ref{maps}, e.g. UGC 3685, UGC 3851, UGC 5414 -Fig. \ref{mapsu7901}-, UGC 6628, UGC
11557);

\emph{(ii)} disks showing a solid body \rc\ have the same velocity
gradient everywhere in the field and thus no peak of velocity
shear can be observed in the \vdm\ (see Appendix \ref{maps}, e.g.
UGC 6419, UGC 7853 -Fig. \ref{mapsu7901}-);

\emph{(iii)} asymmetries in the \Ha\ distribution can induce an
offset velocity dispersion peak (hence misclassified as
``perturbed rotators'') since the resulting \vdm\ is the
combination of \vf\ shears weighted by the \Ha\ monochromatic flux
(see Appendix \ref{model} and Appendix \ref{maps}, e.g.
UGC 4393, UGC 5316, UGC 5789 -Fig. \ref{mapsu7901}-);

\emph{(iv)} galaxies with a patchy \Ha\ emission seem to have a
continuous emission once projected at high redshift from which can result
peculiar \vfs\ and \vdms\ (see Appendix \ref{maps}, e.g. UGC 10310 in Fig. \ref{mapsu7901});

\emph{(v)} a central hole (or ring) in the flux distribution can
be completely blurred depending on the actual size of the galaxies
(see Appendix \ref{maps}, e.g. UGC 3382, UGC 4820 -Fig. \ref{mapsu7901}-, UGC 5045);

\emph{(vi)} the presence of a strong bar can induce very peculiar
\vfs\ with an apparent \pa\ of the major axis completely biased (see
Appendix \ref{maps}, UGC 5556 being the most impressive case -Fig. \ref{mapsu7901}-);

\emph{(vii)} using only broad band images, very close pairs (see
Appendix \ref{maps}, e.g. UGC 5931 \& UGC 5935 in Fig. \ref{mapsu7901}) can appear as a
single galaxy with two main clumps. Kinematical data are helpful
to distinguish single galaxies from systems composed by two or
more galaxies. A paired galaxies system or even a compact group of
galaxies may look like a single perturbed galaxy when they are in
fact composed of distinct galaxies in interaction or just seen
close in projection on the sky plane. Reciprocally, chaotic single
galaxies composed by bright clumps may look like multiple systems.
In a given \FOV, multiple galaxies can be identified using the
discontinuities in the velocity gradients, the variation of the
major axis \pa\ and the possible multiple components along the
line-of-sight in the line profiles (e.g. \citealp{Amram:2007}).
Velocity discontinuities are obvious when the different galaxies
are rotating in apparent opposite directions but are also visible
when the galaxies are rotating with the same apparent spin. Within
a given pixel, multiple components in the line profiles can be
identified by the relative difference in velocities and often also
by difference in flux ratio. In the case of UGC 5931/35 the
\vf\ looks disturbed even though the actual \vf\ is more regular,
the \pa\ of the major axis is biased and the velocity dispersion
signature of a rotating disk is partly lost.

In addition to these effects, this classification cannot be used for galaxies with a high local velocity dispersion since the peak in the \vdm\ is smoothed.

\subsection{IMAGES classification}

GIRAFFE observations, in the frame of the IMAGES program
\citep{Yangetal:2008,Neicheletal:2008,Puechetal:2008,Rodrigues:2008}, provided a sample of 63 galaxies (including
those of \citealp{Flores:2006} and \citealp{Puech:2006}) ranging
from $z=0.4$ to $z=0.75$ representative of the population of
emission line galaxies more massive than $1.5 \times 10^{10}M_{\odot}$
(see Figure 6 in \citealp{Yangetal:2008}).
In this sample, \citet{Yangetal:2008} found 32\%\ of regular ``rotating disks''. A lower limit of the number of ``anomalous kinematics (pertubed and complex)'' galaxies can be given considering that absorption line galaxies are not perturbed. \citet{Yangetal:2008} estimated that absorption line galaxies represent 40\%\ of the total population of galaxies at $z\sim0.6$.
Thus, taking into account all the galaxies (emission and absorption line galaxies) in that redshift
range, these authors found that at least $41\pm7$\% of them have
anomalous kinematics (not relaxed), including $26\pm7$\% with
complex dynamics (not simply pressure or rotationally supported).
The merger hypothesis is favored by these authors to explain this
complex dynamics. Even if the condition of projection of the local
GHASP sample of galaxies presented in this paper is built to match
the SINFONI observations rather the GIRAFFE ones,
a comparison between local galaxies and galaxies at intermediate
redshift (IMAGES/GIRAFFE) may also be done. Indeed, the seeing
conditions (without AO) are statistically the same,
the sizes of the galaxies do not dramatically differ between
redshift $z=1.7$ and $z=0.6$ (at $z=1.7$, 1\arcsec $\sim$ $8.6$
kpc and at $z=0.6$, 1\arcsec $\sim$ $6.7$ kpc, see Figure \ref{scale}), the main
difference is the sampling of the seeing on the CCD, the one of
SINFONI ($0.125$\arcsec) being approximately four times higher than
that of GIRAFFE ($0.52$\arcsec). Nevertheless, the spectral
sampling is higher in GIRAFFE ($22-30$\kms) than in SINFONI
($67-160$\kms) but lower than in GHASP ($\sim17$\kms).
On the one hand, from the comparison between IMAGES and GHASP, it can be concluded that
actual disks in rotation with emission lines at intermediate redshift look like local
disks in rotation projected at high redshift but the absence of perturbed disks in the
local sample does not allow to conclude if perturbed disks at
intermediate redshift look like perturbed local galaxies.
On the other hand, due to the items developped in section \ref{classificationkin} (ordered from the most to the least relevant), we have shown that 30\%\ of the rotating disks may be misclassified using the classification given by \citet{Flores:2006}. At high redshift, this is particularly critical for galaxies where noise in the outer parts of the velocity field causes off-center dispersion peak. The ``corrected'' number of rotating disks in IMAGES sample of 63 galaxies may be underestimated by a factor 1.4. In other words, the fraction of rotating disks found in IMAGES may pass from 32\%\ (see above) to 44\%.
Reciprocally, the fraction of galaxies with anomalous kinematics for the total population, including absorption and emission line galaxies, may thus be lowered from 41\%\ (see above) to 33\%. This gives a lower limit to the fraction of galaxies having anomalous kinematics. Indeed, it is likely that a fraction of absorption line galaxies ara perturbed and also have anomalous kinematics.
In addition, based on the observed dynamics in the IMAGES survey
and the possible misclassification due to the faint spatial
sampling (no AO and large pixel scale) combined to the small spatial
coverage (due to the small sizes of the galaxies) and the low SNR
in some cases, the anomalous kinematics and even the complex
dynamics for several galaxies could be due to unrelaxed gas disk
without involving, in all the cases, a merger. Indeed,
\citet{Liang:2006} estimated that the gas content in intermediate
galaxies at $z\sim0.6$ was twice larger than in galaxies at
the current epoch and that one cannot exclude transient
episodes of intense gas accretion making the disk unstable during a
relatively short period.

To conclude, the kinematical classification made by \citet{Flores:2006} is relevant for a reasonable fraction of rotating disks, assuming that the local velocity dispersion is lower than the rotation velocity. However, low velocity gradient in the \vf, solid body shape for the \rc, flux asymmetries in the \ha\ distribution and other asymmetries like strong bars could cause the IMAGES sample to look more pertubed than it actually is.


\section{Fitting method}
\label{fitting_method}

\subsection{General model}
\label{fitting_method_general}

To recover the actual kinematic parameters (those from high
resolution data) through the degenerate blurred data cube, it is
absolutely necessary to model the blurred data.
Models consisting in a thin planar disk have
been used to retrieve (i) the projection parameters (inclination $i$ with respect to the line of sight,
\pa~of the major axis $PA$ and systemic velocity $V_{sys}$) and (ii) the kinematical parameters (center of rotation, rotation velocity and local velocity dispersion $\sigma$ both as a function of the radius). No hypothesis is done on the nature of the gravitational support (rotation or pressure). The only assumption we do is that the gaseous disk is infinitely thin without any supposition on the amplitude of the velocity dispersion.
To constrain the kinematical parameters, this general model allows the use of the blurred \vfs\ alone, the blurred \vdms\ alone or the combination of both.

\subsection{Method used}
\label{fitting_method_used}

In the following, we have only used the blurred \vfs. A discussion on this choice is provided in section \ref{fitting_method_dispersion}.
The \vf\ is supposed to be
axisymmetric and the \rc\ is described by two parameters: the
maximum velocity of the model $V_t$ and a transition radius $r_t$.
To avoid any $a~priori$ shape for the
\rc\ describing the redshifted data, we have tested four different
models of \rc\ in order to evaluate which ones describe at best the
data. These four models all have two free parameters ($r_t$ and $V_t$). We have chosen \rcs\ that have been used for such studies in the literature and which may have rising,
flat or decreasing shape: (1) an exponential disk as used for the SINS sample \citep{Cresci:2009}; (2) an
isothermal sphere as used in mass models \citep{Spano:2008}; (3) a model described by an inner linear slope
to reach $V_t$ and a plateau after $r_t$ (referred hereafter
as ``flat model'') as used for OSIRIS data \citep{Wright:2007,Wright:2009}; and finally (4) a model described by an
arctangent function as used for the IMAGES sample \citep{Puechetal:2008}. The two first models may have a physical
meaning, the two last are well known to fit \rcs\ of local
galaxies. Except for the arctangent model, the maximum velocity of the model $V_t$ is reached at the transition radius $r_t$. Ideally, to increase the flexibility of the fit, it
should be useful to use a \rc\ described by three parameters,
but the addition of one more parameter makes the fit difficult to
converge since the number of free parameters is already of the
same order than the number of data measurements. These models are
described in Appendix \ref{model} (section \ref{rcmodels}) and
illustrated in Figure \ref{models} (using $r_t=11~kpc$,
$V_t=190~km~s^{-1}$ and $i=45$\degr).
In Appendix \ref{rcz}, we have over-plotted the four models to the
\rcs\ (exponential disk in red, isothermal sphere in green, ``flat
model'' in black, and arctangent function in blue).
Thus, the global model contains seven parameters ($i$, $PA$, $V_{sys}$, $V_t$, $r_t$ and the center coordinates).
They are determined from a Levenberg-Marquardt nonlinear least-squares $\chi^2$ minimization \citep{Press:1992} and the statistical errors of the fits have been used (see Tables \ref{table_modzexp} to \ref{table_modzata}).
Since a simple thin rotating
disk model is not suited for the description of highly inclined
disks, we set an upper limit of $80$\degr~to the inclination.
Moreover, due to the degeneracy between the velocity and the
inclination, we set a minimum inclination of $10$\degr~to avoid
unrealistically high rotation velocities. It results that 16
galaxies have been excluded from the fitting and thus only 137 out
of the 153 galaxies of our sub-sample have been used for the
studies presented hereafter.

\subsection{Computing method and limitations}
\label{fitting_method_computing}

From the model parameters previously defined, a high resolution
\vf~model is created. Then, the seeing has to be taken into
account. In order to do that, ideally, one should know the high resolution
line flux map and create a high resolution data cube.
Indeed, the line flux weights the contribution of each high resolution spatial element.
From observations, it is not yet possible to know the high resolution line flux map. One solution is to use flux distribution models. However, the GHASP local dataset shows that such assumption is abusive since some galaxies display rings, asymmetry or holes. An other solution would be to perform deconvolution from the observed maps. In this study, we simply use the low
resolution line map that we interpolate. The method we adopted is more robust than deconvolution technics, but will not recover holes, rings, asymmetries, etc. However, the seeing blur will decrease their effect.
Creating a model data cube is a time consuming task. It is possible to avoid the creation of high resolution data cubes by assuming that the \Ha\ line is locally well described by a gaussian.
This formalism enables to compute directly the blurred \vf\ and
\vdm\ from the seven parameters of the model and is equivalent to generate high resolution data cubes that also need the same assumption. Analytical details are presented in Appendix \ref{model}.
In equation \ref{dispersion_lowres}, giving the expression of the blurred velocity dispersion, the first term represents the local velocity dispersion contribution whereas the second term corresponds to a velocity shear feature induced by beam smearing effects.

\begin{figure}
\begin{center}
\includegraphics[width=8.5cm]{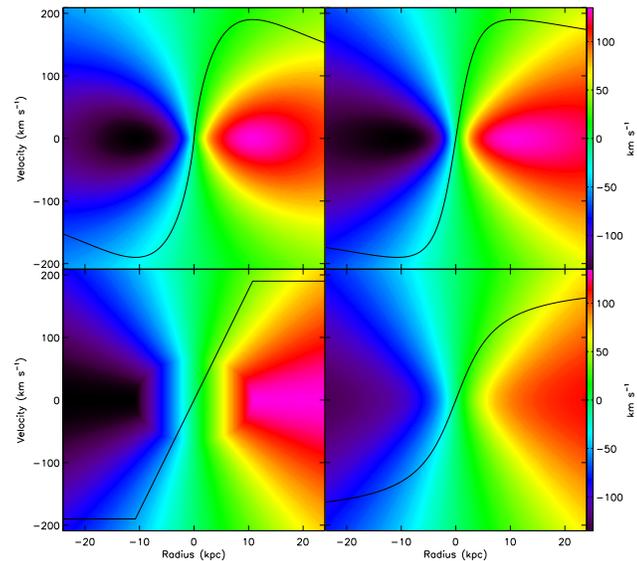}
\end{center}
\caption{High resolution \rcs\ (black curve) superimposed on
\vfs\ (color image) of the four models used. From left to right and
from top to bottom: exponential disk, isothermal sphere, ``flat''
and arctangent models. The radius (x-axis) is common for the four
\vfs\ and the four \rcs. The \vfs\ scale is given by the rainbow
scale on the right side of the images. The velocity amplitude of
the \rcs\ is given by the scale on the left side of the y-axis.}
\label{models}
\end{figure}

\subsection{Local velocity dispersion maps}
\label{fitting_method_dispersion}

To constrain the kinematical parameters, the generic model presented in section \ref{fitting_method_general} allows the use of the blurred \vfs\ alone, the blurred \vdms\ alone or the combination of both.
In the forthcoming analysis, the kinematical model has been constrained using the blurred \vfs\ only. Indeed, the blurred \vdms\ do not add any constraining power, thus, adding a dispersion parameter to the model is not necessary to fit the data. In a second step, the model has been used to correct the beam smearing effects in the \vdm\ (see Appendix \ref{model}).

To demonstrate that the use of the \vdm\ is not necessary to constrain the kinematical parameters, we have attempted to combine it to the \vf\ in order to retrieve the parameters of the model. In order to model the expected local \vdm\, an additional hypothesis concerning the physical nature of the velocity dispersion is needed.
We may choose the local velocity dispersion to be constant (i.e. the same value everywhere in the plane of the galaxy). This hypothesis, being a possibility since it is mainly what is observed in the GHASP sample (\citealp{Epinat:thesis}, \citeauthor{Epinat:2009}, in preparation), leads to a satisfying agreement with the parameters of the local sample.
However, if this method works for the GHASP sample, this is mainly due to the fact that, for nearby galaxies, the velocity shear is high with respect to the local velocity dispersion and the \snr\ is high. This might not be the case for distant galaxies for which the \snr\ is lower and for which the physical nature of the velocity dispersion is unknown. In addition, even if the method using an unique and constant velocity dispersion works, it not necessary since
%
%
%
(i) this parametrical approach needs the introduction of one or more parameters to describe the local \vdm\ (radial and azimuthal dependencies, etc.); (ii) the projection parameters and the velocity gradient can be recovered using the \vf\ alone; (iii) the constant velocity dispersion could also be retrieved from the velocity field only (see equation \ref{dispersion_lowres}); (iv) the velocity shear cannot be constrained efficiently when lower than the local velocity dispersion and (v) from a technical point of view, the low \snr\ affects more strongly the velocity dispersion (second order momentum) than the velocity (first order momentum) and this would lead to larger uncertainties, in particular for the velocity determination.

To summarize, we favor the method using the \vf\ alone since it allows to avoid any \textit{a priori} hypothesis on the local velocity dispersion. The velocity dispersions are corrected from beam smearing effect using the parameters of the model.

%

\subsection{Residual maps of nearby and high redshift galaxies}

\Vfs\ and \rcs\ of low redshift galaxies exhibit a large range of
shapes and despite a large number of attempts, no ``universal''
\rc\ is adequate to describe the large variety and complexity of
velocity gradients of rotationally supported galaxies. In nearby
spirals observed at high spatial and spectral resolutions, typical
deviations of $\sim10-20$\kms\ caused by non circular motion
(spiral arms, bar, etc.) are locally observed
\citep{Sofue:2001,Epinat:2008b,Epinat:2008a}. Subtracting
model describing galaxies dominated by circular motions from the GHASP data thus lead to mean residuals equal to
zero and r.m.s. lower than 20\kms\
\citep{Epinat:2008b,Epinat:2008a}.
The velocities observed in the residual \vfs\ of both nearby and projected samples have typically the same amplitude. This indicates that the method does not create artefacts.
\section{Analysis}
\label{analysis}

\subsection{Beam smearing parameter}
\label{BeamSmearingParameter}

Since \citet{Burbidge:1975}, it is known that the turnover radius
of a \rc~for a given galaxy differs if determined from optical
line or from HI 21 cm line studies.  This is due to the large
beams generally used in 21 cm line observations. This
artifact may induce spurious effects, for instance, in the
determination of the luminous and dark matter distributions and on
the internal shape and properties of dark haloes (e.g.
\citealp{Blais-Ouellette:1999}). A suitable parameter to
characterize the effect of the beam on radio HI data is the ratio
R/b, i.e. the ratio between the (Holmberg) radius $R$ of a galaxy
and the half-power beamwidth $b$. Mimicking this definition given
by \citet{Bosma:1978} suitable for HI data, we define hereafter
the so-called ``beam smearing parameter'' $B$, ratio between the
optical radius of a given galaxy and the seeing FWHM $s$ during
its observation (see Table \ref{table_z} in Appendix \ref{table} for $B$ values):
\begin{equation}
B=\frac{D_{25}/2}{s}
\label{b_def}
\end{equation}
Following \citet{Bosma:1978}, a ``believable'' \rc\ in the HI may
be obtained from a 2D velocity field when $B$ is greater or equal to
seven.
This criterion quantifies the spatial sampling needed to model the rotation of a galaxy. Thus, it may be exported to any sampling problem, independently on the nature of the probed component (neutral or ionized gas). In others words, the \rc\ must contain at least seven
independent measurements on both sides of the galaxy.

Thanks to the advent of AO, leading to a resolution
of typically 0.1\arcsec, we will find $B^>_\sim10$, for a galaxy
with a size $\geq2$\arcsec. Thus, the determination of the
kinematical parameters of the galaxy such as the dynamical center,
its inclination, position angle and its maximum rotational
velocity ($V_c^{max}$) becomes reliable. When B is large enough,
$V_c^{max}$ may be computed from the rotation curve rather than
from the width of the central velocity dispersion, in the center
of the galaxy.

\citet{Yangetal:2008} estimated that galaxies extending over less
than six spatial pixels may lead to a less robust kinematical
classification than for more extended galaxies. This is the case
for compact galaxies having their half light radius ($\sim 1~kpc$) within one
GIRAFFE pixel (0.52\arcsec). The same authors estimated that with
a median spatial coverage of nine pixels at \SNR~$>4$ the
classification is robust and unambiguous.

The beam smearing parameter $B$ in the projected sample ranges from $0.8$ to $8.4$ (see Table \ref{table_z}), but half of them has $B<2.4$. In the next sections we will show that an acceptable agreement between high and low resolution \rcs\ is only given for $B>6-7$. Nevertheless, $B\geq 2-3$ allows the determination of the position angle of the major axis and of the maximum rotation velocity.

\subsection{Galaxy projection parameters determination}
\label{proj_param}

\begin{table*}
\caption{Successfulness of the four $z=1.7$ models to recover $z=0$ actual parameters.}
\label{table_success}
 \begin{tabular}{c|c|ccccc|c|cccc}
\hline
& \bf{\textbar} &\multicolumn{5}{c|}{Successfulness$^{~(1)}$} & \bf{\textbar} & \multicolumn{4}{c}{RMS$^{~(2)}$} \\
& \bf{\textbar} & $i^{~(a)}$  & $PA^{~(b)}$ & $V_c^{max~(c)}$ & $\sigma^{~(d)}$ & RC$^{~(e)}$ & \bf{\textbar} & $i^{~(a)}$  & $PA^{~(b)}$ & $V_c^{max~(c)}$ & $\sigma^{~(d)}$ \\
Model & \bf{\textbar} & \%\ & \%\ & \%\ & \%\ & \%\ & \bf{\textbar} & \degr  & \degr & \kms & \kms \\
\hline

Exponential disk        & \bf{\textbar} & 27 & 29 & 26 & 15 & 37 & \bf{\textbar} & 15 & 5.8 & 24.9 & 8.8 \\
Isothermal sphere       & \bf{\textbar} & 10 & 8  & 15 & 10 & 12 & \bf{\textbar} & 15 & 5.8 & 22.7 & 8.5 \\
``Flat model''          & \bf{\textbar} & 51 & 39 & 41 & 51 & 29 & \bf{\textbar} & 14 & 5.7 & 22.9 & 8.0 \\
Arctangent             & \bf{\textbar} & 12 & 24 & 18 & 24 & 22 & \bf{\textbar} & 16 & 5.8 & 21.9 & 7.9 \\
\hline
\multicolumn{12}{l}{$(1)$: Percentage of galaxies better described by each model.}\\
\multicolumn{12}{l}{$(2)$: RMS between true and fitted parameters for each model.}\\
\multicolumn{12}{l}{$(a)$: Kinematical inclination.}\\
\multicolumn{12}{l}{$(b)$: Kinematical position angle of the major axis.}\\
\multicolumn{12}{l}{$(c)$: Maximum velocity. The RMS is computed from the relative difference between}\\
\multicolumn{12}{l}{the maximum velocities at $z=1.7$ and $z\sim0$ ($\Delta V_c^{max}/V_c^{max}$).}\\
\multicolumn{12}{l}{$(d)$: Local velocity dispersion.}\\
\multicolumn{12}{l}{$(e)$: \Rc\ shape agreement quantified by the residuals $\Delta V_c^{mean}$ between the actual}\\
\multicolumn{12}{l}{\rc\ at $z=0$ and the model \rc\ at $z=1.7$ (cf. section \ref{section_rcshape}).}\\

 \end{tabular}
\end{table*}

In this paragraph, the four models described in section
\ref{fitting_method} have been tested to recover the different
kinematical parameters at high redshift discussed hereafter (Tables \ref{table_modzexp} to \ref{table_modzata}). The
quality of the models at $z=1.7$ is tested by their ability to
retrieve the parameters at $z=0$ (given in Table \ref{table_modz0}). Table \ref{table_success}
presents the percentage of galaxies which are better described by
these different models.
It shows that the ``flat model'' is the one that statistically has the best recovery of almost all the parameters.
Since the difference with the other models is small in terms of the RMS, it could be that the ``flat model'' recovers the parameters best because it somehow yields more robust fit.
However, this may also reflect the flat general trend of nearby galaxy \rcs\ outside the inner solid body part. Indeed, the exponential disk and isothermal sphere \rc\ models are decreasing beyond $r_t$ while the arctangent is rising. From the nowadays observed \rcs\ of high redshift galaxies, there is no evidence for decreasing or rising \rcs. A fraction of the \rcs\ are still rising at the last observed point but this is probably because the maximum rotation velocity is not reached. This effect is even worse due to beam smearing effects and moreover to the fact that high redshift galaxies are probably smaller.
In the following sections, the plots only show the results obtained using this model.

\subsubsection{The center}
\label{analysis_center}

In nearby galaxies for which high resolution data are available, the determination of
the center is very sensitive to the method used to find it.
The center may be fixed by the
morphology, i.e. the position of the galaxy nucleus seen on high
resolution broad-band images in the near infrared or even in the
optical. Alternatively it can be computed using the kinematics and
becomes very sensitive to asymmetries in the \rc, especially in
its solid body domain. In this case, it is computed by making the
central regions of the \rc~as symmetric as possible. In best fit
model techniques based on least square computations (e.g. ROTCUR
in GIPSY package, \citealp{Begeman:1987}) the position of the
center may strongly depend on the value of the other kinematical
parameters as well as on asymmetries in moment maps ($m=1$ effects
like lopsidedness). The kinematical center may thus be offset by
$\sim 1~kpc$ with respect to the
morphological center \citep{Hernandez:2005b,Chemin:2006}. For nearby
galaxies, the offset may be much larger than the seeing (up to
60\arcsec), and thus may not be explained by spatial
resolution effects. The shift between
the center position  of the galaxy determined from the
photometry and from the kinematics is clearly a function of the
morphological type of the galaxy. The strongest discrepancies
occur for later type spirals for which the morphological center is
not always easy to identify \citep{Hernandez:2005b}.

In addition to the previous spurious effects, in
\hz\ data, the determination of the center is strongly affected by
the low spatial resolution, the size of the seeing disk being
equal to several $kpc$.  Indeed, due to the small number of
independent velocity measurements in the \vf~compared to the large
number of free kinematical parameters, whatever the model used is,
best fit models cannot converge to fix the center.  Due both to
the low spatial resolution and to the apparent small size of the
disk due to flux detection limitation (or intrinsic small size
since, in the cold dark matter scenario, the first objects
originated from gravitational collapse of the initial fluctuations
are smaller), \rcs\ for \hz\ galaxies tend to show solid body shapes
and thus do not display a clear turnover,
even if we observe a plateau at high spatial resolution.
This effect makes almost impossible the
determination of the position of the center using either the
method of symmetrization of \rcs~or best fit models.

To determine the position of the center, the central peak induced
by the inner velocity gradient observed in the \vdms~is not more
helpful than the \ha~intensity maps at the same spatial
resolution.
For galaxies with $B^>_\sim3$ for which the \rc\ shows a slope break, the center may be found from the \vfs.
Moreover, actual \hz~galaxies seem to show large local
velocity dispersions (see paragraph \ref{velocity_dispersion_analysis}),
which makes even more difficult to distinguish the velocity
dispersion peak.

In this work, as it has been done for instance in
\citet{Epinat:2008b}, the center of the \vfs\ chosen to compute
the \rcs\ has been fixed \emph{a priori} to match the morphological
centers (nuclei) from high resolution images. This method could
easily be applied to real \hz\ data using for instance HST imagery.

In conclusion, due to the lack of spatial resolution, photometric centers from high resolution broad-band images should be used because kinematical ones are not reliable.

\subsubsection{The inclination}

\begin{figure}
\begin{center}
\includegraphics[width=8.5cm]{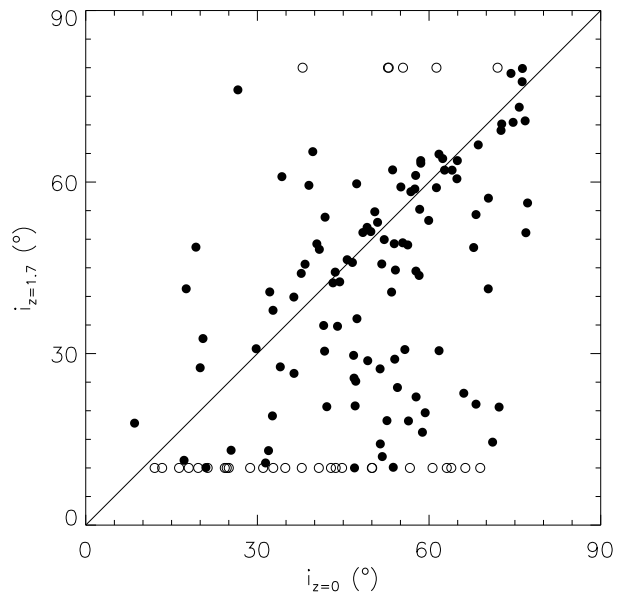}
\end{center}
\caption{Kinematical inclination computed at $z=1.7$ using a
``flat model'' vs actual kinematical inclination evaluated at
$z=0$. Each circle represents a galaxy. The open ones are galaxies
which are stacked to the boundaries allowed for any fit (10 and 80\degr).
The line indicates $y=x$.} \label{inclination}
\end{figure}

The determination of the inclination of a galaxy disk with respect
to the plane of the sky is a key parameter since it fixes the
amplitude of the maximum rotation parameter $V_c^{max}$. It is a
critical kinematical parameter to determine for \hz~galaxies. For
instance, a disk rotating at $V_c^{max}=200$\kms~inclined by
$35$\degr~with respect to the plane of the sky, might be confused
with a disk rotating at $V_c^{max}=160$ or $270$\kms~if the
inclination is respectively overestimated by 10\degr ($25$\degr)
or underestimated by 10\degr ($45$\degr). Thus, wrong
determinations of the inclination increase the dispersion of
$V_c^{max}$ hence, for instance, the scatter in the \TF~relation.\\
\\
\textit{Kinematical inclination}\\
Due to the degenracy between the inclination and the maximum rotation velocity in kinematical projection models,
the inclination is
probably the most difficult parameter to recover,
even for high resolution kinematical data of local galaxies \citep{Palunas:2000,Epinat:2008b}.
Morphological inclination measured on high
spatial resolution images is in global agreement with the
kinematical inclination but with a rather large scatter.
Figure \ref{inclination} presents the comparison between the kinematical
inclinations derived from high resolution \vfs~on the local data
in \citet{Epinat:2008b} and those obtained from the fit to the
redshifted dataset derived using the ``flat model''. A high scatter is
observed.  The four models lead to the same uncertainties in the
determination of the inclination but the ``flat model'' enables to
determine an inclination for $77$\% of the sample while the three
other models recover an inclination only for $58\pm2$\% of the
sample (``flat model'' provides less galaxies with inclination set
to the extreme values 10\degr~and 80\degr~compared to the other
models). It is also the one which statistically provides the best estimate of the inclination (see Table \ref{table_success}). The four models lead to a RMS between true and fitted inclinations equal to
$15\pm1$\degr~and a median equal to $8\pm1$\degr, which means that
the inclination can only be recovered with an error lower than
$\sim8$\degr\ in 50\% of cases. The standard deviation and the
median are also smaller for the ``flat model'' than for the other
models, when considering only the 70 galaxies for which the four models
recover an inclination.

\begin{figure}
\begin{center}
\includegraphics[width=8.5cm]{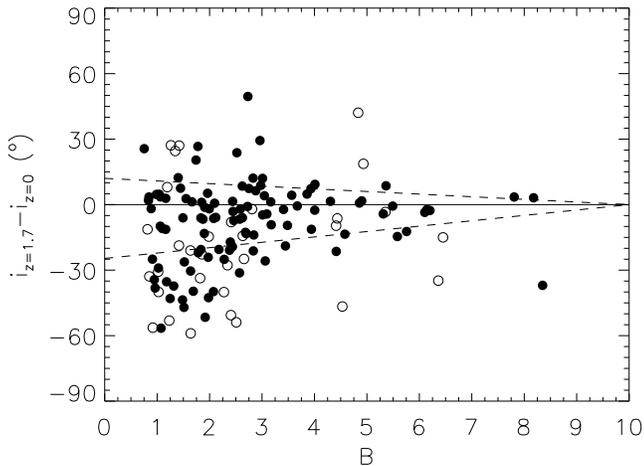}
\end{center}
\caption{Difference in the kinematical inclination between actual
$z=0$ galaxies and simulated galaxies at $z=1.7$ using a ``flat
model'' vs the beam smearing parameter $B$. Each circle represents
a galaxy. The open ones are galaxies which are stacked to the
boundaries allowed for any fit (10 and 80\degr). The two dashed lines represent the mean positive and negative errors.}
\label{inclination_bsp}
\end{figure}

Figure \ref{inclination_bsp} shows that, for \hz~galaxies,
the scatter in the determination of the kinematical inclination
decreases when the beam smearing parameter $B$ increases. It seems
that two regimes are observed depending on $B$ below or above 3.
The scatter around
$i_{z=1.7}-i_{z=0}=0$ is very large for $B^<_\sim3$ and clearly smaller
when $B^>_\sim3$. Moreover, for $B^>_\sim3$, we clearly observe
that the blurring of the data induces underestimated inclinations in
average. This may be explained by the fact that
the isovelocity lines are more ``open" for low values of $B$ (see
discussion in section \ref{biasesproj}). Making the assumption
that \emph{(i)} the discontinuity observed between two regimes is
mainly due to numerical instabilities (as suggested by the statistical error bars,
not plotted for clarity), \emph{(ii)} the actual inclination may
be recovered for $B^>_\sim10$ and \emph{(iii)} the accuracy in the
determination of the inclination could reasonably be well
quantified by a linear function of $B$. Two linear fits (represented by
the two dashed lines on Figure \ref{inclination_bsp}) have been made
to model the mean error on the
underestimate and on the overestimate of the inclination
respectively. The fits avoid the galaxies for which the
inclination has been stacked to its lower of higher boundary
(open symbols). The equations are:\\
If $0<B<10$,
$$i_{z=1.7}-i_{z=0}=-1.2\times B+12~~~\text{for}~~~i_{z=1.7}-i_{z=0}>0$$
\begin{equation}
i_{z=1.7}-i_{z=0}=+2.5\times B-25~~~\text{for}~~~i_{z=1.7}-i_{z=0}<0
\end{equation}
While if $B\geq10$,
\begin{equation}
i_{z=1.7}-i_{z=0}=0
\end{equation}

These formulae are helpful in the case the inclination is determined from the kinematics because they provide the error bars as a function of the beam smearing parameter $B$.\\
\\
\textit{Morphological inclination}\\
Due to the small angular size of high redshift galaxies, the determination of morphological inclination needs to take into account the seeing. Programs widely used like SEXTRACTOR \citep{Bertin:1996}, or like any two-dimensional Gaussian fit, provide axis length measurements that need to be corrected for beam smearing in order to compute the inclination. Models taking into account seeing effects, such as GIM2D \citep{Simard:2002} or GALFIT \citep{Peng:2002} have been developped to compute morphological parameters.
In order to model the effect of the seeing,
we have created two sets of high resolution models of thin inclined galactic disks using an exponential disk surface brightness radial profile with a disk scale length $R_d$ and a flat luminosity function truncated at $R_{opt}=3.2R_d$.
The disk scale length $R_d$ has been set to various physical lengths (2, 3, 4, 5 and 6 kpc) to see the evolution when the beam smearing parameter $B=R_{opt}/s$ varies and the disks have been inclined from 10\degr\ to 80\degr\ with a step of 10\degr.
\par
We have projected these models at redshift $z=1.7$ using a seeing of $0.5$\arcsec\ and a pixel size of $0.125$\arcsec, as we did with our kinematical data.
The axis lengths were determined on both projected and high resolution images using Gauss2dfit IDL routine as the FWHMs of the 2D gaussian function. This fitting procedure gives very accurate results on high resolution images whatever the luminosity profile is, but the lengths are not identical.
The effect of the seeing is very well reproduced, for all inclinations, disk scale lengths and luminosity profiles by assuming that the measured major and minor axis $a_m$ and $b_m$ are quadratically overestimated by a fraction $C$ of the seeing FWHM $s$:
\begin{equation}
\cos{i}=\frac{b}{a}=\sqrt{\frac{b_m^2-C^2\times s^2}{a_m^2-C^2\times s^2}}
\end{equation}
where $a$, $b$ and $i$ are respectively the actual major axis, small axis and disk inclination.
The fraction $C$ almost does not depend on the luminosity profile. Thus for an exponential luminosity profile, $C=1.014\pm0.002$, and for a flat profile, $C=1.015\pm0.010$, which is in both cases very close to 1. The better accuracy obtained for the exponential distribution reflects the fact that an exponential distribution is better described by a gaussian distribution than the flat distribution. Since \emph{(i)} the high resolution image can be well reproduced by a 2D gaussian function and \emph{(ii)} blurring the image with the seeing consists in convolving the high resolution image with a 2D gaussian function, it is reasonable that the blurred image is well reproduced by a 2D gaussian function whose measured axis are the quadratic combinations of the true lengths with the seeing.

In addition to beam smearing effects, the presence of large clumps may bias the morphological inclination determination. Indeed, numerical simulations as well as observations show that no more than 5-10 large clumps are seen in a disk of a \hz\ galaxy. In the case where the inclination is measured using the \ha\ image, even if these large clumps are randomly distributed through the disk, they will visually induce a overestimation of the actual disk inclination. One may preferentially use high resolution broad-band imaging tracing the bulk of stars rather than bright stars located in those clumps.
\\
\par
In conclusion, the inclination should be derived from broad-band images, with high
resolution if possible,
in order to better constrain the model and to relax from one unity
the number of free parameters. Ideally, to avoid contamination due
to clumps of star formation in the determination of the
inclination, the inclination of the old stellar disk should be
measured in the near-infrared rest-frame of the galaxy. We have given a simple correction of beam smearing effects to determine the inclination from axis ratio.
When no high resolution imagery is available, we have provided a model to
estimate the uncertainties on kinematical inclination. In the
following sections, we have fixed the inclination to the
kinematical inclination derived at redshift zero.

\subsubsection{The position angle of the major axis}

\begin{figure}
\begin{center}
\includegraphics[width=8.5cm]{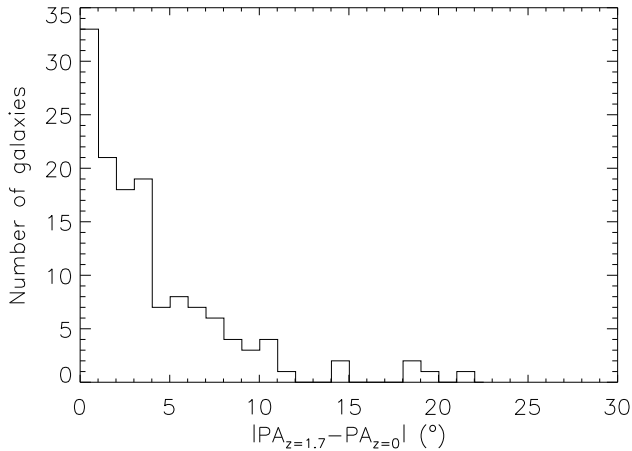}
\end{center}
\caption{Histogram of the difference in kinematical \pas~computed
at $z=0$ and at $z=1.7$ using a ``flat model''.} \label{pa}
\end{figure}

\begin{figure}
\begin{center}
\includegraphics[width=8.5cm]{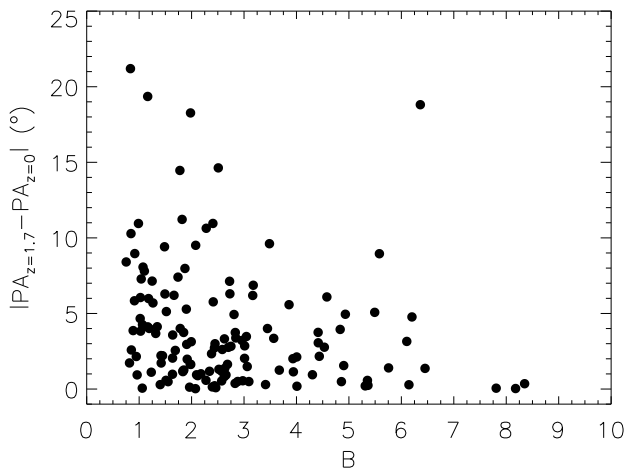}
\end{center}
\caption{Difference in the \pa~between actual $z=0$ galaxies and
simulated galaxies $z=1.7$ using a ``flat model'' vs the beam
smearing parameter $B$. Each circle represents a galaxy.} \label{pa_bsp}
\end{figure}

Similarly to a bad determination of the inclination, a wrong
determination of the position angle of the major axis will lower
the maximum rotation velocity $V_c^{max}$. The use of integral
field spectroscopy enables to compute reliable kinematical \pas~of
the major axis.\\
\\
\textit{For nearby galaxies}\\
The kinematical and photometric position angles have been
compared for the whole GHASP sample \citep{Epinat:2008b}. The
histogram of the variation between kinematical and morphological
position angles indicates that for 57\%\ of the galaxies,
the agreement is better than 10\degr; for 79\%, the
agreement is better than 20\degr\ and the disagreement is larger
than 30\degr\ for 15\%\ of the galaxies.  Most of the
galaxies showing a disagreement in position angles larger than
20\degr\ present a bad morphological determination of the position
angle, a kinematical inclination lower than 25\degr\ or are
specific cases due essentially the presence of bar and spirals
arms. In conclusion, the agreement between morphological and
kinematical \pas~is satisfactory for rotating disks but not very
good for low inclination systems ($i\le25$\degr) and strongly
barred galaxies. In any case, integral field spectroscopy
constitutes the best technique to determine position angles and as
a consequence, rotation curves.\\
\\
\textit{Comparison with projected galaxies}\\
We have compared the kinematical \pas~derived from high resolution
\vfs~\citep{Epinat:2008b}, with those computed from the redshifted
data as illustrated on Figure \ref{pa} on which the ``flat
model'', that gives the best estimate for more galaxies than the
other models (see Table \ref{table_success}), has been used.
Whatever the model used, for more than 70\% of the data set, the
agreement is better than 5\degr. Less than 8\% have a disagreement
larger than 10\degr. When the inclination is a free parameter, the
estimate of \pas~remains as accurate. This is to be pointed out
since a good \pa~estimate is mandatory to recover the true \rc.
The accuracy is even better for large galaxies as seen in Figure
\ref{pa_bsp}: the agreement is better than 5\degr~for more than
78\% of the galaxies with a beam smearing parameter greater than
3. Bars also induce the strongest disagreements, as well as a low
\Ha~extent (e.g. UGC1655).\\
\\
\textit{Signature of non-circular motions}\\
The comparison between morphological and kinematical \pas~at
\hz~should be used to assess the presence of strong bars as well
as other non rotation motions. To be able to do that, accurate
measurements of morphological \pas~are necessary and one should
preferentially use high resolution imaging. Indeed, high
redshift galaxies are less regular and have peculiar and more
disturbed \vfs\ than nearby galaxies like the ones studied in the
GHASP sample. Thus, the signature of these peculiarities should be
quantified through the comparison between morphological and
kinematical \pas.\\
\\
\textit{Low inclination high redshift galaxies}\\
The projected GHASP sample may be used
to test the biases introduced by long slit observations. One may
underline the well known effect that measured rotation velocity
declines as misalignment increases. As already mentioned by
\citet{Weiner:2006} who used a simulated \hz~galaxy from $z=0$
\FP~observation, there is a lack of galaxies with high rotation
for misaligned slits. Clearly, measuring rotation velocity from a
misaligned slit is subject to large errors for galaxies with low
inclination. Thus nearly round shaped galaxies (ellipticity $e <
0.25$) should absolutely be avoided for long slit spectrography.\\
\\

In conclusion, due to their small angular size and beam
smearing effects, high redshift galaxies are poorly sampled and
appear rounder that they really are. From any type of galaxies and
any inclination, the GHASP sample allows us to conclude that the
kinematical position angle of 2D projected \vfs\ are recovered with an accuracy
better than 5\degr\ in more than 70\%\ of the cases, giving a
higher limit taking into account that high redshift galaxies are
intrinsically more disturbed than nearby galaxies. We have
stressed that the position angle of the major axis of low
ellipticity high redshift disks, are better determined
\textit{a posteriori} from 2D \vfs\ than \textit{a priori} from
imagery, as it is done for long slit observations and finally that
the difference between the morphological and the kinematical \pas\
gives a signature of non axi-symmetric motions.

\subsubsection{Systemic velocity}

The determination of the systemic velocities is not fundamental
but it allows to test again the validity of the models.  The
systemic velocities are reasonably well recovered from all the
models: the systemic velocity can be recovered within
$6\pm1$\kms~for half the sample (depending on the model used).
This value is probably an upper limit but we have to keep in mind that this
value remains rather low because the position of the center has
been fixed.

\subsection{Shapes of the \rcs}
\label{section_rcshape}

In cases where $B^>_\sim10$, it may become possible to address the
problem of the shape of the inner density profile in spirals (CORE
vs CUSPY controversy) for \hz~galaxies. This problem remains one
of the five main further challenges to $\Lambda$CDM theory
\citep{Primack:2007}. With the help of AO, for
the largest and brightest galaxies, it will be possible to model
the mass distribution in \hz~galaxies in separating luminous from
dark halo contribution. It is not possible to separate them using the
best data observed without AO (i.e with $B<3$), e.g. for galaxy Q2343-BX610 located
at $z\sim2.2$ \citep{Forster-Schreiberetal:2006}. The
question is nevertheless addressable on high quality data obtained with AO,
e.g. for galaxy BzK-15504 \citep{Genzeletal:2006} for which
$B\sim6$.
Indeed, in our projected sample for which the beam smearing parameter $B$ ranges from $0.8$ to $8.4$, an acceptable agreement between high and low resolution \rcs\ is only given for the three galaxies with $B>7$ (cf. \rcs\ of UGC 01886, UGC 03334 and UGC 03809 in Appendix \ref{rcz}). Five other galaxies having $6<B<7$ already show noticeable differences in the inner parts of their \rc.

In order to quantify the ability to recover the shape of the \rcs~for
\hz~galaxies, we computed the difference between redshifted
\rcs\ and original \rcs\ at $z=0$. To avoid biases due to the
\rc~sampling we recomputed velocities with a radial step of
$0.5~kpc$. On local \rcs, this is achieved by computing the mean
value within radial ranges of $0.5~kpc$ weighted by the number of
bins used to compute the high resolution \rcs~presented in
\citet{Epinat:2008b,Epinat:2008a}. On the \rcs~computed along the
major axis of the redshifted \vf, we interpolate the \rc~within
the required radii. For the models, velocities can be computed at
any radius.

\begin{figure}
\begin{center}
\includegraphics[width=8.5cm]{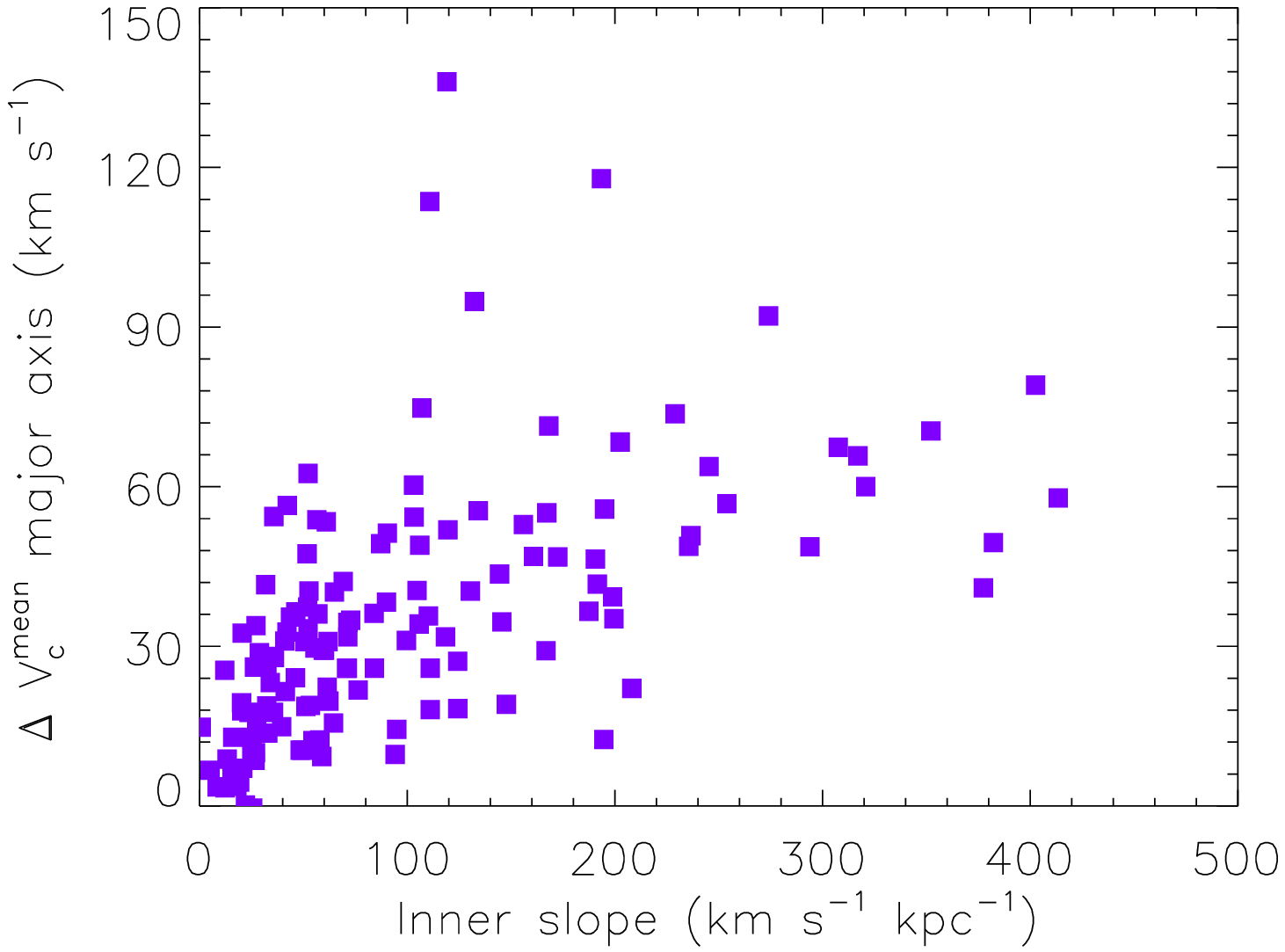}
\includegraphics[width=8.5cm]{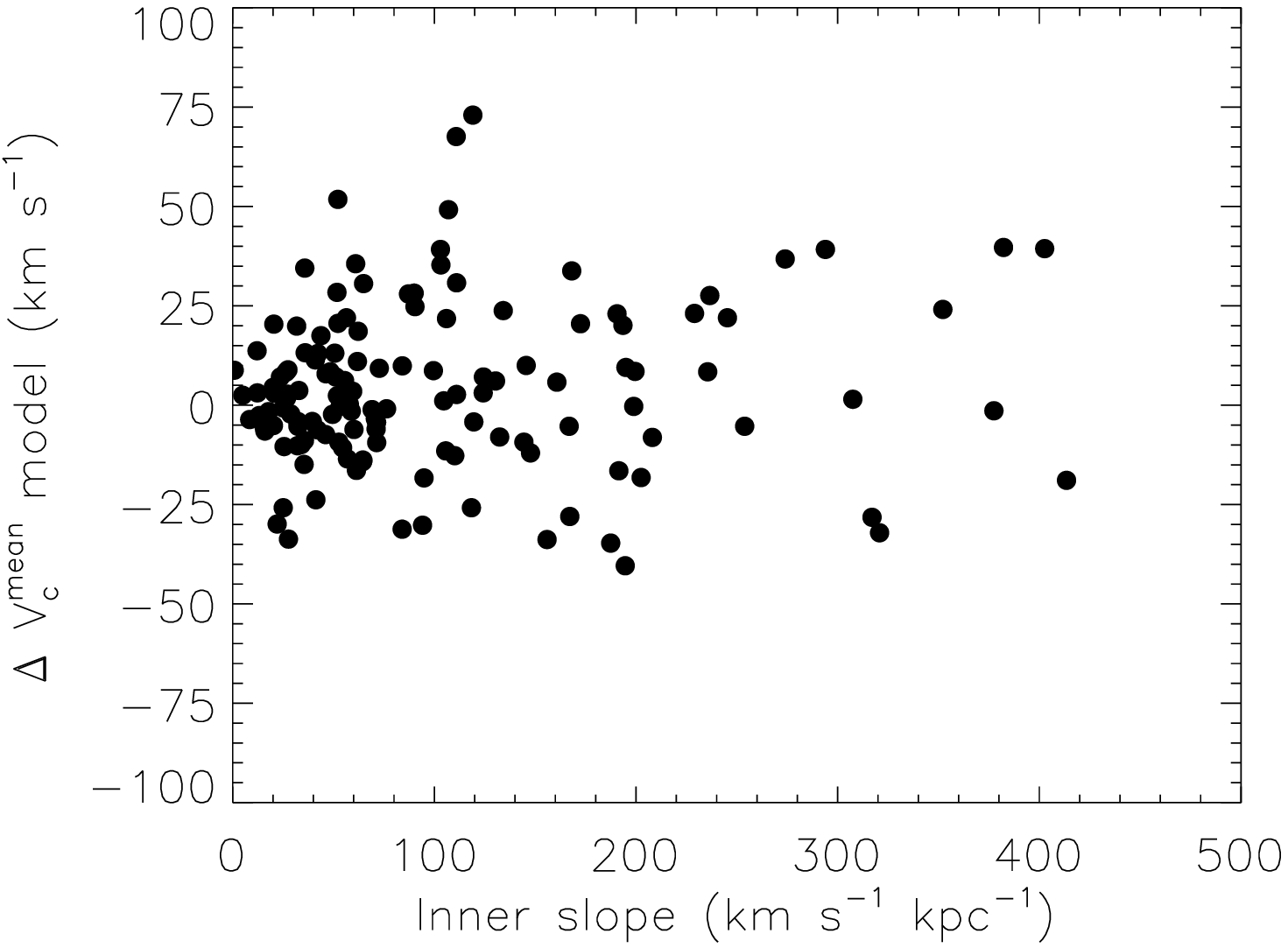}
\end{center}
\caption{Mean rotation velocity difference $\Delta V_c^{mean}$
between actual $z=0$ and different $z=1.7$ \rcs~vs the actual
$z=0$ \rc~inner slope. Top: $\Delta V_c^{mean}$ is the mean difference
between actual $z=0$ and non-corrected $z=1.7$ \rcs.  The $z=1.7$
\rcs~are computed along the major axis of the galaxies. Bottom:
$\Delta V_c^{mean}$ is the mean difference between actual $z=0$ and
model $z=1.7$ \rcs. The $z=1.7$ \rcs~are computed using a ``flat
model''.} \label{rcshape}
\end{figure}

The difference between the \rcs\ is quantified using the parameter
$\Delta V_c^{mean}$ measuring the mean rotation velocity
difference along the whole \rcs:
\begin{equation}
\Delta V_c^{mean} = \frac{\sum_{i=1}^{n_1} \left[ V_0(r_i)-V_z (r_i) \right] + \sum_{j=1}^{n_2} \left[ V_z (r_j)-V_0(r_j) \right] }{n_1+n_2}
\end{equation}
where $r_i$ and $r_j$ are respectively the radii for receding and approaching sides, $V_0$ is the high resolution \rc\ and $V_z$ is the \rc\ of the redshifted data (it can either be the one from the major axis or from the models). The $\Delta V_c^{mean}$ parameter enables to distinguish an overestimate from an underestimate of the \rcs.
Figure \ref{rcshape} (top) shows
that $\Delta V_c^{mean}$ between $z=0$ actual \rcs~and $z=1.7$
non-corrected \rcs~is strongly correlated with the inner slope of
$z=0$ galaxies.
The inner slope has been computed from a fit to the high resolution \rc\ (cf. \citealp{Epinat:thesis}).
$\Delta V_c^{mean}>0$ indicates that the $z=0$
\rcs~always display higher mean rotation velocity than the
projected $z=1.7$ \rcs.  Indeed, galaxies with large inner slopes
are more affected by the beam smearing. Figure \ref{rcshape}
(bottom) displays that $\Delta V_c^{mean}$ between $z=0$ actual
\rcs\ and $z=1.7$ ``flat model'' \rcs\ is on average equal to zero and not
correlated with the inner slope of $z=0$ galaxies. $\Delta
V_c^{mean}$ is positive as much as negative, meaning that the
model respectively overestimates and underestimates the mean \rcs.
The scatter around the axis $\Delta V_c^{mean}=0$ is nevertheless
large (mean errors can be as large as $\pm50$\kms), pointing out the difficulty to retrieve the actual shape of
the \rcs\ even if the general trend is recovered. These comments remain valid for the three other \rc\ models. The
\rc\ shape can thus hardly be used for mass modeling. Using this parameter to quantify the ability to recover the actual shape of \rcs, we find that the exponential disk model statistically gives a better description followed by the ``flat model'' (see Table \ref{table_success}).

We also compared the inner slopes measured from $z=0$ \rcs\ and
from the model \rcs. The scatter around $y=x$ line is very large,
indicating that the use of such models to constrain mass modeling
is not sufficient. AO observations are thus mandatory for mass
modeling.

We find that galaxies with a beam smearing parameter $B^>_\sim3$
tend to present larger $\Delta V_c^{mean}$, but this trend is
not very significant.
This is due to the fact that large galaxies are also the fastest
rotators and thus have larger inner slopes of the \rc.  These
large slopes are often explained by the presence of bulges and are
well known in massive local galaxies dominated in their central
region by the luminous matter distribution. However, without any
high resolution broad band images, it is difficult to assess the
presence of such bulges in \hz~galaxies. Furthermore, the
\rc~shape for \hz~galaxies is unknown and the presence of a bulge
is not mandatory to observe a large inner velocity gradient.

On the other hand, \rcs\ based on large clumps velocity measurements tend to underestimate the tangential velocity of the disk.  These clumps have a radial velocity component which has to be taken into account (Bournaud, private communication). Thus, AO observations are needed to observed disk regions uncontaminated by the blurring of large clumps.

\begin{figure*}
\begin{center}
\includegraphics[width=12cm]{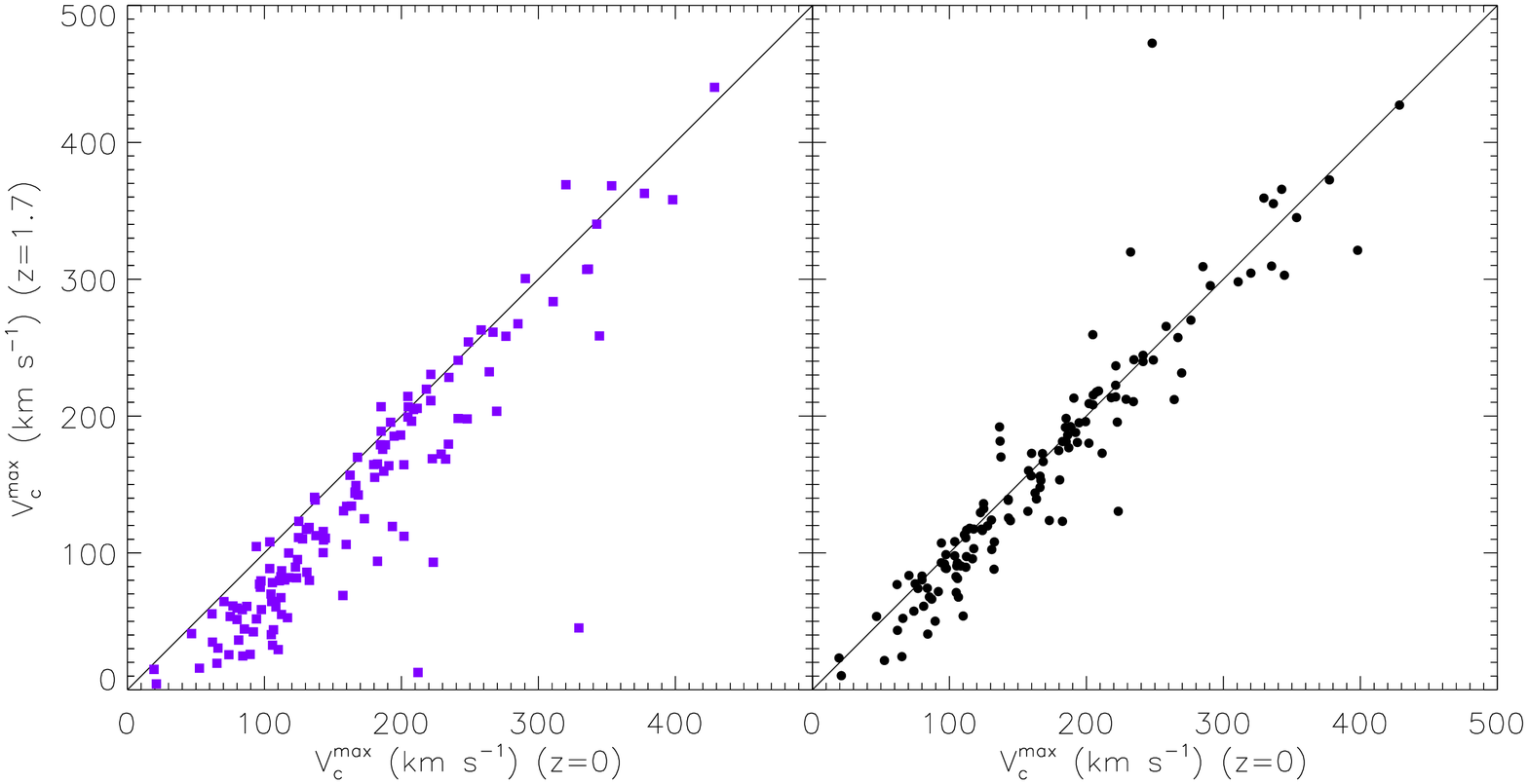}
\includegraphics[width=12cm]{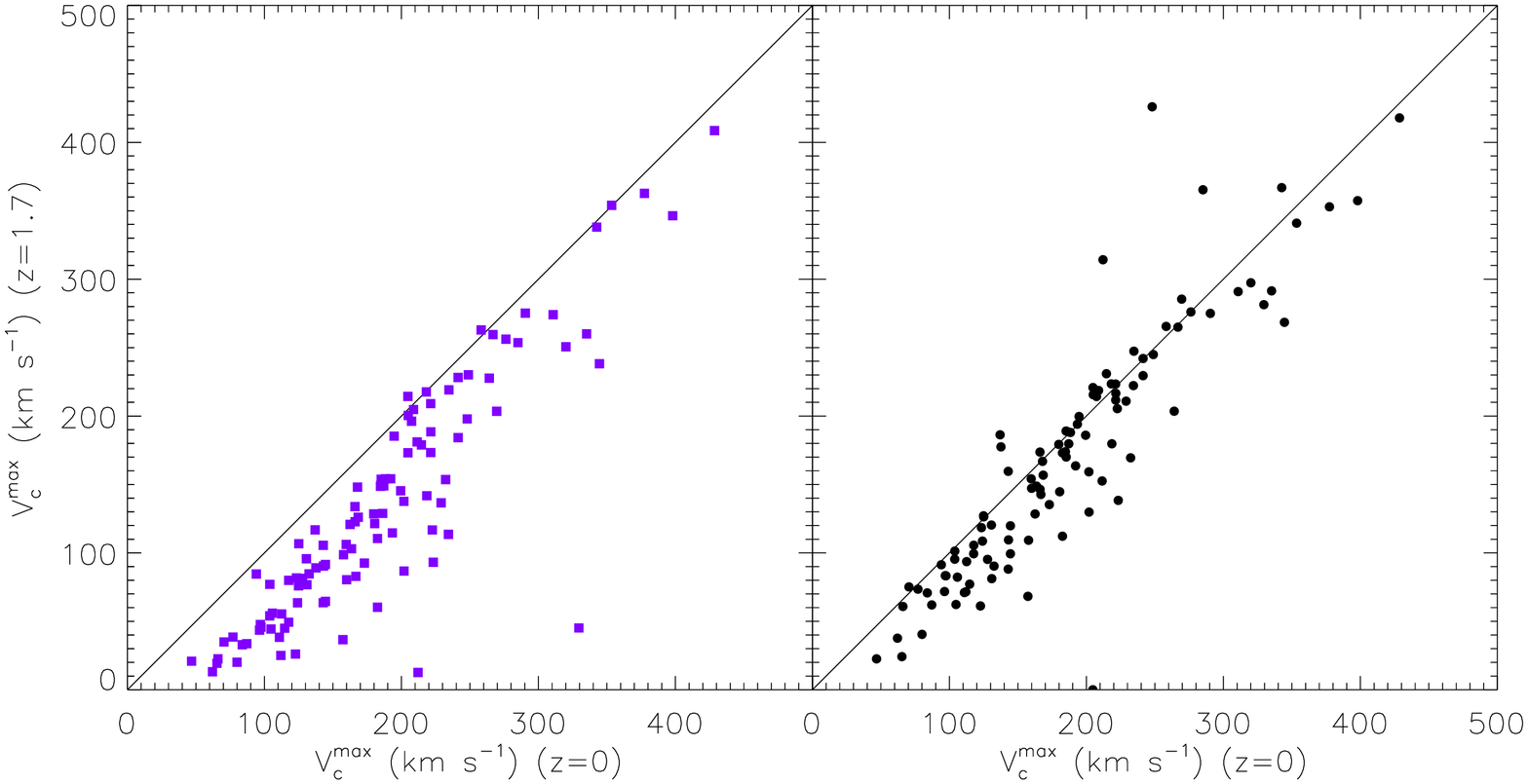}
\end{center}
\caption{Comparison between the maximum rotation velocities at
$z=0$ (x-axis) and $z=1.7$ (y-axis). The blue squares (left
column) and the black dots (right column) represent respectively
the maximum velocities measured along the major axis of the
blurred \vfs~and the maximum deduced from the ``flat model''
fitting.
(Top) \Vfs\ are truncated at diameters $D_{25}$ (along the major axis). (Bottom) \Vfs~are truncated at diameters
$D_{25}/2$.} \label{vmax}
\end{figure*}

We conclude that the distance effect is too important to recover a
reliable \rc~shape, in particular in the inner regions, whatever
the beam smearing is up to at least $6$ (for larger values, our
statistics are very poor), due to the sharper \rc~shapes of large
and massive galaxies. On the other hand, the unknown shape of
\hz~\rcs~makes this comparison very approximative since if the
slope is lower, the \rc~shape should be better recovered.

\subsection{Maximum rotation velocity analysis}
\label{maxvr}

Even without having a complete knowledge of the shape of a \rc,
the first order analysis of kinematical data may allow a
determination of the maximum rotation velocity of the
\rc~$V_c^{max}$. The latter is an important parameter to recover
since it constrains the total amount of dynamical mass of the
galaxies and is used for the analysis of the \TF~relation.

Two methods have been tested to retrieve this parameter. The first
one consists in using the \rc\ computed
along the major axis, which is equivalent to simulate a long slit aligned with the major axis, without taking beam smearing into account. With long slit spectroscopy, it is however possible to use models that take into account the seeing as done, e.g., by \citet{Weiner:2006}, but it is not straightforward to evaluate the contribution due to regions outside the slit. The second one consists in using the \rc\ models that account for the beam smearing. For both, $V_c^{max}$ is estimated from the maximum
amplitude of the \rc\ within the extent of the \vf\ along the major axis.
In Figure \ref{vmax}, we compare the maximum rotation velocity
determined by \citet{Epinat:2008b} with those determined from
these two methods. The model presented in this Figure is the ``flat model''.

\subsubsection{Major axis rotation curve}

The maximum rotation velocity $V_{rc}^{max}$ directly determined from the \rc\ along the
major axis without accounting for beam smearing is systematically underestimated for galaxies with
rotation velocities lower than $\sim150$\kms. The effect is even
worse when we only consider the \rc\ limited to half the optical
radius as it is shown on the Figure \ref{vmax} (bottom).
The maximum rotation velocity derived from \rc\ along the major axis (i.e. equivalent to \rcs\ obtained using long slit
spectroscopy instruments considering a good alignement with the actual \pa) is reliable for galaxies with an optical
radius larger than three times the seeing ($B>3$),
as seen in Figure \ref{deltaVmaxR25} on which the relative difference between the maximum rotation velocities at $z=1.7$ and $z=0$ is plotted as a function of the beam smearing parameter $B$.
The maximum rotation velocities determined from the \rc\ along the major axis are systematically underestimated by more than 25\% for galaxies with $B$ lower than $2.5$.
A correction to recover the actual maximum rotation velocity $V_c^{max}$ depending on the beam smearing parameter $B$ can be applied for smaller galaxies by making the assumption that the \rc\ shape is rather similar for \hz\ and local galaxies:
\begin{equation}
V_c^{max}=\frac{V_{rc}^{max}}{0.1 (\pm 0.2) +0.36~B}
\label{vmax_corr}
\end{equation}
The uncertainty given in parenthesis provides a range of corrections: the lower limit for the correction is given for $0.1+0.2$ and the upper limit for $0.1-0.2$.

This method may be improved in taking into account the beam smearing affecting the data along the major axis. Nevertheless, high redshift galaxies are strongly affected by the slit effect since they are poorly sampled due to their small angular size. Indeed, their angular size measured along the minor axis is comparable to the width of the long slit. In addition, the width of the long slit is usually larger than the seeing. Thus, without additional assumption on the spatial or spectral distribution, a confusion between the velocity and the position remains (slit effect). Moreover, flux distribution outside the slit is not constrained in long slit observations. Thus, assumptions on flux distribution are needed to take into account the contribution of external points to the velocity measurements.

\subsubsection{Velocity field model}

\begin{figure}
\begin{center}
\includegraphics[width=8.5cm]{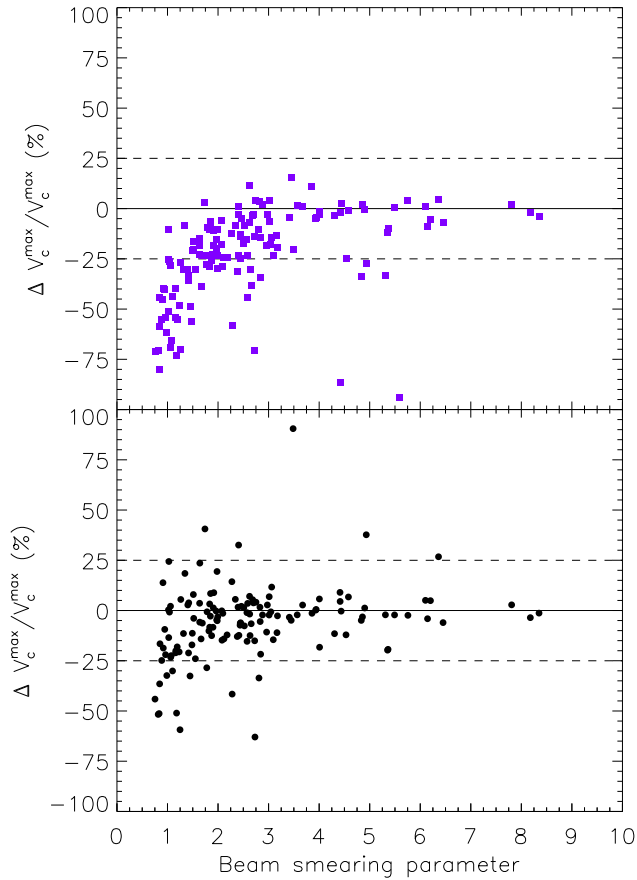}
\end{center}
\caption{Relative difference between the maximum rotation
velocities at $z=1.7$ and $z=0$ as a function of the beam smearing
parameter $B$. The symbols are the same as in Figure \ref{vmax}.}
\label{deltaVmaxR25}
\end{figure}

The model fitting enables to recover more reliable maximum rotation
velocities even for slow rotators (Figure \ref{vmax}).
The four \rc\ models have been compared. We find that the ``flat model'' statistically recovers better values than the other models (see Table \ref{table_success}).
In particular, the three other models give very
large maximum rotation velocities in a few case. In the case of
exponential disk and isothermal halo models, this is due to the
shape of these models that contains a central ``bump''. In the
case of the arctangent function, this is due to the fact that no
plateau is reached unless the transition radius is very small
(e.g. Appendix \ref{rcz}). Due to the beam smearing, the models
are mostly constrained by the inner regions that are more
luminous. Indeed, one can observe the rather good agreement of the
inner slope of the four models for all the \rcs~presented in
Appendix \ref{rcz}.
The difficulty to recover the maximum rotation velocity for slow rotators is due to the fact that these galaxies are in addition intrinsically smaller,
typically of the order of the beam size.
The maximum rotation velocities derived from
model fitting are statistically in good agreement with the actual
maximum rotation velocities. This is especially convincing for the
``flat model'' (used in Figure \ref{deltaVmaxR25}) for which the determination is better than 25\%
even for galaxies with $B$ as low as $1$. The three other models
may overestimate the maximum rotation velocities for some galaxies
with $B$ smaller than $2$.\\
\\

In conclusion, we have stressed that the use of integral field instruments sampling the seeing, enables a more robust
modeling since off-axis points can be taken into account
with less confusion than long slit spectrographs because they have an additional spatial dimension and allow to avoid slit effects.
Moreover, we have shown that, using a simple flat \rc\ to model the disk, the maximum velocity can be recovered with an accuracy better than 25\%, even for galaxies with a beam smearing parameter as low as 1.

\subsection{Velocity dispersion analysis}
\label{velocity_dispersion_analysis}

\subsubsection{Mean velocity dispersion and velocity shear}

The local velocity dispersion $\sigma$ is a non trivial physical parameter to
recover. Indeed, as explained in Appendix \ref{model}, for each
pixel, the measured velocity dispersion $\sigma_1$ is the quadratic combination of the
local velocity dispersion plus a velocity shear feature
induced by beam smearing effects. The velocity shear feature can
however be extracted from the high resolution modeled \vf\ if it
correctly describes the observed \vf.
This requires a good estimate of the spatial PSF.
Theoretically, as it is the
case for the \vf\ modeling, the velocity shear feature of the
\vdm\ also needs the knowledge of a high spatial
resolution line emission map.

\begin{figure}
\begin{center}
\includegraphics[width=8.5cm]{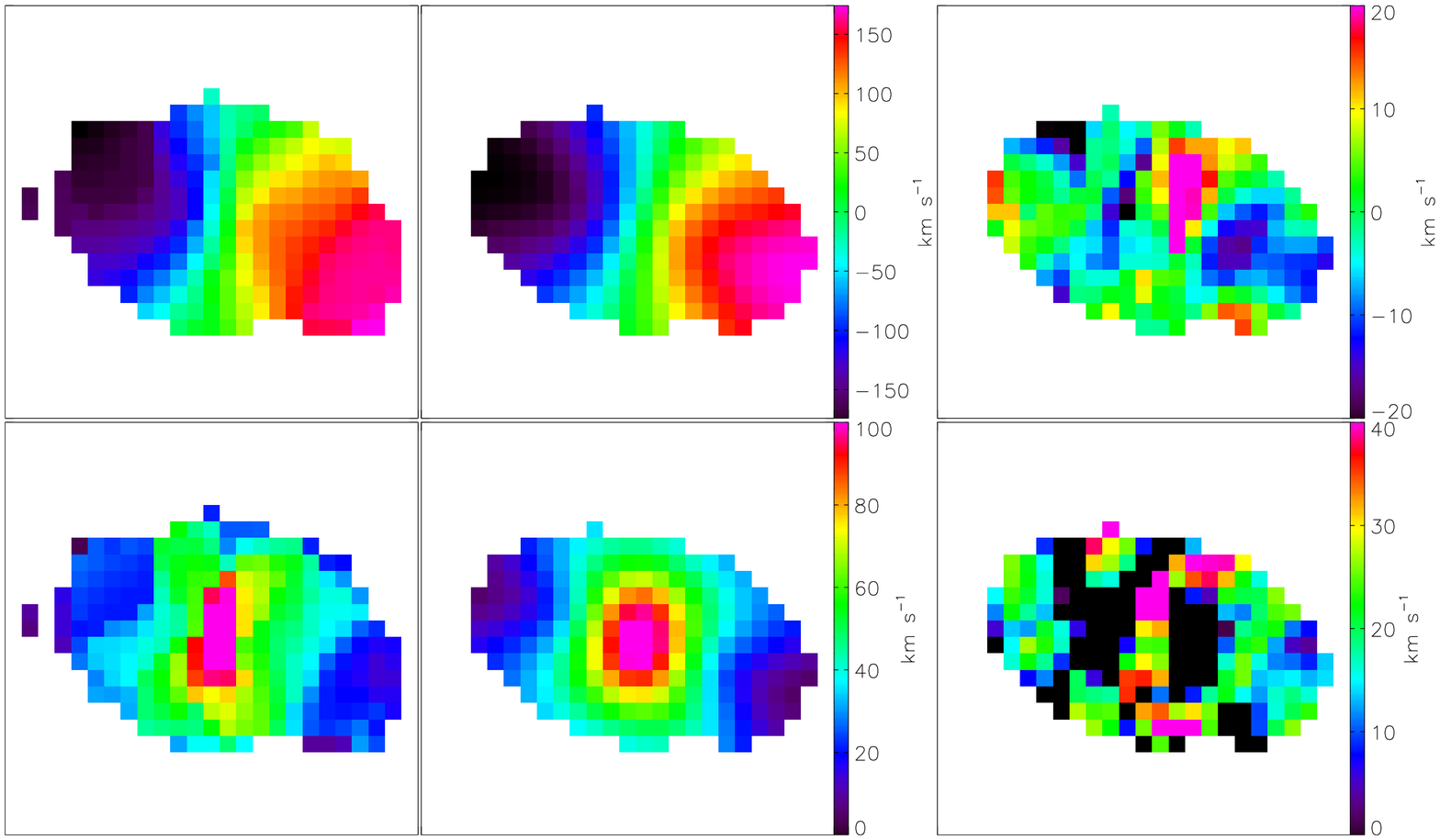}
\end{center}
\caption{Example of comparison between \hz~simulated data (left
column) and \hz~model mimicking the data (middle column) for the
galaxy UGC 7901. A ``flat model'' has been used here. Top
line: \vf. Bottom line: \vdm. The difference between the simulated
\hz~data (left columm) and the model (middle columm) is given for
both the \vf~and the \vdm~(quadratic difference) on the right
column. The velocities are given by the rainbow scales on the
right side of the images.} \label{residualvelocitydispersionmaps}
\end{figure}

The local velocity dispersion component is also affected by
the low spatial resolution. Thus if we consider that the local
velocity dispersion of the gas depends on the gravitational
potential, we have to correct the velocity dispersion component
from this effect. It is thus necessary to use a velocity
dispersion model. In the present study, we avoid this by assuming
that the local gas velocity dispersion is almost constant as
observed in local galaxies
(\citeauthor{Epinat:2009}, in preparation).
Thus, we measure
the local velocity dispersion as the mean value of the
\vdm\ quadratically corrected from the velocity shear term derived
from the \vf\ modeling (illustrated in Figure
\ref{residualvelocitydispersionmaps}).
\citet{Weiner:2006} used both velocity and velocity dispersion to constrain their model, using a constant
velocity dispersion model.
This is an alternative approach from the method used in this paper. This has been discussed
in section \ref{fitting_method}.

The velocity dispersion estimate also depends on the spectral
resolution of the data. Our data that have a very high spectral resolution better
than 10000 are probably not affected by spectral resolution effects.
Spectral resolution effects will be studied in a
forthcoming paper since we aim at probing spatial
resolution effects only in the present study. Another difficulty with this parameter is its
sensitivity to the \snr~that is usually low for \hz~observations.
We do not consider this effect here since our data are not
affected by a low \snr. Moreover, if a constant local velocity
dispersion model is assumed, all the points of the map should have
the same velocity dispersion. There should be at least several
points in the map with a sufficient \snr~in order to do an
accurate measurement.

\begin{figure}
\begin{center}
\includegraphics[width=8.5cm]{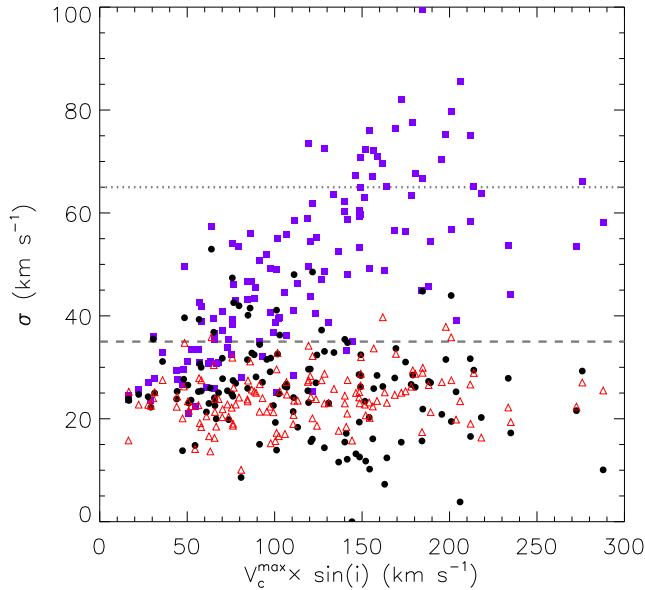}
\end{center}
\caption{Velocity dispersion as a function of projected maximum
velocity measured on $z=0$ \vfs. Each point represents a galaxy.
Blue squares correspond to the seeing-induced central velocity dispersion
measured on $z=1.7$ maps (without applying any corrections).
Red-open triangles represent the mean velocity dispersions
measured on $z=0$ galaxies.  The black dots correspond to the mean
velocity dispersion measured on corrected \vdms~of $z=1.7$
galaxies using a ``flat model''. The grey dashed and dotted lines respectively indicate the mean velocity dispersion in the IMAGES $z\sim0.6$ sample and in a sample of forty-two $1<z<3$ objects observed with OSIRIS and SINFONI.} \label{sigmaZvsVmaxsini}
\end{figure}

We conclude that a model is helpful to disentangle the velocity shear present in the \vdm\ from the local velocity dispersion. This is dramatically true for galaxies with a small beam smearing parameter.

\subsubsection{Velocity dispersions vs rotation velocities}

In Figure \ref{sigmaZvsVmaxsini}, the velocity dispersion is
plotted as a function of the projected maximum velocity (observed
at $z=0$) corrected for the inclination. Red-open triangles
represent the velocity dispersion measurements for the local data.
No dependency is observed with the projected maximum velocity.
The local velocity dispersion does not depend either on the total radial velocity amplitude
of the galaxy, suggesting that it does not depend on the galaxy mass.
Blue squares represent seeing-induced velocity dispersions measured at the center
of $z=1.7$ \vdms\ and show a clear correlation with the projected
maximum velocity. However, a large scatter
is observed. This may be explained by the dependency of rotation curve shape
with the true maximum velocity
\citep{Rubin:1985,Persic:1996,Catinella:2006}, in particular, the
inner gradient is larger for fast rotators. These fast rotators
observed with a low inclination should thus present a higher
central velocity dispersion peak than slower rotators observed
with a high inclination. This trend shows that the central
velocity dispersion gives an indication about the shape of the
inner rotation curve as well as the maximum velocity. The black
dots represent the mean velocity dispersion measured on corrected
\vdms\ using a ``flat model'' being the one that statistically allows the best recovery of the local velocity dispersion (see Table \ref{table_success}). By comparing the error on the corrected velocity dispersion and the beam smearing parameter $B$ (not plotted), we note that the correction is
statistically underestimated for galaxies with $B<2$, and often overestimated for other
galaxies, probably due to both an insufficient resolution for the
line flux maps and the \rc~shape that rises rapidly for faster
rotators. However, the correction is satisfactory since no strong
correlation is seen anymore with the projected maximum velocity.
Moreover, due to its low local velocity dispersion, the GHASP sample provides a strong constrain on the method. Indeed, the velocity shear contribution to the blurred \vdms\ may be negligible for dispersion-dominated galaxies.

In conclusion, we have noticed that the mean gaseous local velocity dispersion does not depend on the mass for nearby galaxies contrarily to the projected sample. We have shown that the model we used for the projected galaxies is suitable to model high redshift kinematics.
Indeed, it allows to remove the unresolved velocity shear contribution due to beam smearing and thus to recover the uniform velocity dispersion observed in nearby galaxies.

\subsubsection{Velocity dispersion estimation used for the IMAGES sample}
\label{vd_images_method}

\citet{Flores:2006} used the minimum observed value in the
\vdms~in order to have an estimate of the local velocity
dispersion.  Indeed, it is necessary to discard from this
measurement all the pixels affected by the velocity shear.

\begin{figure}
\begin{center}
\includegraphics[width=8.5cm]{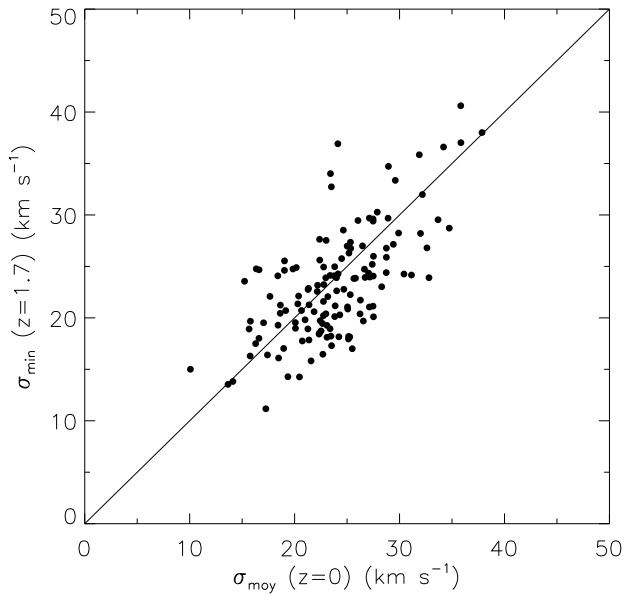}
\end{center}
\caption{Minimum velocity dispersion of $z=1.7$ galaxies as a
function of the mean velocity dispersion at $z=0$.}
\label{sigminvsvmsini}
\end{figure}

To test this method with our reference sample, the
minimum velocity dispersion of the uncorrected and uncut \vdm~has
been compared to the mean velocity dispersion at $z=0$. We find
that using such an estimate, the velocity dispersion is
underestimated by a mean factor around $2$. We obtained a good
agreement by estimating the velocity dispersion as the mean of the
points with the 20\% lowest velocity dispersion values on the
redshifted \vdm, not limited to the optical radius. The comparison
between this estimate and the mean velocity dispersion at $z=0$
is presented in Figure \ref{sigminvsvmsini}. Such an estimate has
been motivated by the fact that $z=0$ velocity dispersion
fluctuates from one region to another one. Moreover, the velocity
dispersion is often slightly lower in the outer parts of galaxies
(\citeauthor{Epinat:2009} in preparation), thus, the mean velocity dispersion at $z=0$
should be larger than the outer velocity dispersion. Such velocity
dispersion measurements depend on the map extent, and thus on the
signal-to-noise ratio of \hz~observations. This makes the
estimator sensitive to this parameter in particular for rotation-dominated galaxies. Indeed, for dispersion-dominated galaxies, the local velocity dispersion may be easier to recover as the velocity shear is relatively less important.
Note that this comparison
with FLAMES/GIRAFFE data only indicates a trend since the sampling
used in the present paper does not match the sampling of IMAGES dataset.

The conclusion is that this method allows to estimate the local velocity dispersion without the help of any model. However, it is very sensitive to both the \snr\ and to the radial extent of the galaxy.

\subsubsection{Unresolved beam and line-of-sight effects}

The GHASP sample of local galaxies allows to test if high local
velocity dispersions observed at high redshift may be the result
of the integration, within a seeing disk, along the line-of-sight
of individual HII regions.

From an observational point of view, in GHASP disks, the very central regions being excluded, the typical size of a bin for which the \ha~emission has a \SNR~$\sim7$ ranges from less than $0.1~kpc$ (one pixel) for the intense \ha~knots to $\sim0.5~kpc$ for the most diffuse regions \citep{Epinat:2008b,Epinat:2008a}.

The local velocity dispersion within those bins
(\SNR~$\sim7$) ranges from $10$ to $30$\kms. When these local
galaxies are projected at high redshift, a seeing disk
($\sim0.5$\arcsec) may thus contain more than $100$ bins, mixing
along the line-of-sight the velocity components of several tens to
several hundreds of individual regions. Taking into account the
local velocity dispersion at $z=0$ and the number of regions
integrated within a seeing disk at high redshift, even with the
high spectral resolution of GHASP (which is not reached by far by any IFU
spectrographs on 8-10 meter-class telescopes), the different
components are undiscernible in the spectrum along the line-of-sight.

In conclusion, one has no way to know if the large local velocity
dispersions seen in \hz\ galaxies are due to very large, extended
and massive clumps or, at the opposite, to the addition and the superposition along
the line-of-sight within a seeing disk of a large amount of
individual smaller clumps.
This should be adressed using high resolution observations of the luminosity distribution (HST, AO or future JWST imaging).

\subsubsection{Velocity dispersion evolution with the redshift}

In order to study the evolution of the velocity dispersion with the redshift, we have compared the GHASP local sample with IMAGES sample (at $z\sim 0.6$) and with $z>1$ samples observed with SINFONI and OSIRIS.
We have estimated the minimum value of the \vdm\ for each of the 63 galaxies of the IMAGES sample following \citet{Flores:2006} and we have used the velocity dispersion values given by the authors for 42 $1<z<3$ galaxies observed with SINFONI (SINS and MASSIV pilot run) and OSIRIS.

At $z=0$, the GHASP sample used in this paper provides a mean local velocity dispersion of $24\pm5$\kms. The mean local velocity dispersion for the whole IMAGES sample is $35\pm18$\kms\ while it reaches $\sim65\pm25$\kms\ for $1<z<3$ galaxies (as illustrated by the dashed and dotted lines in Figure \ref{sigmaZvsVmaxsini}).

Moreover, it is interesting to notice that the mean local velocity dispersion in the IMAGES sample does not significantly differ for ``rotating disk'' ($37\pm 10$\kms), ``perturbed rotation'' ($34\pm24$\kms) and ``complex kinematics'' ($35\pm17$\kms). Using an alternative approach to compute the local velocity dispersion (excluding the central hot region and weighting by the \snr\ after a 1-sigma bootstrapping), \citet{Puech:2007} estimated slightly higher values ($\sim45$\kms), but the previous conclusion does not change.
This may indicate that different physical mechanisms (cosmological gas accretion, galaxy accretions, turbulence generated by self-gravity and/or star formation, etc.) may occur for galaxies having different histories and however lead to velocity dispersions having typically the same value.
Alternatively, considering that a fraction of IMAGES ``perturbed rotators'' may be classified as ``rotating disks'' (see section \ref{classificationflores}), this could explain why no clear difference in the velocity dispersion is observed between both categories.

Contrarily to what is observed for nearby galaxies (see Figure \ref{sigmaZvsVmaxsini}), as already noticed in \citet{Epinat:2009c}, for $z>1$ galaxies, the maximum rotation velocities decreases when the local velocity dispersion increases (if we exclude the values from \citealp{Law:2009}, since the latter are limited to the very inner part of the galaxies). Indeed, for the local GHASP sample, when no correction for beam smearing is applied, the velocity dispersion increases with the maximum rotation velocity. When the correction is applied, the velocity dispersion of these galaxies is not correlated to the maximum rotation velocity.

In conclusion, we note a clear and continuous increase of the local velocity dispersion with the redshift. This indicates an evolution of the galactic dynamics through the ages, from eleven Gyr ($z\sim2.5$) to six Gyr ($z\sim0.6$) up to now ($z\sim0$). This might be due to the evolution of the dynamical support (dispersion towards rotation via e.g. violent relaxation processes) and/or to the evolution of non-circular motions (instabilities due to the presence of bars, etc.) and/or chaotic motions (turbulence, energy injection due to high star formation rates and/or AGNs).

\subsection{Gravitational support} \label{section_gravitationalsupport}

The ratio of the maximum circular rotation velocity $V_c^{max}$
and the local velocity dispersion $\sigma$ measures the nature
of the gravitational support of a system in equilibrium. A high
circular velocity compared to velocity dispersion
($V_c^{max}/\sigma>1$) is the signature of a rotation-dominated
gravitational support, whereas a lower ratio
($V_c^{max}/\sigma<1$) is the signature of a dispersion-dominated
system, as it is the case for elliptical galaxies.
For nearby spirals, characteristic values for the local gaseous velocity
dispersion $\sigma$ range from 10 to 40\kms\ (see Figure \ref{sigmaZvsVmaxsini}).

\begin{figure}
\begin{center}
\includegraphics[width=8cm]{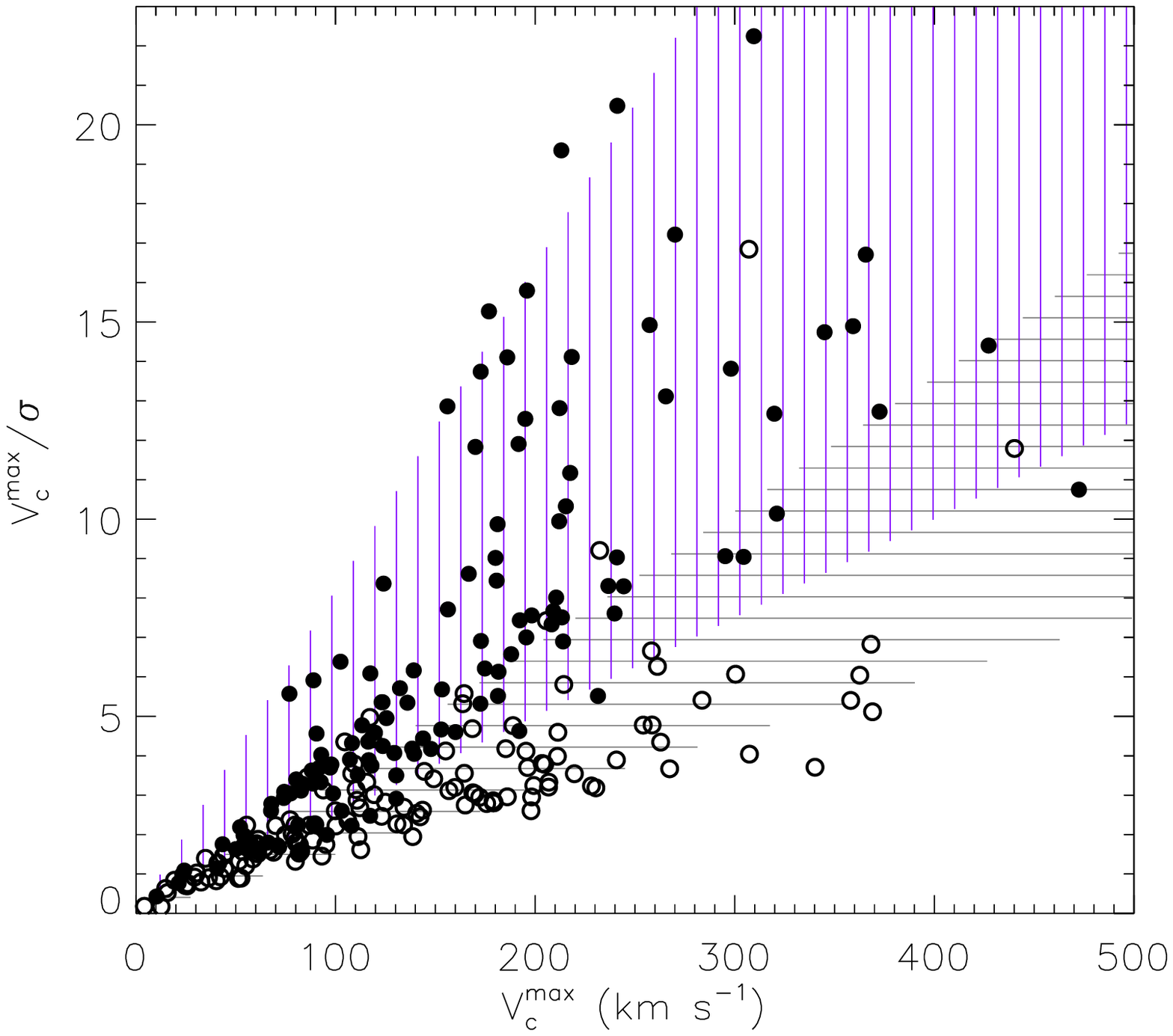}
\includegraphics[width=8cm]{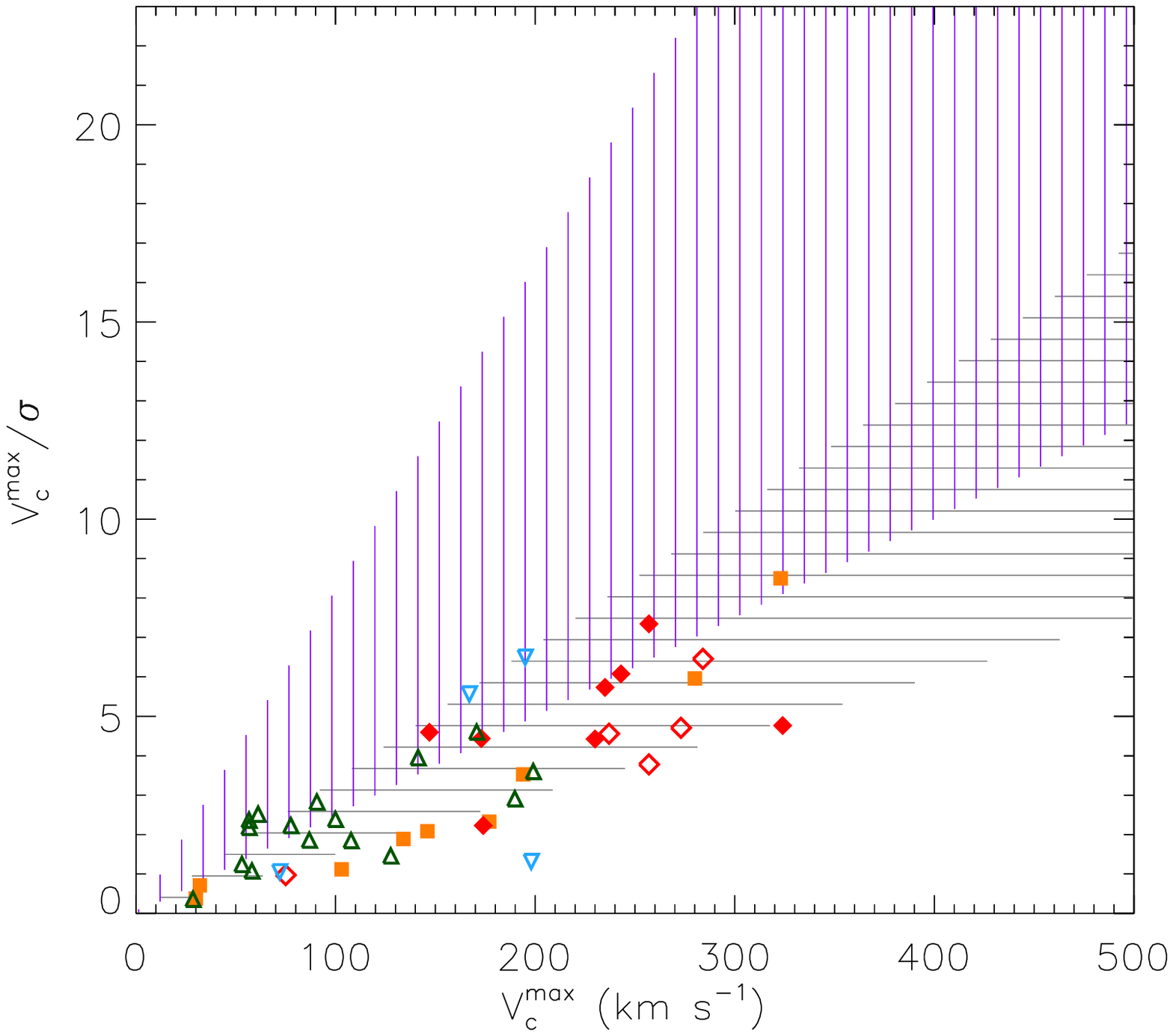}
\end{center}
\caption{Ratio between the maximum rotation velocity and the mean velocity dispersion as a function of the maximum rotation velocity.
\textbf{Top.} GHASP projected sub-sample. Circles: values uncorrected for the beam smearing. 85\%\ of these circles are within the grey horizontal hatchings zone.
Dots: values corrected for the beam smearing using the rotating disk modeling. 85\%\ of these dots are within the purple vertical hatchings zone. The grey and purple hatchings are reported in the Bottom figure for reference.
\textbf{Bottom.} Observed high redshift galaxies. Open symbols correspond to observations using AO. Red rhombuses: SINS rotating disks at $z\sim2$ \citep{Cresci:2009}. Orange squares: MASSIV pilot run galaxies at $z\sim 1.5$ \citep{Epinat:2009c}. Green triangles: $z\sim 3$ \citet{Law:2009} OSIRIS observations. Blue upside down triangles: rotating disks in \citet{Wright:2007,Wright:2009} at $z\sim 1.5$ observed with OSIRIS.} \label{gravitationalsupport}
\end{figure}

An interesting possible probe of the state of dynamical galaxy evolution is
given by the behavior of this ratio $V_c^{max}/\sigma$ with the
redshift.
To test the beam smearing effects on the measure of the ratio $V_c^{max}/\sigma$,
we have computed it for the projected GHASP sub-sample and
compared it to actual distant galaxies.
The ratio $V_c^{max}/\sigma$ as a function of $V_c^{max}$ has been
plotted in Figure \ref{gravitationalsupport} for both local (top) and high redshift observed galaxies (bottom).

We have plotted GHASP local galaxies projected at $z=1.7$ using dots. These points have been computed using the maximum velocity $V_c^{max}$ derived from kinematics modeling and the velocity dispersion $\sigma$ corrected for beam smearing effects as explained in section \ref{velocity_dispersion_analysis}.
Since the GHASP sample is mainly composed of rotation-dominated galaxies, our sub-sample shows values of the ratio $V_c^{max}/\sigma$ lower than 2 only for very slow rotators ($V_c^{max}<100$\kms) and values ranging from $5$ to $20$ for rotators ranging from $100$ to $400$\kms.
This ratio is strongly correlated with $V_c^{max}$ for local galaxies (slope $0.048~km^{-1}~s$),
but with a large scatter. The correlation is expected since the velocity dispersion of the gas is rather constant with the maximum velocity for local galaxies, and the large scatter is due to the difficulty to recover both $V_c^{max}$ and $\sigma$ for the projected galaxies because of beam smearing effects.
We have also plotted with circles the values without any correction for the beam smearing: the maximum velocity being computed from the rotation curve along the major axis and the velocity dispersion being estimated as the mean of the uncorrected velocity dispersion maps. The corresponding regions in which 85\%\ of the points are lying have been reported on both plots of Figure \ref{gravitationalsupport} using vertical purple and grey horizontal hatchings. They respectively refer to beam smearing corrected and uncorrected points (see sections \ref{maxvr} and \ref{velocity_dispersion_analysis}). Since the uncorrected maximum velocity and velocity dispersion are respectively underestimated and overestimated, grey and purple areas show only a small overlap. This grey area represents the ``worst'' estimation for the redshifted dataset.

From long slit spectrography, line-of-sight kinematic line widths
($\sigma$) of several hundreds good-quality measurements galaxies at $z\sim 1$,
\citet{Weiner:2006} roughly divided their sample into rotation
($V_c^{max}/\sigma>1$) and dispersion-dominated galaxies
($V_c^{max}/\sigma<1$). Dispersion-dominated galaxies are blue,
mostly irregular and are not elliptical galaxies.  These authors
conclude that these galaxies probably have a disordered kinematics
that is integrated over by the seeing.

\citet{Forster-Schreiberetal:2006} found that their three best
rotators candidates at $z\sim2$ show $V_c^{max}/\sigma\sim2-4$, they concluded
that these very gas rich disks are dynamically hot, geometrically
thick and unstable to global star formation and fragmentation.
These authors argue that these observations may be described by
simulations \citep{Immeli:2004a} of gas-rich disks in which clumpy
fragmentation disks are unstable and star forming clumps evolve by
fuelling the galaxy center by dynamical friction and finally form
a central bulge on $\sim1$ Gyr time scale.
\citet{Genzeletal:2008} and \citet{Cresci:2009} extended the SINS sample first described by \citet{Forster-Schreiberetal:2006} to 13 rotating disks candidates for which local velocity dispersion has been measured. They found $V_c^{max}/\sigma\sim1-6$ with a mean value of 4.6.

Using 16 galaxies in the same range of redshift ($z\sim2$), \citet{Law:2009} found $V_c^{max}/\sigma\sim0.1-1$ with a mean value of 0.5. These values are notably different from the SINS sample. As mentioned in section \ref{zorglub}, the mean radius of the galaxies observed by \citet{Law:2009} is eight times lower than for SINS galaxies. Moreover, their parameters (maximum velocity and velocity dispersion) are not corrected for the beam smearing and for inclination. Indeed, their maximum rotation velocity is the half of the whole shear whereas the velocity dispersion is flux-weighted, i.e. dominated by the inner regions. Similarly to \citet{Genzeletal:2008}, \citet{Law:2009} concluded that the high velocity dispersion they observe may be neither a merger nor a disk, but rather the result of instabilities related to cold gas accretion.

In the redshift range of $1.2<z<1.7$, a compilation of 13 galaxies classified as rotators extracted from \citet{Wright:2007}, \citet{Wright:2009} and \citet{Epinat:2009c} provides $V_c^{max}/\sigma\sim0.4-8.5$ with a mean value of 3.1. These values are comparable to the ones found at higher redshifts. \citet{Epinat:2009c} argued that, considering the samples presently available, several processes may drive galaxy evolution. For instance, for their perturbed rotators, it is  not straightforward to disentangle whether the high velocity dispersion is the result of gas accretion or gas rich minor mergers.

\citet{Bournaud:2008} found $V_c^{max}/\sigma\sim1-2$ (not
corrected for inclination) for a clumpy galaxy at $z=1.6$ and
proposed an internal fragmentation formation scenario of a
gas-rich primordial disk becoming unstable. In comparing their
SINFONI observations to numerical models, \citet{Bournaud:2007}
concluded that (i) complex morphology can result from the internal
evolution of an unstable gas-rich disk galaxy and (ii)
irregularities in the \vf~($\sim$ several tens of\kms) can be
explained by clump-clump interactions that cause the individual
velocity of each clump to differ significantly from the initial
rotation velocity.

In Figure \ref{gravitationalsupport} (bottom), we have over-plotted the points corresponding to real \hz\ galaxies. The values of these points have been corrected for beam smearing.
Green triangles correspond to the galaxies observed by \citet{Law:2007,Law:2009} at $z\sim 3$ with OSIRIS. These authors did not provide the inclination of the disks, thus we have used a mean statistic inclination of 45\degr\ to compute the maximum rotation velocity in the galaxy plane. Moreover, we have corrected these velocities for beam smearing effect using the lower limit given in equation \ref{vmax_corr}. This correction only provides lower values for $V_c^{max}$ since, due to the very small extent of these objects, the plateaus may not be reached. The local velocity dispersion has also been estimated from the velocity dispersion maps using the estimation given in section \ref{vd_images_method} instead of the flux-weighted velocity dispersion provided in \citet{Law:2009} uncorrected for beam smearing effects.
Blue upside down triangles are the rotating disks also observed with OSIRIS by \citet{Wright:2007,Wright:2009} at $z\sim1.5$. Orange squares are galaxies part of MASSIV pilot run \citep{Epinat:2009c} and red rhombuses correspond to SINS rotating disks \citet{Cresci:2009}, both observed with SINFONI. Open
symbols correspond to AO corrected observations. SINFONI setup in
natural seeing observation has a pixel size of 0.125\arcsec\ and a
seeing around 0.5\arcsec. Observations using AO use a pixel size
of 0.05\arcsec~with a seeing up to 0.2\arcsec. SINFONI spectral
resolution is 4500 (70\kms) in K band, 3000 (100\kms) in H band
and 1900 (160\kms) in J band. OSIRIS observations are using AO
devices in order to have a seeing up to 0.1\arcsec~and use a
pixel size of 0.05\arcsec~with a spectral resolution of 3600
(85\kms).

All these authors observed that, for a given circular velocity, $V_c^{max}/\sigma$ is lower for \hz\ galaxies than
for local galaxies.
Several of them corrected for beam smearing but others did not.
Since beam smearing artificially causes lower values of $V_c^{max}/\sigma$, the values for the projected sample obtained without correcting for the beam smearing effects provide a lower limit for nearby disks.
Figure \ref{gravitationalsupport} shows that $V_c^{max}/\sigma$ values for \hz\ galaxies, derived taking into account the seeing, are below this lower limit. This is thus a strong evidence for dynamical evolution between $z > 1.5$ and $z\sim 0$.

We observe that the correlation for \hz\ galaxies has a lower slope
($0.014~km^{-1}~s$) than for local galaxies, showing that \hz\ galaxies
are more dispersion-dominated.
Ideally, to probe the gravitational potential, dispersion measurements should be done on the non collisional stellar component rather than on the collisional gas component.
However, stellar kinematics are unreachable at \hz\ with the current instrumentation.
We could also notice that
it is possible that the extent of \hz~galaxies may be lower than
for \lz~galaxies and thus the maximum velocity may be missed.
Another care is that it is probable that \hz~surveys have
selection biases.
Moreover, all these \hz~observations have a much lower spectral
resolution than the local sample. We expect from this low spectral
resolution that velocity dispersion correction is less accurate
and thus that velocity dispersion measurements have much larger
uncertainties.

\subsection{Tully-Fisher relation}
\label{section_tf}

\begin{table*}
\caption{Fits of local and distant \TF\ relation.}
\label{table_tf}
\begin{tabular}{lcccl}
\hline
Sample & Band & Slope $a~^{(a)}$ & Zero point $b~^{(a)}$ & Comment \\
\hline
GHASP local whole sample              & B & $-7.2\pm1.2$   & $-3.97~^{(d)}$ & Free slope, from \citet{Epinat:2008b} \\
GHASP local sub-sample, RD$~^{(b)}$ $+$ PR$~^{(c)}$     & B & $-6.2\pm2.2$   & $-6.23~^{(d)}$ & Free slope \\
GHASP projected sub-sample, RD $+$ PR & B & $-5.2\pm1.9$   & $-8.68~^{(d)}$ & Free slope \\
GHASP local sub-sample, RD            & B & $-7.4\pm1.7$   & $-3.50~^{(d)}$ & Free slope \\
GHASP projected sub-sample, RD        & B & $-6.6\pm1.4$   & $-5.38~^{(d)}$ & Free slope \\
GHASP projected sub-sample, RD        & B & $-7.4$         & $-3.67\pm0.10$ & Fixed slope \\
SDSS local subsample                  & K & $-6.88\pm0.57$ & $-6.54\pm1.33$ & Free slope, from \citet{Puechetal:2008} \\
IMAGES $z\sim0.6$, RD                 & K & $-7.24\pm1.04$ & $-5.07\pm2.37$ & Free slope, from \citet{Puechetal:2008} \\
IMAGES $z\sim0.6$, RD                 & K & $-6.88$        & $-5.88\pm0.09$ & Fixed slope, from \citet{Puechetal:2008} \\
\hline
\end{tabular}
\\
(a): \TF\ relation: $M=a\times \log{V_c^{max}}+b$.
(b): RD refers to rotating disks.
(c): PR to perturbed rotators.
(d): Error bar on the zero point is not provided for the GHASP sample when the slope is free since it is very sensitive to changes in the slope. Keeping the slope fixed would lead to errors around $0.1$.\\
\end{table*}

The \TF\ relation is a way to constrain galaxy-formation models as
well as to probe the dynamical stability of galaxies.

We computed the B-band \TF\ relation for both local and redshifted
GHASP sub-sample. As it has been done in \citet{Epinat:2008b,Epinat:2008a}, the
\TF\ relation has been computed as the mean relation from the one
obtained using a fit on absolute magnitudes as dependent variable
and the one obtained using a fit on velocities as dependent
variable. The errors are estimated as the difference between those
two fits.

The GHASP sample is limited to rotating disks. It does not contain strongly interacting galaxies nor galaxies supported by random pressure. Nevertheless, as discussed in section \ref{classificationflores}, once projected and following the classification done by \citet{Flores:2006}, the kinematics of some galaxies may resemble to perturbed rotators.
We first did not exclude any galaxy since we want to compare the
scatter using the same projection parameters for both. Moreover,
at high redshift, all the galaxies could be interpreted as
rotating disks since they all present a velocity gradient at high
redshift. \citet{Epinat:2008b} found a slope of $-7.2\pm1.2$ from
the whole GHASP sample but excluding several galaxies because of
their low inclinations (that induce strong uncertainties), or
because their maximum velocity is probably not reached. From the
local sub-sample defined in the present paper, the slope is
estimated to $-6.2\pm2.2$. This lower slope (although within the errors) can be explained by
selection biases (see section \ref{subsample}): we only excluded the smallest galaxies, but not
those that do not reach their maximum velocity or those with an
inclination lower than $25$\degr. In particular, low mass galaxies
hardly reach their maximum velocity within the optical radius, and
\citet{Epinat:2008b} have shown that the fastest rotators present a
lower slope since they are less luminous than expected from local
\TF\ relation. The \TF\ relation derived from the redshifted dataset
is still lower ($-5.2\pm1.9$) but compatible with the one derived from the
local sub-sample within the error bars (see Table \ref{table_tf}). This lower slope can be
explained by beam smearing effects. Indeed, as shown in
section \ref{maxvr}, the maximum velocity is more difficult to recover for
slow rotators than for fast rotators for which the model is better constrained. The scatter for those two slopes are rather similar
and larger than the one derived by \citet{Epinat:2008b}, due to
selection effects, in particular, the inclination inducing the
strong scatter.

Absolute K-band magnitudes have been obtained for the IMAGES sample \citep{Flores:2006}. Unfortunately, K-band photometry for the GHASP sample is not available, thus,
we compare the B-band \TF\ relation for the GHASP sample to the K-band \TF\ relation for the IMAGES sample
\citep{Puechetal:2008}. This induces color-luminosity biases (e.g. \citealp{Sakai:2000,Verheijen:2001}). This data obtained with FLAMES/GIRAFFE instrument have a spectral resolution very similar to that of the GHASP redshifted data.
Let's notice however that GIRAFFE has a lower spatial resolution ($0.8$\arcsec~seeing and $0.52$\arcsec/pixel) than our simulations and a small spatial extent (6 by 4 spaxels).

Since a simple magnitude correction
consists in adding a given value, the slopes should be comparable.
Our slope is found to be lower than theirs (see Table \ref{table_tf}). Their
magnitude range however is tighter than ours as shown in Figure
\ref{histo_masses} and our sample thus contains lower luminosity
systems for which the maximum velocity has not been derived with
confidence. The slope determination is also highly dependent on
the fitting method used. Surprisingly, assuming no evolution effects, we would have expected from IMAGES data a \TF\ relation with a lower slope due to the instrumental spatial resolution differences between both samples. Indeed, the slope of the \TF\ relation obtained from the projected GHASP sample is lower than the one obtained from high spatial resolution \rcs\ (see Table \ref{table_tf}). In addition, we have checked that this effect remains true whatever the magnitude range, in particular when the faintest galaxies are excluded.

\begin{figure}
\begin{center}
\includegraphics[width=8.5cm]{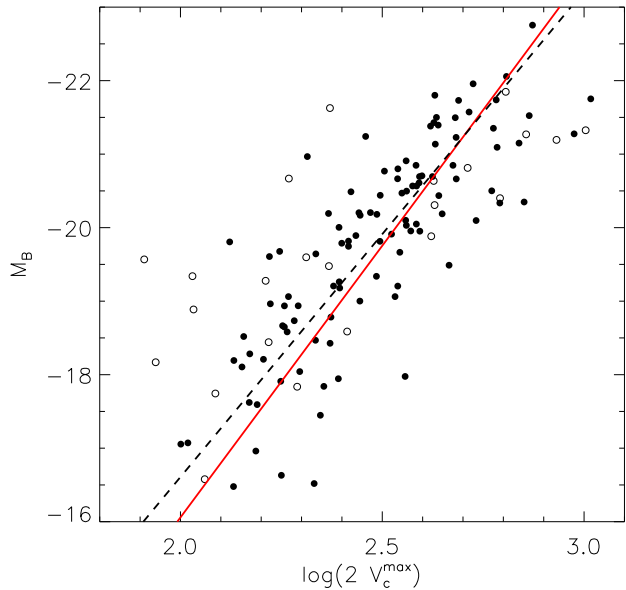}
\end{center}
\caption{\TF~relation at $z=0$ -- red line -- compared with \TF~relation computed at
$z=1.7$ -- black dots and black dashed line --.  Open symbols
correspond to galaxies that would not be classified as rotating
disks at redshift 1.7. The two linear regressions (red line and
black dotted line) are computed using only galaxies classified as
rotating disks.} \label{TF}
\end{figure}

In order to derive the \TF\ relation in the same conditions as
\citet{Puechetal:2008}, we also computed a \TF~relation using only
galaxies that would be classified as ``rotating disks''. Indeed,
we noticed on the \TF~relation plots that some of the galaxies
that \citet{Puechetal:2008} classified as ``complex kinematics'' and ``perturbed rotators'' have
their counterparts in our sub-sample, corresponding to galaxies
that would be misclassified. By using ``pertubed rotators'' and ``rotating disks'' to derive
their slope, they would probably have found a lower slope. Indeed, this is the trend that we observe when we use the whole GHASP sub-sample (see Table \ref{table_tf}). In
Figure \ref{TF} GHASP galaxies classified as ``rotating disks''
correspond to full points, and galaxies misclassified are
displayed as open circles. The red continuous and black dashed
lines correspond to the \TF~relation computed when using only galaxies that
would be classified as ``rotating disks'' respectively for local
and redshifted samples. These determinations are consistent with
local determination: from the non redshifted sample we find a
slope of $-7.4\pm1.7$ and for the redshifted sample we find a
slope of $-6.6\pm1.4$ that is in agreement with the one
derived at redshift $z\sim 0.6$ by \citep{Puechetal:2008}.
However, the trend is still to find a lower slope for the
redshifted sample, although the difference is lower than the statistical error bars. The scatter also then becomes lower. Indeed,
using only galaxies that would be classified as rotating disks
imply that most of the low inclination systems are excluded as
well as those with a solid body rotation curve for which the
maximum is not determined with confidence. The use of the GHASP
sample may indicate that the differentiation between ``rotating
disks'' and ``perturbed rotators'' could be incorrect since GHASP
galaxies misclassified as ``perturbed rotators'' are actually
``rotating disks'' and have the same behavior in our \TF\ relation
that in $z\sim 0.6$ relation. However, this classification enables
in fact to exclude galaxies for which the lack of spatial
resolution induces biases in the parameters determination.

Since \citet{Puechetal:2008} used K-band magnitudes, we cannot
compare directly the \TF\ zero point. However, the comparison
between local and \hz~zero point can be probed. The \TF\ zero point
$b$ is defined by \citet{Puechetal:2008} as
\begin{equation}
M=a\times \log{V_c^{max}}+b
\label{tf_eq}
\end{equation}
In order to compare the zero point of their $z\sim 0.6$ sample and
local value (from a SDSS sub-sample), they fixed the slope of the
\hz~\TF~to the local slope. Thus, they found a galaxy brightening
of $0.66\pm0.14$ magnitude from $z\sim0.6$ to $z=0$ that would
indicate that galaxies double their stellar mass between these two
epochs. They conclude that rotating disks observed at $z\sim0.6$
should be rapidly transforming their gas into stars. We did the
same comparison on our local and redshifted galaxies by fixing the
slope to $-7.4$ found from the local galaxies classified as
``rotating disks''.
We found that beam smearing effects cannot account for the brightening suggested by \citet{Puechetal:2008} since it leads to
a difference of the zero point in \TF\ relation equivalent to a loss of
brightness of $0.17\pm0.10$ magnitude.
Considering that the
uncertainty is of the same order of magnitude than the difference
of zero point, we conclude that both local and redshifted zero
points are compatible. This result would indicate that the signicative evolution of the \TF\ zero point with respect to the error bars observed by \citet{Puechetal:2008} is not accounted for by beam smearing effects.
\\
\\
\noindent
In this section, we have discussed:
\begin{itemize}
 \item The comparison between the local \TF\ relation for the GHASP sample with the \TF\ relation for the GHASP sample projected at \hz. The conclusions are that the slope of the \TF\ relation is lowered by beam smearing effects within the error bars and that the zero point of this relation is not significantly modified. This supports that the evolution of the zero point in the \TF\ relation, observed by \citet{Puechetal:2008}, cannot be explained by beam smearing effects.\\
\\
 \item The comparison between the slope of the \TF\ relation for the GHASP sample projected at \hz\ with the slope of the \TF\ relation for the IMAGES sample. The conclusion is that the slopes of the \TF\ relations derived from the GHASP sample projected at \hz\ and from the IMAGES sample are compatible within the errors. Nevertheless, the latter comparison is limited (i) by the distributions in mass or in velocity of both samples that do not match, (ii) by the fact that the magnitude in the GHASP sample is computed in the B-band whereas it is computed in the K-band for the IMAGES sample and (iii) by the fitting methods used to compute the \TF\ relation coefficients that are not exactly the same.
\end{itemize}
\section{Discussion} \label{discussion}

A common explanation for massive galaxies having
irregular kinematics and high nuclear gas fraction is that they
may have undergone major mergers of gas-rich galaxies. On the
other hand, models linking cosmological simulations to
galaxy evolution (e.g. \citealp{Ocvirk:2008,Dekel:2009}) proposed
a scenario where galaxies at $z\sim2$ accrete significant amounts
of cold gas which form unstable gaseous disks.
\citet{Immeli:2004a}, followed by other authors,
relaunched an old debate in suggesting that, in early-stage
galactic disks, efficient gas cooling could have led to high cold
gas fractions, which then fragmented due to self-gravity and
collapsed to form a nuclear starburst.  The kinematics of the
brightest nebular emitting regions may be relatively featureless
and may dominate the entire line emitting regions through the
galaxy up to observable radii. The absence of shear may be a
transient effect vanishing to further lower surface brightness
ionized gas at later evolutionary stage.

\subsection{Formation and evolution of high redshift gaseous disk}

For rotating disks, at a resolution of a few kiloparsecs, it is
challenging to know whether the large velocity dispersion observed
in high redshift galaxies is due to (i) cold gas
accretion, (ii) gas rich minor merger accretion events (e.g.
\citealp{Semelin:2002}) or (iii) wet major mergers. The three
scenarios may fuel the disk in fresh gas.  Scenario (i)
might have two origins, internal or external to the galaxy.
Indeed, huge reservoirs of gas, gravitationally bound to the
galaxy \citep{Pfenniger:1994}, may fuel the galaxies in cold gas
as well as cold gas accretion flowing from the intergalactic
medium.  Both accretion mechanisms may lead to gas instabilities,
cloud fragmentation and collapse, thus finally to strong starburst
activity \citep{Immeli:2004b,Bournaud:2008}.
The existence of a disk in rotation does not prove that it has been formed by continuous gas accretion. Indeed, if the initial spins of the progenitors are not too different, old wet major merger events may produce a rotating disk after a timescale $>0.5$ Gyr (shorter than the nowadays timescale) indistinguishable from a disk formed by the two other mecanisms.
Alternatively, the presence of large reservoirs of gas around disk galaxies
\citep{Daddi:2008} indicates that large amounts of gas are
available to fuel the star formation.
If minor mergers (10:1
to 50:1) occur with a high frequency, relaxation processes eject
the pre-existing stars from the disk to the spheroid or to the
thick disk.
From an observal point of view, these stars are undistinguishable from the ones belonging to the thin disk.
The formation of a spheroid or a thick disk will not perturb significantly
the disk kinematics and its signature would be difficult to
detect directly.
Nevertheless, the stabilization of the disk due to these structures should be indirectly observable: they will diminish the star formation inducing less
sub-structures as \Ha\ and UV clumps.
A gaseous or stellar disk stable to all axisymmetric perturbations
requires the Toomre's parameters $Q>1$ \citep{Safronov:1960,Toomre:1964}.
Giant star-forming clumps observed in \hz\ galaxies, in which the star
formation is as high as $100-1000 M_{\odot}~yr^{-1}$, require high
turbulent speeds and a dense disk with few stars in a spheroid \citep{Bournaud:2009}.
The formation of these clumps requires that most of
the stars and the gas lie in a rotating disk during the clump formation,
otherwise $Q>1$, the disk is stable and massive clumps do not form.
Indeed, halo as well as stellar speroid stabilize the disk and make the
disk too stable to allow giant clumps to develop.
The distribution, size and mass of these clumps may be considered as indirect indicators of the disk formation history.
At the opposite,
smooth diffuse gas accretion is not supposed to be efficient to
form a stellar spheroid and instabilities dominate the disks and
are observable through deep imaging addressing the formation of
clumps. For very minors mergers (e.g. mass ratio $>$ 100:1), the
dwarf galaxies are dislocated by the tidal field once they
experience the gravitational field of the main galaxy. Torn by
tidal field, this kind of accretion resembles very much to diffuse
gas accretion. For a baryonic mass galaxy of $10^{10}M_{\odot}$,
these satellites have masses lower than $10^8M_{\odot}$. If these
galaxies exist, they are not detectable in observations or in
numerical simulations, due to the lack of spatial resolution.

Large turbulence in the neutral gas disk could be provided by energy dissipation due to rapid
external gas accretion (cosmological filaments, outer disk gas
reservoirs). The huge sizes, masses and velocity dispersions of
star-forming clumps still need to be understood.
\citet{Taniguchi:2001} favored a multiple merger origin similar to
what is observed in several local compact groups.
\citet{Noguchi:1999}, \citet{Immeli:2004a,Immeli:2004b}, \citet{Bournaud:2007,Bournaud:2008} proposed that, resulting from Jeans
instabilities, primordial gaseous disk could fragment into several
dense clumps. If the gas accretion is large and fast enough, the
disk may become unstable leading to clumps formation with Jeans
length-scale of $1-2$ kpc and Jeans mass-scale of
$\sim10^{8-9}M_{\odot}$. Unfortunately, the kpc-scale turbulence
in the neutral atomic gas component (or the molecular gas via its
CO content) has never been observed at high redshift. The presence
of large clumps indicates nevertheless that it should be higher
than in the local universe.

Disks observed at high redshift may be
short-lived and not the precursors of today disk galaxies.
Indeed, new pictures emerge in the literature in which young thick disks
form by cold flows \citep{Dekel:2009,Keres:2009} and other
types of diffuse gas accretion \citep{Semelin:2005}, bulges form by
internal and clump-driven evolution \citep{Elmegreen:2008,Genzeletal:2008}, and the thin disk forms later by further smooth accretion
(e.g., \citealp{Bournaud:2002}).
Models discussed in \citet{Bournaud:2007,Bournaud:2008} predict that
velocity shear tracing the initial gas rotation should be observed
but with high velocity dispersion as shown by the observations of
high redshift galaxies. In simulations including external gas
accretion \citep{Bournaud:2008}, the relatively thin initial disk
($700$ pc) becomes thicker ($\sim1-2$ kpc). This is due to
gravitational heating processes linked to clump formation
processes. Stars formed in clumps constitute the thick disk or merge in
the central bulge. The gas which has not been transformed in stars
during the clump phase cools down and falls down in the
pre-existing thin stellar disk.

The standard model indicates that massive galaxies formed earlier,
thus having accreted their mass earlier and having been unstable at
higher redshifts. As a consequence, their clumps should have been dispersed
in the bulge or in the thick disk at earlier epochs than for less
massive galaxies. Indeed, a bulge component seems to be already
present in the most massive galaxies in the SINS sample.

\subsection{What do observations of high redshift galaxies show?}

\subsubsection{Seeing-limited observations}

Observations with a seeing disk of $\sim4~kpc$ ($\sim 0.5$\arcsec) do not allow to sample the internal substructures of high redshift galaxies. Nevertheless, they are easier to obtain than AO ones. On the one hand, they do not require natural or laser guide star and, on the other hand, the larger pixel scale allow to observe the disk outskirts which have a lower surface brightness.

The bulk of SINS galaxies have been observed without AO (51 out of 63 galaxies) with a seeing disk of $\sim0.6$\arcsec.
Many SINS galaxies are bright and large, they have been
selected from previous long slit observations
\citep{Erb:2003,Erb:2006} on the basis of consequent velocity
shear and/or velocity dispersion. 
Among the 51 galaxies, 14 of them are classified as rotating disks \citep{Cresci:2009}. These authors did not point out specific conclusions linked to the absence of AO and invoke the need for gas accretion to form disk as suggested by the predictions of the lastest N-body/hydrodynamical simulations of disk formation and evolution (e.g. \citealp{Dekel:2009}).

In the frame of the MASSIV program, nine galaxies have been observed with a mean seeing of 0.65\arcsec\ in the redshift range $1.2<z<1.6$ during a pilot run. \citet{Epinat:2009c} found that six of them are compatible with rotators. They distinguished two rotating disks and four perturbed rotators showing a high velocity dispersion. For the MASSIV program, a special care is given in the selection of the targets.
No definitive conclusion can yet be drawn. In particular for their perturbed rotators, they concluded that the high velocity dispersion may be the signature of gas accretion as well as gas rich minor mergers.

In conclusion, the SINS survey and the MASSIV pilot run reach roughly the same conclusion that almost a third of high redshift galaxies has rotation-dominated disks, another third has dispersion-dominated disks while the last third is composed of merging galaxy candidates. This is consistent with previous results obtained with long slit spectroscopy data by \citet{Weiner:2006} and \citet{Kassin:2007} who found one third of dispersion-dominated disks from more statistically complete samples.
Forthcoming integral field spectroscopy data, like the MASSIV sample, will help in distinguishing between the various processes of galaxy formation acting at these redshifts.
Indeed, the MASSIV sample has been selected from the VIMOS-VLT Deep Survey \citep{Le-Fevreetal:2005}, which is both statistically representative of the overall population and volume complete, based on the measured masses and on-going star formation rate (\citeauthor{Contini:2008} et al., in preparation). 

\subsubsection{Adaptive optics observations}

Adaptive optics observations allow to reach the kpc-scale which is a large improvement to analyze the internal kinematics of high redshift galaxies.

In the frame of the SINS program, twelve galaxies have been observed with SINFONI assisted by AO \citep{Forster-Schreiberetal:2009}.
Only five of them, classified as rotating disks, have been published up to now \citep{Genzeletal:2008}.
The first observation by \citet{Genzeletal:2006} of a high redshift galaxy (BzK
15504 at $z=2.38$) with an angular resolution of 150 mas ($\sim1~kpc$)
exhibits a resolved velocity shear which is nevertheless not well fitted by a
simple disk model. With AO, the line-of-sight velocity dispersion remains
high at high radii ($\sigma\sim 60-100$\kms) and the residual
velocity map between the observed \vf~and the model (best-fitting
exponential disk) shows deviations larger than 100\kms.
\citet{Genzeletal:2006} argued that it may be explained by radial
gas inflows fuelling the central AGN.

The galaxy 1E0657-56 at $z=3.2$ being strongly lensed, SINFONI observations without AO of this object lead to a spatial resolution of
$\sim200~pc$ in the source plane, even better than AO resolution for non lensed galaxies. The \pvm\ within the central
kpc of this galaxy looks like a rotating L$_\star$ nearby spiral galaxy
\citep{Nesvadba:2006} suggesting that, in some cases at least, a
significant amount of mass could be already in place on small physical
scale at $z\geq3$.

OSIRIS instrument also assisted by AO has been used by \citet{Law:2007,Law:2009} and \citet{Wright:2007,Wright:2009} for observing a total of 25 high redshift galaxies.
\citet{Law:2009} provide 16 galaxies at $z\sim2-3$, including at most five rotating disks, resolved with a PSF $\sim110-150$ mas. These authors concluded that, even for galaxies showing clear velocity gradients, rotation may not be the dominant mechanism of physical support. They refuted a simple bimodal disk/merger classifiaction scheme but underlined the dynamical importance of cold gas accretion.
At lower redshifts ($\sim1.5$), \citet{Wright:2007,Wright:2009} have observed nine galaxies, four of them have been classified as rotating disks. Among these four cases, two look like local disks while due to their high velocity dispersion, the two other ones look more like unstable disks.

\subsubsection{High velocity dispersion in high redshift galaxies: comparison with local galaxies}

We have shown in this work that, although the seeing-limited
observations of intermediate and \hz\ galaxies (from $z\sim 0.4$ to $z\sim 3$) suffer from significant beam
smearing effects, it is not sufficient to explain the increase of velocity dispersion with the redshfift.
Moreover, AO observations of \hz\ galaxies
reaching the lower limit of the kpc-scale also display a high velocity
dispersion.
This unambiguously indicates a clear and continuous dynamical evolution in disk galaxies through the last eleven Gyr.
Three mechanisms act simultaneously and are responsible for gaseous velocity dispersion: turbulence due to local gravity, feedback linked to star formation processes and infall in the potential well of the galaxy. It is a challenging question to quantify the contribution of each processes.

Galaxies at earlier stages of evolution are observed to be very
different from nowadays galaxies. Their high star formation rate
of $\sim10^{2-3}M_{\odot}$ per year has no equivalent in the local
universe. This high star formation rate could be fuelled by large
amounts of neutral and molecular gas. Theoretical calculations as
well as observational evidence show that molecular cloud-cloud
collision account for a substantial fraction of the star
formation in the Galaxy (e.g. \citealp{Tan:2000},
\citealp{Sato:2000} and references therein). In \hz\ galaxies, high
velocity dispersions up to $100$\kms\ are observed in the warm
phase of the gas on several kpc-scale,
the velocity dispersion of the
cold phase of the gas and of the stellar component being not
observable. Very extended (as large as $\sim1~kpc$) and massive
star forming clumps ($\sim10^{8-9}M_{\odot}$) are observed at
redshift $\geq1$, (e.g. \citealp{Elmegreen:2007} and references
therein). Corresponding high-mass clumps do not exist in local
galaxies, even in high star forming objects, where their masses do
not exceed $\sim10^6M_{\odot}$. On deca/hecto pc-scale, the
typical velocity dispersion of the cold gas phase in the
interstellar medium of local galaxies is of the order of $5$\kms:
even in massive molecular clouds observed in local HII regions
(e.g. $\sim2.10^5M_{\odot}$ in NGC 7538, \citealp{Minn:1975}), the
internal velocity dispersion does not exceed $\sim5$\kms. The
formation of OB-stars associations leads to the ionization of
smaller clumps, the so-called HII regions. Local gaseous
velocity dispersion in nearby galaxies within HII regions spans
only between 10 and 30 \kms\ (see section \ref{velocity_dispersion_analysis} and \citealp{Weiner:2006}).
During the strongest phase of star formation, mainly
due to supernova activity, the ionized gas component is more
turbulent and its velocity dispersion higher although not as high as
observed in high redshift galaxies.

The mean rotation velocity of the ionized gas component may be similar
to the unobserved neutral gas component (atomic or molecular).
The spatial distribution of the ionized gas is more clumpy and its velocity
dispersion higher than the neutral gas. It follows the distribution and dynamics of young
stars and the stellar winds induced by them.
Strong supernovae winds, large bubbles in expansion increase the velocity
dispersion of the ionized gas and participate to the turbulent motions linked to star formation processes.
However, turbulent motions observed through the ionized gas component, on kpc-scale
structures, probably cannot be explained by star formation processes only.
External mechanisms, like cosmological gas accretion, combined to local self-gravity are needed to provide additional energy to sustain the high velocity dispersions \citep{Lehnert:2009} and thus, should be present also in the neutral gas component.
In nearby galaxies, a fraction of the local velocity dispersion observed in the gas is due to turbulence
linked to mass density contrasts generated for instance by $m=2$ pertubations (spiral arms). At higher redshift, since gas density is higher, this turbulence linked to local gravity may increase and it will not indicate that the disk is unstable.

\subsubsection{Large clumps observed in high redshift galaxies: huge HII regions or conglomerate of small clumps?}

Galaxies are increasingly clumpy with redshift \citep{Conselice:2005}.  A large fraction of their luminous mass (up to 30\%) and optical light (up to 50\%) is confined to a few kpc-size clumps \citep{Elmegreen:2005}. These clumps are probably formed inside the galactic disk rather than entered from outside in a merger \citep{Bournaud:2009}. They are specific to \hz\ galaxies that do not have spirals, nor bulges or exponential profiles. These clumps tell us about galaxy evolution and could be progenitors of modern spiral disks.

If these gaseous clumps are gravitationally bound and dynamically
relaxed, they may trace the gravitational potential as it
should be shown by the hidden stellar velocity dispersion. If one
considers that the clumps observed at high redshift proceed from
molecular clouds gravitationally bound as massive as
$\sim10^9M_{\odot}$, extrapolation of the scaling relation of
\citet{Larson:1981} leads to internal velocity dispersion not
higher that $\sim25$\kms.  To reach internal velocity dispersions
$\sim100$\kms, like those observed in clumps at high redshift,
clump progenitors should have the mass of a massive galaxy
($\sim10^{12}M_{\odot}$). These huge clumps may not be gravitationally linked systems or may be bound
but not in equilibrium. The scaling relations of \citet{Larson:1981} are not valid to describe the physics of these clumps.

Many numerical works have been concerned with collision between
so-called high-mass clouds which nevertheless do not excess
$\sim10^3M_{\odot}$ (e.g. \citealp{Chapman:1992}). These clouds
are obviously much less massive than the large clumps observed at
high redshift. Nevertheless, one might expect a collision between
two high-mass clouds to consist of many smaller-scale collisions
between the clouds of lower mass of which the clumps are composed
maybe down to $\sim10~M_{\odot}$ (e.g. \citealp{Kitsionas:2008}).
These clumps may be the result of a high star-formation occurring
at this stage but most massive clumps predicted in models
simulating disk instability have a velocity dispersion
$\sim20-30$\kms~\citep{Immeli:2004a} or
$40-50$\kms~\citep{Bournaud:2008}.  In other words, the mean local
velocity scatter around circular motions (i.e. the dispersion in
the rest frame of a disk in circular motion) expected from
simulations ranges from $20$ to $50$\kms. The circular rotation of the
clumps is given by the mean potential well of the galaxy
(disk+dark halo) but clump-clump (2-body gravitational)
interaction induces their velocity dispersion with an
amplitude lower than $50$\kms. The high velocity dispersion
observed on 2D \vdms\ may be due to integration through the
line-of-sight within the size of the unresolved observed beam.
Very strong winds due to supernova activity may also increase the
local velocity dispersion in the ionized gas component.

\subsubsection{Large clumps at kpc-scale resolution}

In this paper, we do not focus on merging systems but only on rotators for which three mechanisms of formation are possible: major mergers, minor mergers and gas accretion.
%
%

%
The assembly of galaxies at redshifts $z\sim1-2$ which have clumps embedded in what appears to be a disk is unlikely to be mostly driven by hierarchical merging of smaller galaxies.
Indeed, the formation of giant clumps (with the masses, sizes and elongations typically observed) in massive and highly turbulent disks, requires that the dominant process of mass assembly be some smooth accretion of cold and diffuse gas \citep{Bournaud:2009}.
This is consistent with the picture in which young thick disks form by cold flows \citep{Dekel:2009,Keres:2009}. However, the actual physical nature and the characterization of these clumps request further attention before disentangling different mechanisms occurring at different epoch of galaxy assembly.

The nature of the massive clumps observed in \hz\ galaxies
is well established from imaging \citep{Conselice:2005,Elmegreen:2005}. Nevertheless, the lack of spatial resolution does not allow to fully characterize them. Indeed, large clumps with large velocity dispersions might be composed of several unresolved smaller clumps. Even if it is well established that the sizes, the luminosities, the velocity dispersion and thus the masses of these clumps are large with respect to the clumps observed in lower redshift galaxies, it may not be excluded that their ``oversized'' geometric and kinematics properties is due to the lack of resolution and to the fact that they are not resolved.  In any cases, their physics is poorly understood and need higher spatial resolution to be modeled through numerical simulations.
The high velocity dispersion in \hz\ galaxies may be due to the blended
kinematics of neighboring, self gravitating clouds. The low
spatial resolution (limited by the seeing) combined with the low
spectral resolution make difficult the deconvolution by both spatial and spectral
instrumental PSF. Thus these large star
forming clumps observed at high redshift maybe consist of the
conglomerate of unresolved smaller scales clumps. In that case,
the large velocity dispersion observed on kpc-scale in high
redshift galaxies may thus be the result of the velocity
dispersion of the different small clouds composing the unresolved
clumps, rather than a local velocity dispersion within a large
individual star-forming clump. Sub-kpc data are needed to observe
sub-clumps in order to know if they are gravitationally bound or just
spatial resolution effects.

Only high resolution observations may enable to compute \vfs\ and \rcs\ uncontaminated by the blurring of the data (see section \ref{section_rcshape}).
IFU AO observations provide in one shot a kpc-scale sampling both on the morphology and on the kinematics. These data are needed to sample the \vf, \rc\ and \vdm, as discussed in section \ref{analysis}, and to recognize the disk formation mechanisms printed in the morphology and the kinematics.
\section{Conclusion}
\label{conclusion}

Due to the lack of spatial resolution, consequence of their large
distance, observations of galaxies at high redshift are affected
by beam smearing effects. The different moment maps (intensity
maps, \vfs, \vdms, etc.) as well as the one dimensional plots (line
profiles, \rcs, etc.) are severally blurred on kpc-scale. For
instance, beam smearing effects completely modify the shape of the
\rcs~in inducing artificially a solid body shape trend (i.e. a
lower inner slope and a higher outer slope than real).

In this work, in order to study the biases induced by beam
smearing effects existing in observation of high redshift
galaxies and to provide new tools and recipes to analyse \hz\ galaxies, we have used 3D data cubes for a large sample of local
galaxies. This sample of nearby galaxies consists of 153 objects
observed with Fabry-Perot technics belonging to the GHASP
sample. We have simulated observations of this sample at redshift $1.7$ and have attempted to recover hidden information from
the blurred \vfs~and \vdms~using simple kinematical models.
The conclusions can be summarized through the different items as follows:\\
\par
\emph{(I)} The analysis led in this work enables us to test the
validity of high redshift dynamical classification made by
\citet{Flores:2006,Yangetal:2008} to distinguish rotating disks from mergers. We
have shown that, using this classification, most of the rotating
disks are correctly classified but we have also pointed out that around 30\% of disk galaxies would be misclassified as perturbed
rotators, or even complex kinematics.
This may lower the fraction of galaxies with anomalous or perturbed kinematics in the IMAGES sample from 41\%\ to 33\%.
This work will be further
completed in projecting at high redshift a local sample of
galaxies showing complex kinematics (mergers, close binaries,
compact groups, blue compacts galaxies) in order to evaluate the
fraction of these systems which would be misclassified as rotating
disks.\\
\par
\emph{(II)} This sample was used to test the relevance of
recovering the actual dynamical parameters of high redshift
galaxies (inclination, \pa~of the major axis, center, maximum
rotation velocity, etc.) taking into account the lack of spatial
resolution quantified by a ``beam smearing parameter'' $B$,
ratio between the optical galactic radius ($D_{25}/2$) and the
seeing FWHM. Actual observations generally lead to a
$B$ parameter lower or equal to 2-3 without AO (e.g. \citealp{Forster-Schreiberetal:2006}) or even $B^>_\sim6$ when using AO (e.g. \citealp{Genzeletal:2008}).
The ``recipes'' to recover the dynamical parameters are the following:
\vspace{-0.25cm}
\begin{itemize}
\item the position of the kinematical center is poorly constrained
by the kinematics and should be fixed using high resolution
broad-band images. When no clear center can be deduced from
sub-kpc images of high redshift galaxies, as it is often the case,
the position of the center is strongly affected by beam smearing
effects. The determination of the center estimated by
the symmetrization of the \rcs~is not reliable for galaxies showing a
solid body shape at all radii, which is likely the case for galaxies with $B^<_\sim3$;
\item the inclination needs to be constrained by high
resolution morphologies since the agreement is better between high
resolution morphological inclinations and high resolution
kinematical inclinations than between high redshift and
low redshift kinematical inclinations. The beam smearing however needs to be taken into account. In addition to the possible corrections already discussed in the literature (e.g. \citealp{Simard:2002,Peng:2002}), a simple way to do it consists in correcting half light radius major and minor axis by subtracting quadratically the seeing. The uncertainties in the determination
of the kinematical inclination can be quantified by a linear function of the beam
smearing parameter $B$;
\item the \pa\ of the major axis is recovered with an accuracy better than 5\degr~for 70\% of the sample
using a 2D \vf\ using simple rotating disk models, even with a rather low spatial resolution ($B\sim2$);
\item the observed velocity dispersion of the gas is strongly
correlated with the velocity shear of the galaxy, especially in
the inner regions. The local velocity dispersion $\sigma$ can be statistically recovered \emph{(i)} by
subtracting quadratically the \vdm\ model deduced from the
\vf~modeling although with a large scatter or \emph{(ii)} by considering regions with the lowest values that are the less affected by beam smearing. The larger the local velocity dispersion is, the weaker the above correlation is;
\item the maximum velocity is statistically fairly well recovered for
galaxies larger than three times the seeing in radius (i.e. with $B>3$), even if
this limit probably depends on the unknown high redshift shape of
the \rcs. For galaxies with $B<3$, we provide a correction of the maximum velocity as a function of $B$. The use of a simple \vf~modeling enables to recover
statistically the maximum velocity with an error lower than 25\%\ in almost any case.
We have also shown that a simple model of \rc\ consisting of a
solid body part and a flat plateau statistically gives better estimates of the maximum velocity compared to exponential disk, isothermal sphere or arctangent \rc\ models;
\item the local GHASP sub-sample of galaxies was also used
to test different \rc\ models to recover the actual \rcs,
i.e. unaffected by beam smearing effects.  A direct comparison
between actual high resolution data ($z=0$) and various  models
was done in this purpose. In average, the various models are
able to recover the general trend of the actual $z=0$ \rcs\ but the
scatter around the mean difference in the \rcs\ ($\Delta
V_c^{mean}$) is large, pointing out the difficulty to retrieve the
actual shapes.
Moreover, observations having a value of $B^<_\sim3$ do not allow suitable beam smearing
corrections to recover the rough shape of the \rc\ whatever the model used. In order to be able
to address problematics linked to the shape of the \rc\ (e.g. CORE vs CUSPY controversy about the inner density profile
in spirals) for \hz~galaxies, $B^>_\sim10$ are necessary.
\end{itemize}
\par
\emph{(III)} Finally, this sample of local and evolved galaxies
projected at high redshift has been compared to samples of actual high redshift
galaxies observed using integral field capabilities (SINFONI, OSIRIS, GIRAFFE) to disentangle evolution effects from
distance effects. By applying the same methods of analysis on both projected and observed
samples, a relative comparison can be done to probe the
kinematical evolution of galaxies, since the same observational
biases exist in both samples. Our results suggest \emph{(i)} that the trend in the evolution of the \TF\ relation observed by \citet{Puechetal:2008} is not due to beam smearing effects
and \emph{(ii)}
that, except if no beam smearing correction is done on actual
high redshift data, the high local velocity dispersion observed in
high redshift galaxies cannot be reproduced in the local projected
sample. This unambiguously means that, at the opposite of local
evolved galaxies, it exists from redshifts $z\sim 3$ to $z\sim 1$ at least a population
of disk galaxies for which a large fraction of the dynamical
support is not only due to rotation but also to velocity
dispersion. At $z\sim 0.6$, galaxies show intermediate velocity dispersions between local and higher redshift galaxies.
This demonstrates a strong and continuous dynamical evolution in disk galaxies through the last eleven Gyr ($z\sim 2.5$).
This conclusion is relevant at least for some galaxies
among the relatively small sample of high redshift galaxies
observed using IFU to date. Indeed, one cannot exclude important observation
biases in the selection of the targets which was dictated by the
feasibility of the observations rather than by strong considerations on
the representativity of a given epoch by a set of galaxies
correctly selected using for instance luminosity or mass
functions. For a given observing time, multi-slit spectroscopy enables to observe larger samples than IFU techniques. However, due to the low spatial coverage per galaxy, long slit data does not allow a complete kinematical analysis.

The low numerical value for ($V_c^{max}/\sigma\sim1-2$) is a convincing evidence for the existence of a population of thick and transient turbulent gas disks in \hz\ galaxies. However, the large turbulence is a consequence of the large amount of gas which induces feedback, local gravitational disturbances and infall processes.  It does not prove that the disk is formed by continuous gas accretion rather than by frequent wet minor mergers or old wet major mergers \citep{Robertson:2008}.
If these thick disks are still seen later, they may be transformed into bulges and central galactic black hole.
On the other hands, it has to be surveyed if the \hz\ galaxies observed to date are representative of their epoch of formation or, alternatively, if the sample is biased by selection effects.
These open questions justify the MASSIV on-going program (\citeauthor{Contini:2008} in preparation, \citealp{Epinat:2009c}, \citealp{Queyrel:2009}) dealing with galaxies ranging from $z\sim 1.0$ to $z\sim 1.8$ and selecting the targets using criteria making them representative of given epochs.

The spatial resolution reached by AO observations enables to reduce significantly the beam smearing effects.
In this paper, the limits of the determination of kinematical parameters for high redshift galaxies observed under seeing limited conditions have been discussed: morphological and kinematical AO observations in the redshift range $0.5<z<3$ are essential to discuss the different scenarios of mass assambly and galaxy evolution.

In forthcoming works, the effects of spectral resolution and of the noise will be
studied using the GHASP sample and local disturbed
disk galaxies as compact groups galaxies (\citeauthor{Torres:2009} in preparation), blue compact star forming galaxies, strongly barred galaxies, mergers, close binaries will be compared to high redshift galaxies using the same methods presented in this paper.

The data used for this work will be available
in a database under construction containing Fabry-Perot
data \url{http://fabryperot.oamp.fr/}, enabling
to retrieve directly from the database redshifted datacubes with a
given seeing, pixel size and spectral resolution.  This database
will also contain data from several other Fabry-Perot surveys
(barred galaxies; galaxies in clusters, in compact groups; blue compact galaxies, etc.).

\section*{acknowledgements} {We thank David R. Law for kindly providing us before publication their velocity and velocity dispersion maps \citep{Law:2009}. We thank Fr\'ed\'eric Bournaud for discussion. We also thank the referee, Benjamin Weiner, for careful reading of the manuscript and useful comments that helped to improve the paper.}

\bibliographystyle{mn2e}
\bibliography{biblio}

\appendix

\section{The model}
\label{model}
\subsection{Real light distribution}
We note $S(x,y,\lambda)$ the spectral distribution of light at position $(x,y)$
 at wavelength $\lambda$. This spectral distribution contains continuum ($C$) and line emission ($L$):
\begin{equation}
S(x,y,\lambda)=L(x,y,\lambda)+C(x,y)
\end{equation}
The line flux or monochromatic flux is defined by equation \ref{monofluxdef}:
\begin{equation}
M(x,y)=\int_\lambda L(x,y,\lambda)d\lambda
\label{monofluxdef}
\end{equation}
The velocity (first moment of the line) is defined by equation \ref{velocitydef}:
\begin{equation}
V(x,y)\equiv\overline{V(x,y)}=\frac{\int_\lambda L(x,y,\lambda)v(\lambda)d\lambda}{M(x,y)}
\label{velocitydef}
\end{equation}
And finally, the local velocity dispersion (second moment of the line) is defined by equation \ref{velocitydispdef}:
\begin{equation}
\sigma(x,y)^2\equiv\overline{V(x,y)^2}-\overline{V(x,y)}^2
\label{velocitydispdef}
\end{equation}
where
\begin{equation}
\overline{V(x,y)^2}= \frac{\int_\lambda L(x,y,\lambda)v(\lambda)^2d\lambda}{M(x,y)}
\end{equation}
These are ideally the quantities that one wants to estimate. However this is not obvious as spectral \PSF\ and spatial \PSF\ are not Dirac distributions, and because instruments sample the light distributions through pixels and spectral channels.

\subsection{Spectral \PSF~and sampling effects}

The effect of the spectral \PSF~is a convolution with the spectrum:
\begin{equation}
S_1(x,y,\lambda)=L\otimes_\lambda PSF_\lambda+C
\end{equation}
$PSF_\lambda$ being the spectral \PSF. The spectral \PSF~can be considered constant within the wavelength range.
Since the continuum does not vary with wavelength (by definition) it can be considered as null.

Spectral sampling is equivalent to convolving the spectrum with a ``door'' function:
\begin{equation}
S_2(x,y,\Lambda_i)=\int_{\Lambda_i-\Delta \lambda/2} ^{\Lambda_i+\Delta \lambda/2} S_1 (x,y,\lambda)d\lambda
\end{equation}
As spectral channels are contiguous, and because the spectral \PSF~does not introduce any loss in flux,  the monochromatic flux can be expressed as:
\begin{equation}
M(x,y)=\sum_i S_2(x,y,\Lambda_i)
\end{equation}
By assuming that the rectangle method is giving a good estimate of integrals, that is true only when the spectral resolution (\PSF~and sampling) enables to oversample the line, the following equations can then be written:
\begin{equation}
\sum_i S_2(x,y,\Lambda_i)v(\Lambda_i)\approx \int_\lambda S(x,y,\lambda)v(\lambda)d\lambda
\end{equation}
\begin{equation}
\sum_i S_2(x,y,\Lambda_i)v(\Lambda_i)^2\approx \int_\lambda S(x,y,\lambda)v(\lambda)^2d\lambda
\end{equation}
This is the first approximation.
It enables to deduce:
\begin{equation}
V(x,y)\equiv \overline{V(x,y)}=\frac{\sum_i S_2(x,y,\Lambda_i)v(\Lambda_i)}{M(x,y)}
\end{equation}
\begin{equation}
\overline{V(x,y)^2}=\frac{\sum_i S_2(x,y,\Lambda_i)v(\Lambda_i)^2}{M(x,y)}
\end{equation}
and then to express the velocity dispersion as in equation \ref{velocitydispdef}.

\subsection{Spatial \PSF~and sampling effects}

The spatial \PSF, noted $\PSF_{xy}$ is due to diffraction limit (Airy disk) as well as to seeing conditions. However, the induced defaults have to be compared with velocity variations.

\subsubsection{Spatial \PSF~effects}
The effect of the spatial \PSF~is a 2D convolution with the images:
\begin{equation}
S_3(x,y,\Lambda)=S_2(x,y,\Lambda)\otimes_{xy}\PSF_{xy}
\end{equation}
One can measure 
\begin{equation}
M_0(x,y)=\sum_i S_3(x,y,\Lambda_i)
\end{equation}
And deduce from analytical computing that
\begin{equation}
M_0=M\otimes_{xy}\PSF_{xy}
\end{equation}
The measurement of the moments is also biased by this convolution :
\begin{equation}
\overline{V_0^\alpha}=\frac{\left[\overline{V^\alpha} M\right]\otimes_{xy} \PSF_{xy}}{M_0}
\label{v_seeing}
\end{equation}
By combining equation \ref{velocitydispdef} and \ref{v_seeing} we deduce the square of the blurred velocity dispersion before sampling:
\begin{equation}
\begin{split} 
\sigma_0^2=\frac{\left[\sigma^2 M\right]\otimes_{xy} \PSF_{xy}}{M_0}\\
+\frac{\left[\overline{V}^2 M\right]\otimes_{xy} \PSF_{xy}}{M_0}-\left(\frac{\left[\overline{V} M\right]\otimes_{xy} \PSF_{xy}}{M_0}\right)^2
\end{split} 
\end{equation}

\subsubsection{Sampling effects}

Spatial sampling is equivalent to convolve each frame with a 2D ``door'' function. Thus the measured spectrum is:
\begin{equation}
S_4(X,Y,\Lambda)=\int_{X-\Delta x/2}^{X+\Delta x/2} \int_{Y-\Delta y/2}^{Y+\Delta y/2} S_3(x,y,\Lambda)dxdy
\end{equation}
To have lighter notations, the notation $\int_{pix}dxy$ is used instead of $\int_{X-\Delta x/2}^{X+\Delta x/2} \int_{Y-\Delta y/2}^{Y+\Delta y/2}dxdy$.
The measured quantities are noted with the index $1$ ($M_1$, $V_1$, $\overline{V_1^\alpha}$, $\sigma_1$).

The observed flux is:
\begin{equation}
M_1(X,Y)=\sum_i S_4(X,Y,\Lambda_i)
\end{equation}
from which is deduced the link with the real monochromatic flux, by assuming that the spatial \PSF\ does not depend neither on the wavelength nor on the position:
\begin{equation}
M_1 (X,Y)=\int_{pix}  M\otimes_{xy} \PSF_{xy} dxy
\end{equation}
In other words, the measured flux is the sum of the \PSF\ convolved flux in one pixel.
Within the same hypothesis, we deduce the observed momenta:
\begin{equation}
\overline{V_1^\alpha}(X,Y)=\frac{\int_{pix}  \left[M\overline{V^\alpha}\right]\otimes_{xy} \PSF_{xy} dxy}{M_1}
\end{equation}
and thus, the expression of the observed velocity:
\begin{equation}
V_1 (X,Y)\equiv\overline{V_1^1}(X,Y)=\frac{\int_{pix}  \left[M\overline{V}\right]\otimes_{xy} \PSF_{xy} dxy}{M_1}
\end{equation}
and the square of the observed velocity dispersion:
\begin{equation}
\begin{split} 
\sigma_1^2 (X,Y)\equiv \overline{V_1(X,Y)^2}-\overline{V_1(X,Y)}^2 =\frac{\int_{pix}\left[\sigma^2 M\right]\otimes_{xy} \PSF_{xy}dxy}{M_1}\\
+\frac{\int_{pix}\left[\overline{V}^2 M\right]\otimes_{xy} \PSF_{xy}dxy}{M_1}-\left(\frac{\int_{pix}\left[\overline{V} M\right]\otimes_{xy} \PSF_{xy}dxy}{M_1}\right)^2
\end{split} 
\label{dispersion_lowres}
\end{equation}

\subsection{Comments}

The previous set of equations is obtained with a very few hypothesis. It enables to understand why low resolution makes kinematical studies critical, in particular at high redshift.
Moreover, it can be used as the basis to write kinematical models: it is possible to avoid the modeling of a data cube in order to gain computing time and resources. Modeling a \vf~is sufficient providing that we make some hypothesis on the flux distribution. Indeed, even if the flux distribution is known at the observing resolution, in the previous set of equations, we see the need for high resolution flux map. Ideally, high resolution narrow band observations should be provided to improve the modeling (using Tunable Filters for instance on space telescopes). HST data could also be used but making the approximation that the maps are tracing the gas distribution.

These equations also enable to disentangle resolution effects from real dispersion features in the \vdms. Indeed, equation \ref{dispersion_lowres} presents a natural decomposition in two terms: a local velocity dispersion one and a velocity shear one due to the beam smearing.
By using a satisfying \vf~model, unresolved velocity gradient can be subtracted quadratically from the \vdm. The remaining term is thus the local dispersion convolved with the spatial \PSF. This term contains the signature of the spectral \PSF.
In particular, by making the hypothesis that the local velocity dispersion $\sigma$ is constant, what seems to be the case for the gaseous component for local galaxies, then the expression is simplified:
\begin{equation}
\begin{split} 
\sigma_1^2=\sigma^2+\frac{\int_{pix}\left[\overline{V}^2 M\right] \otimes_{xy} \PSF_{xy} dxy}{M_1}\\
-\left( \frac{\int_{pix} \left[ \overline{V} M\right] \otimes_{xy} \PSF_{xy} dxy}{M_1} \right)^2
\end{split} 
\end{equation}
In the case one wants to constrain models with the \vdm, a velocity dispersion model has to be built.

\subsection{Rotation curve models}
\label{rcmodels}

Four models are used in this paper. These four models are only described with two parameters having the same physical signification: the maximum velocity $V_t$ of the function, and the radius at which it is reached $r_t$ (hereafter called transition radius) except for the arctangent model since the maximum velocity is reached at infinity.
The transition radius is constrained to measure at least one pixel.

\subsubsection{First model: exponential disk}
This model describes a galaxy whose luminosity profile is fit with an exponential law, and for which the gravitation potential is uniquely due to the stars (no dark matter halo). It is a Freeman disk.
\begin{equation}
V(r)=\frac{r}{r_0}\sqrt{\pi G \Sigma_0 r_0\left(I_0 K_0 - I_1 K_1 \right)}
\end{equation}
Where $r_0$ is the exponential radius, $\Sigma_0$ is the central disk surface density, $I_i$ and $K_i$ are the i-order modified Bessel function evaluated at $0.5r/r_0$. The maximum velocity $V_t\sim0.88\sqrt{\pi G \Sigma_0 r_0}$ is reached at $r_t\sim2.15r_0$.
This model is the one used in \citet{Forster-Schreiberetal:2006}.

\subsubsection{Second model: isothermal sphere}
This model describes the \rc~due to an isothermal sphere dark matter halo. \citet{Spano:2008} have shown that this model is the best fit model for local galaxies.
\begin{equation}
V(r)=\sqrt{4\pi G\rho_0 r_c^2 \left[\frac{r_c}{r}\ln\left(\frac{r}{r_c}+\sqrt{1+\frac{r^2}{r_c^2}}\right)-\frac{1}{\sqrt{1+\frac{r^2}{r_c^2}}} \right]}
\end{equation}
Where $r_c$ is the core radius and $\rho_0$ is the central halo density. The maximum velocity $V_t\sim0.54\sqrt{4\pi G \rho_0 r_c^2}$ is reached at $r_t\sim2.92r_c$.
This model should be used when the contribution of the stars to the gravitation potential is negligible (i.e. for LSB galaxies).

\subsubsection{Third model: ``flat model''}

This model does not describe any classical mass distribution. However, it can describe correctly lots of observed \rcs~of local galaxies, in particular those reaching a plateau.
\begin{eqnarray}
V(r)=V_t\frac{r}{r_t} \mathrm{, for~} r<r_t,\\
V(r)=V_t \mathrm{, for~} r \geq r_t.
\end{eqnarray}
This model is that used in \citet{Wright:2007}.

\subsubsection{Fourth model: arctangent}

This model is used by \citet{Puechetal:2008}. The \rc\ is described by an arctangent function. Since the maximum velocity is reached asymptotically for an infinite radius, the transition radius $r_t$ is defined as the radius for which the velocity reaches 70\% of the asymptotic velocity $V_t$:
\begin{eqnarray}
V(r)=V_t\frac{2}{\pi} \arctan{\frac{2r}{r_t}}
\end{eqnarray}
This function is rather similar to the ``flat model'' but is smoother. Moreover, the plateau is not clearly reached, thus it is more likely a \rc~with an increasing plateau.



\section{Tables}
\label{table}
\onecolumn
\begin{longtable}{ccccccccc}
\caption{Galaxy parameters at $z=0$}\\
\hline
Galaxy & $i_{z=0}~^{(a)}$ & $PA_{z=0}~^{(b)}$ & $D_{25}/2~^{(c)}$ & $M_B~^{(d)}$ & $V^{max}_c~^{(e)}$ & $\sigma~^{(f)}$ & $V^{max}_c/\sigma$ & $S_{in}~^{(g)}$ \label{table_modz0} \\
 & \degr\ & \degr\ & $kpc$ & mag & $km~s^{-1}$ & $km~s^{-1}$ & & $km~s^{-1}~kpc^{-1}$  \\
\hline
\endfirsthead

\caption[]{\textit{continued}}\\
\hline
Galaxy & $i_{z=0}~^{(a)}$ & $PA_{z=0}~^{(b)}$ & $D_{25}/2~^{(c)}$ & $M_B~^{(d)}$ & $V^{max}_c~^{(e)}$ & $\sigma~^{(f)}$ & $V^{max}_c/\sigma$ & $S_{in}~^{(g)}$ \\
 & \degr\ & \degr\ & $kpc$ & mag & $km~s^{-1}$ & $km~s^{-1}$ & & $km~s^{-1}~kpc^{-1}$  \\
\hline
\endhead

\hline
\endfoot

\hline
\multicolumn{9}{l}{(a): Inclination from \citet{Epinat:2008b}.}\\
\multicolumn{9}{l}{(b): Position angle of the major axis from \citet{Epinat:2008b}.}\\
\multicolumn{9}{l}{(c): Optical radius from the RC3 catalog \citep{de-Vaucouleurs:1992} or from the HyperLeda database (refereed}\\
\multicolumn{9}{l}{by an asterisk $^*$, \citealp{Paturel:2003}).}\\
\multicolumn{9}{l}{(d): B-band magnitude from the HyperLeda database \citep{Paturel:2003}.}\\
\multicolumn{9}{l}{(e): Maximum velocity from \citet{Epinat:2008b}.}\\
\multicolumn{9}{l}{(f): Mean local velocity dispersion.}\\
\multicolumn{9}{l}{(g): Inner slope of the \rc. No value is provided when no high resolution rotation curve is computed}\\
\multicolumn{9}{l}{(edge on galaxies, \citealp{Epinat:2008b,Epinat:2008a}).}\\
\endlastfoot
UGC 89 & $   33 \pm    13$ & $  177 \pm     4$ & $ 18.5 $ & $-21.5$ & $  343 \pm   117$ & $   30 \pm    17$ & $ 11.4 \pm   7.6$ & $  317 \pm    26$ \\
UGC 94 & $   42 \pm     5$ & $   94 \pm     2$ & $ 17.2 $ & $-20.4$ & $  209 \pm    21$ & $   23 \pm    15$ & $  9.1 \pm   6.0$ & $  118 \pm     7$ \\
UGC 508 & $   25 \pm     7$ & $  123 \pm     1$ & $ 25.9 $ & $-21.8$ & $  553 \pm   127$ & $   22 \pm    16$ & $ 25.1 \pm  19.2$ & $  321 \pm    24$ \\
UGC 528 & $   21 \pm    14$ & $   52 \pm     3$ & $  3.5 $ & $-19.6$ & $   84 \pm    52$ & $   24 \pm    11$ & $  3.5 \pm   2.7$ & $  103 \pm     8$ \\
UGC 763 & $   54 \pm     6$ & $  117 \pm     3$ & $  7.1 $ & $-18.9$ & $  104 \pm    11$ & $   23 \pm    13$ & $  4.5 \pm   2.6$ & $   34 \pm     2$ \\
NGC 542 & $   90 \pm     1$ & $  143 \pm     9$ & $  9.9^*$ & $-19.5$ & $  125 \pm     8$ & $   28 \pm    16$ & $  4.5 \pm   2.6$ & $-$ \\
UGC 1249 & $   90 \pm     1$ & $  150 \pm     9$ & $  6.9 $ & $-18.3$ & $   65 \pm     8$ & $   15 \pm    15$ & $  4.3 \pm   4.4$ & $-$ \\
UGC 1256 & $   76 \pm     2$ & $   73 \pm     2$ & $  7.2 $ & $-18.9$ & $  105 \pm     9$ & $   17 \pm    13$ & $  6.2 \pm   4.8$ & $   29 \pm     1$ \\
\end{longtable}
\twocolumn

\onecolumn
\begin{landscape}
\begin{longtable}{c|c|ccc|ccccc}
\caption{Exponential disk model on the sample projected at $z=1.7$}\\
\hline
Galaxy$~^{(a)}$ & $i_{z=1.7}~^{(b)}$ & $PA_{z=1.7}~^{(c)}$ & $r_t~^{(d)}$ & $V_t~^{(e)}$ & $V^{max}_c~^{(f)}$ & $\sigma~^{(g)}$ & $V^{max}_c/\sigma$ & $\Delta V^{mean}_c~^{(h)}$ & $S_{in}~^{(i)}$ \label{table_modzexp} \\
 & \degr\ & \degr\ & $kpc$ & $km~s^{-1}$ & $km~s^{-1}$ & $km~s^{-1}$ & & $km~s^{-1}$ & $km~s^{-1}~kpc^{-1}$  \\
\hline
\endfirsthead

\caption[]{\textit{continued}}\\
\hline
Galaxy$~^{(a)}$ & $i_{z=1.7}~^{(b)}$ & $PA_{z=1.7}~^{(c)}$ & $r_t~^{(d)}$ & $V_t~^{(e)}$ & $V^{max}_c~^{(f)}$ & $\sigma~^{(g)}$ & $V^{max}_c/\sigma$ & $\Delta V^{mean}_c~^{(h)}$ & $S_{in}~^{(i)}$ \\
 & \degr\ & \degr\ & $kpc$ & $km~s^{-1}$ & $km~s^{-1}$ & $km~s^{-1}$ & & $km~s^{-1}$ & $km~s^{-1}~kpc^{-1}$  \\
\hline
\endhead

\hline
\endfoot

\hline
\multicolumn{10}{l}{(a): Galaxies for which no parameter is provided are those for which the fit was not possible.}\\
\multicolumn{10}{l}{(b): Inclination deduced from the fit. The error is a statistical error and thus gives a lower limit. The asterisk $^*$ indicates that the}\\
\multicolumn{10}{l}{inclination was stacked to one boundary.}\\
\multicolumn{10}{l}{(c): Position angle of the major axis deduced from the fit with the inclination fixed to the value in Table \ref{table_modz0}. The error is a statistical}\\
\multicolumn{10}{l}{error and thus gives a lower limit.}\\
\multicolumn{10}{l}{(d): Scale length radius of the model (defined in Appendix \ref{model}) deduced from the fit with the inclination fixed to the value in Table \ref{table_modz0}.}\\
\multicolumn{10}{l}{The error is a statistical error and thus gives a lower limit.}\\
\multicolumn{10}{l}{(e): Velocity of the model (defined in Appendix \ref{model}) deduced from the fit with the inclination fixed to the value in Table \ref{table_modz0}. The error is}\\
\multicolumn{10}{l}{a statistical error and thus gives a lower limit.}\\
\multicolumn{10}{l}{(f): Maximum velocity of the model within $R_{last}$ (see Table \ref{table_z}). The error is the same as in column (e) and thus not take into account}\\
\multicolumn{10}{l}{the uncertainty on the inclination.}\\
\multicolumn{10}{l}{(g): Local velocity dispersion (beam smearing corrected).}\\
\multicolumn{10}{l}{(h): Mean difference of the model \rc\ with the actual \rc\ at $z=0$.}\\
\multicolumn{10}{l}{(i): Inner slope of the \rc\ from the model. The asterisk $^*$ indicates that only the slope is constrained by the model.}\\
\endlastfoot
UGC 89 & $  10^*$ & $ 169\pm   1$ & $ 4.3\pm  0.1$ & $401\pm   1$ & $  401$ & $  21\pm  20$ & $ 19.4$ & $  -22$ & $  492$ \\
UGC 94 & $  22\pm   1$ & $  91\pm   1$ & $ 4.9\pm  0.1$ & $232\pm   1$ & $  232$ & $  16\pm  15$ & $ 14.5$ & $  -10$ & $  250$ \\
UGC 508 & $  10^*$ & $ 121\pm   1$ & $10.2\pm  0.1$ & $549\pm   1$ & $  549$ & $  27\pm  16$ & $ 20.2$ & $   19$ & $  308$ \\
UGC 528 & $  10^*$ & $  54\pm   3$ & $\infty$ & $\infty$ & $   44$ & $  35\pm   2$ & $  1.3$ & $   33$ & $  127^*$ \\
UGC 763 & $  18\pm   9$ & $ 119\pm   1$ & $ 4.4\pm  0.1$ & $103\pm   1$ & $  103$ & $  27\pm   5$ & $  3.8$ & $   -5$ & $  122$ \\
NGC 542 & $-$ & $-$ & $-$ & $-$ & $-$ & $-$ & $-$ & $-$ & $-$ \\
UGC 1249 & $-$ & $-$ & $-$ & $-$ & $-$ & $-$ & $-$ & $-$ & $-$ \\
UGC 1256 & $  10^*$ & $  67\pm   1$ & $ 7.3\pm  0.7$ & $ 96\pm   2$ & $   96$ & $  25\pm   5$ & $  3.8$ & $   -4$ & $   73$ \\
\end{longtable}
\end{landscape}
\twocolumn

\onecolumn
\begin{landscape}
\begin{longtable}{c|c|ccc|ccccc}
\caption{Isothermal sphere model on the sample projected at $z=1.7$}\\
\hline
Galaxy$~^{(a)}$ & $i_{z=1.7}~^{(b)}$ & $PA_{z=1.7}~^{(c)}$ & $r_t~^{(d)}$ & $V_t~^{(e)}$ & $V^{max}_c~^{(f)}$ & $\sigma~^{(g)}$ & $V^{max}_c/\sigma$ & $\Delta V^{mean}_c~^{(h)}$ & $S_{in}~^{(i)}$ \\ \label{table_modziso}
 & \degr\ & \degr\ & $kpc$ & $km~s^{-1}$ & $km~s^{-1}$ & $km~s^{-1}$ & & $km~s^{-1}$ & $km~s^{-1}~kpc^{-1}$  \\
\hline
\endfirsthead

\caption[]{\textit{continued}}\\
\hline
Galaxy$~^{(a)}$ & $i_{z=1.7}~^{(b)}$ & $PA_{z=1.7}~^{(c)}$ & $r_t~^{(d)}$ & $V_t~^{(e)}$ & $V^{max}_c~^{(f)}$ & $\sigma~^{(g)}$ & $V^{max}_c/\sigma$ & $\Delta V^{mean}_c~^{(h)}$ & $S_{in}~^{(i)}$ \\
 & \degr\ & \degr\ & $kpc$ & $km~s^{-1}$ & $km~s^{-1}$ & $km~s^{-1}$ & & $km~s^{-1}$ & $km~s^{-1}~kpc^{-1}$  \\
\hline
\endhead

\hline
\endfoot

\hline
\multicolumn{10}{l}{(a): Galaxies for which no parameter is provided are those for which the fit was not possible.}\\
\multicolumn{10}{l}{(b): Inclination deduced from the fit. The error is a statistical error and thus gives a lower limit. The asterisk $^*$ indicates that the}\\
\multicolumn{10}{l}{inclination was stacked to one boundary.}\\
\multicolumn{10}{l}{(c): Position angle of the major axis deduced from the fit with the inclination fixed to the value in Table \ref{table_modz0}. The error is a statistical}\\
\multicolumn{10}{l}{error and thus gives a lower limit.}\\
\multicolumn{10}{l}{(d): Scale length radius of the model (defined in Appendix \ref{model}) deduced from the fit with the inclination fixed to the value in Table \ref{table_modz0}.}\\
\multicolumn{10}{l}{The error is a statistical error and thus gives a lower limit.}\\
\multicolumn{10}{l}{(e): Velocity of the model (defined in Appendix \ref{model}) deduced from the fit with the inclination fixed to the value in Table \ref{table_modz0}. The error is}\\
\multicolumn{10}{l}{a statistical error and thus gives a lower limit.}\\
\multicolumn{10}{l}{(f): Maximum velocity of the model within $R_{last}$ (see Table \ref{table_z}). The error is the same as in column (e) and thus not take into account}\\
\multicolumn{10}{l}{the uncertainty on the inclination.}\\
\multicolumn{10}{l}{(g): Local velocity dispersion (beam smearing corrected).}\\
\multicolumn{10}{l}{(h): Mean difference of the model \rc\ with the actual \rc\ at $z=0$.}\\
\multicolumn{10}{l}{(i): Inner slope of the \rc\ from the model. The asterisk $^*$ indicates that only the slope is constrained by the model.}\\
\endlastfoot
UGC 89 & $  10^*$ & $ 170\pm   1$ & $ 2.7\pm  0.1$ & $406\pm   1$ & $  406$ & $  20\pm  17$ & $ 20.8$ & $  -43$ & $  478$ \\
UGC 94 & $  22\pm   1$ & $  91\pm   1$ & $ 3.6\pm  0.1$ & $231\pm   1$ & $  231$ & $  15\pm  13$ & $ 15.2$ & $  -21$ & $  201$ \\
UGC 508 & $  10^*$ & $ 121\pm   1$ & $ 7.0\pm  0.1$ & $552\pm   1$ & $  552$ & $  27\pm  16$ & $ 20.2$ & $   -3$ & $  246$ \\
UGC 528 & $  10^*$ & $  54\pm   3$ & $ 1.1\pm  0.1$ & $ 44\pm   2$ & $   44$ & $  35\pm   2$ & $  1.3$ & $   30$ & $  130$ \\
UGC 763 & $  18\pm   8$ & $ 119\pm   1$ & $ 4.2\pm  0.2$ & $101\pm   1$ & $  101$ & $  27\pm   5$ & $  3.7$ & $   -7$ & $   76$ \\
NGC 542 & $-$ & $-$ & $-$ & $-$ & $-$ & $-$ & $-$ & $-$ & $-$ \\
UGC 1249 & $-$ & $-$ & $-$ & $-$ & $-$ & $-$ & $-$ & $-$ & $-$ \\
UGC 1256 & $  10^*$ & $  67\pm   1$ & $ 7.4\pm  0.7$ & $ 94\pm   2$ & $   94$ & $  25\pm   5$ & $  3.7$ & $   -3$ & $   40$ \\
\end{longtable}
\end{landscape}
\twocolumn

\onecolumn
\begin{landscape}
\begin{longtable}{c|c|ccc|ccccc}
\caption{``Flat model'' on the sample projected at $z=1.7$}\\
\hline
Galaxy$~^{(a)}$ & $i_{z=1.7}~^{(b)}$ & $PA_{z=1.7}~^{(c)}$ & $r_t~^{(d)}$ & $V_t~^{(e)}$ & $V^{max}_c~^{(f)}$ & $\sigma~^{(g)}$ & $V^{max}_c/\sigma$ & $\Delta V^{mean}_c~^{(h)}$ & $S_{in}~^{(i)}$ \label{table_modzslp} \\
 & \degr\ & \degr\ & $kpc$ & $km~s^{-1}$ & $km~s^{-1}$ & $km~s^{-1}$ & & $km~s^{-1}$ & $km~s^{-1}~kpc^{-1}$  \\
\hline
\endfirsthead

\caption[]{\textit{continued}}\\
\hline
Galaxy$~^{(a)}$ & $i_{z=1.7}~^{(b)}$ & $PA_{z=1.7}~^{(c)}$ & $r_t~^{(d)}$ & $V_t~^{(e)}$ & $V^{max}_c~^{(f)}$ & $\sigma~^{(g)}$ & $V^{max}_c/\sigma$ & $\Delta V^{mean}_c~^{(h)}$ & $S_{in}~^{(i)}$ \\
 & \degr\ & \degr\ & $kpc$ & $km~s^{-1}$ & $km~s^{-1}$ & $km~s^{-1}$ & & $km~s^{-1}$ & $km~s^{-1}~kpc^{-1}$  \\
\hline
\endhead

\hline
\endfoot

\hline
\multicolumn{10}{l}{(a): Galaxies for which no parameter is provided are those for which the fit was not possible.}\\
\multicolumn{10}{l}{(b): Inclination deduced from the fit. The error is a statistical error and thus gives a lower limit. The asterisk $^*$ indicates that the}\\
\multicolumn{10}{l}{inclination was stacked to one boundary.}\\
\multicolumn{10}{l}{(c): Position angle of the major axis deduced from the fit with the inclination fixed to the value in Table \ref{table_modz0}. The error is a statistical}\\
\multicolumn{10}{l}{error and thus gives a lower limit.}\\
\multicolumn{10}{l}{(d): Scale length radius of the model (defined in Appendix \ref{model}) deduced from the fit with the inclination fixed to the value in Table \ref{table_modz0}.}\\
\multicolumn{10}{l}{The error is a statistical error and thus gives a lower limit.}\\
\multicolumn{10}{l}{(e): Velocity of the model (defined in Appendix \ref{model}) deduced from the fit with the inclination fixed to the value in Table \ref{table_modz0}. The error is}\\
\multicolumn{10}{l}{a statistical error and thus gives a lower limit.}\\
\multicolumn{10}{l}{(f): Maximum velocity of the model within $R_{last}$ (see Table \ref{table_z}). The error is the same as in column (e) and thus not take into account}\\
\multicolumn{10}{l}{the uncertainty on the inclination.}\\
\multicolumn{10}{l}{(g): Local velocity dispersion (beam smearing corrected).}\\
\multicolumn{10}{l}{(h): Mean difference of the model \rc\ with the actual \rc\ at $z=0$.}\\
\multicolumn{10}{l}{(i): Inner slope of the \rc\ from the model. The asterisk $^*$ indicates that only the slope is constrained by the model.}\\
\endlastfoot
UGC 89 & $  19\pm   1$ & $ 171\pm   1$ & $ 1.0\pm  0.1$ & $366\pm   1$ & $  366$ & $  22\pm  18$ & $ 16.7$ & $  -28$ & $  363$ \\
UGC 94 & $  21\pm   1$ & $  91\pm   1$ & $ 1.0\pm  0.1$ & $218\pm   1$ & $  218$ & $  15\pm  13$ & $ 14.1$ & $  -26$ & $  224$ \\
UGC 508 & $  10^*$ & $ 122\pm   1$ & $ 1.0\pm  0.1$ & $519\pm   1$ & $  519$ & $  28\pm  16$ & $ 18.6$ & $  -32$ & $  517$ \\
UGC 528 & $  10^*$ & $  54\pm   3$ & $ 1.1\pm  0.1$ & $ 41\pm   2$ & $   41$ & $  35\pm   2$ & $  1.1$ & $   39$ & $   38$ \\
UGC 763 & $  24\pm   6$ & $ 119\pm   1$ & $ 1.2\pm  0.3$ & $ 98\pm   1$ & $   98$ & $  26\pm   5$ & $  3.8$ & $  -10$ & $   83$ \\
NGC 542 & $-$ & $-$ & $-$ & $-$ & $-$ & $-$ & $-$ & $-$ & $-$ \\
UGC 1249 & $-$ & $-$ & $-$ & $-$ & $-$ & $-$ & $-$ & $-$ & $-$ \\
UGC 1256 & $  78\pm   6$ & $  67\pm   1$ & $ 3.0\pm  0.2$ & $ 90\pm   2$ & $   90$ & $  25\pm   5$ & $  3.6$ & $   -2$ & $   30$ \\
\end{longtable}
\end{landscape}
\twocolumn

\onecolumn
\begin{landscape}
\begin{longtable}{c|c|ccc|ccccc}
\caption{Arctangent model on the sample projected at $z=1.7$}\\
\hline
Galaxy$~^{(a)}$ & $i_{z=1.7}~^{(b)}$ & $PA_{z=1.7}~^{(c)}$ & $r_t~^{(d)}$ & $V_t~^{(e)}$ & $V^{max}_c~^{(f)}$ & $\sigma~^{(g)}$ & $V^{max}_c/\sigma$ & $\Delta V^{mean}_c~^{(h)}$ & $S_{in}~^{(i)}$ \label{table_modzata} \\
 & \degr\ & \degr\ & $kpc$ & $km~s^{-1}$ & $km~s^{-1}$ & $km~s^{-1}$ & & $km~s^{-1}$ & $km~s^{-1}~kpc^{-1}$  \\
\hline
\endfirsthead

\caption[]{\textit{continued}}\\
\hline
Galaxy$~^{(a)}$ & $i_{z=1.7}~^{(b)}$ & $PA_{z=1.7}~^{(c)}$ & $r_t~^{(d)}$ & $V_t~^{(e)}$ & $V^{max}_c~^{(f)}$ & $\sigma~^{(g)}$ & $V^{max}_c/\sigma$ & $\Delta V^{mean}_c~^{(h)}$ & $S_{in}~^{(i)}$ \\
 & \degr\ & \degr\ & $kpc$ & $km~s^{-1}$ & $km~s^{-1}$ & $km~s^{-1}$ & & $km~s^{-1}$ & $km~s^{-1}~kpc^{-1}$  \\
\hline
\endhead

\hline
\endfoot

\hline
\multicolumn{10}{l}{(a): Galaxies for which no parameter is provided are those for which the fit was not possible.}\\
\multicolumn{10}{l}{(b): Inclination deduced from the fit. The error is a statistical error and thus gives a lower limit. The asterisk $^*$ indicates that the}\\
\multicolumn{10}{l}{inclination was stacked to one boundary.}\\
\multicolumn{10}{l}{(c): Position angle of the major axis deduced from the fit with the inclination fixed to the value in Table \ref{table_modz0}. The error is a statistical}\\
\multicolumn{10}{l}{error and thus gives a lower limit.}\\
\multicolumn{10}{l}{(d): Scale length radius of the model (defined in Appendix \ref{model}) deduced from the fit with the inclination fixed to the value in Table \ref{table_modz0}.}\\
\multicolumn{10}{l}{The error is a statistical error and thus gives a lower limit.}\\
\multicolumn{10}{l}{(e): Velocity of the model (defined in Appendix \ref{model}) deduced from the fit with the inclination fixed to the value in Table \ref{table_modz0}. The error is}\\
\multicolumn{10}{l}{a statistical error and thus gives a lower limit.}\\
\multicolumn{10}{l}{(f): Maximum velocity of the model within $R_{last}$ (see Table \ref{table_z}). The error is the same as in column (e) and thus not take into account}\\
\multicolumn{10}{l}{the uncertainty on the inclination.}\\
\multicolumn{10}{l}{(g): Local velocity dispersion (beam smearing corrected).}\\
\multicolumn{10}{l}{(h): Mean difference of the model \rc\ with the actual \rc\ at $z=0$.}\\
\multicolumn{10}{l}{(i): Inner slope of the \rc\ from the model. The asterisk $^*$ indicates that only the slope is constrained by the model.}\\
\endlastfoot
UGC 89 & $  10^*$ & $ 172\pm   1$ & $ 1.0\pm  0.1$ & $389\pm   1$ & $  274$ & $  24\pm  20$ & $ 11.6$ & $  -15$ & $  490$ \\
UGC 94 & $  21\pm   1$ & $  91\pm   1$ & $ 1.0\pm  0.1$ & $232\pm   1$ & $  163$ & $  17\pm  14$ & $  9.4$ & $  -17$ & $  302$ \\
UGC 508 & $  10^*$ & $ 122\pm   1$ & $ 1.0\pm  0.1$ & $536\pm   1$ & $  378$ & $  28\pm  16$ & $ 13.5$ & $  -22$ & $  679$ \\
UGC 528 & $  10^*$ & $  54\pm   3$ & $ 1.1\pm  0.1$ & $ 48\pm   2$ & $   33$ & $  36\pm   2$ & $  0.9$ & $   40$ & $   57$ \\
UGC 763 & $  14\pm  11$ & $ 119\pm   1$ & $ 1.1\pm  0.1$ & $107\pm   1$ & $   76$ & $  27\pm   5$ & $  2.8$ & $  -10$ & $  126$ \\
NGC 542 & $-$ & $-$ & $-$ & $-$ & $-$ & $-$ & $-$ & $-$ & $-$ \\
UGC 1249 & $-$ & $-$ & $-$ & $-$ & $-$ & $-$ & $-$ & $-$ & $-$ \\
UGC 1256 & $  10^*$ & $  67\pm   1$ & $ 2.7\pm  0.5$ & $107\pm   4$ & $   75$ & $  26\pm   4$ & $  2.9$ & $   -4$ & $   51$ \\
\end{longtable}
\end{landscape}
\twocolumn

\onecolumn
\begin{longtable}{cccccccc}
\caption{Parameters computed without beam smearing correction for the sample projected at $z=1.7$.}\\
\hline
Galaxy & $B~^{(a)}$ & $V^{max}_{rc}~^{(b)}$ & $\sigma_{cen}~^{(c)}$ & $\sigma_{min}~^{(d)}$ & $V^{max}_{rc}/\sigma_{cen}$ & $\Delta V^{mean}_c~^{(e)}$ & $R_{last}~^{(f)}$ \label{table_z} \\
 & & $km~s^{-1}$ & $km~s^{-1}$ & $km~s^{-1}$ & & $km~s^{-1}$ & $kpc$  \\
\hline
\endfirsthead

\caption[]{\textit{continued}}\\
\hline
Galaxy & $B~^{(a)}$ & $V^{max}_{rc}~^{(b)}$ & $\sigma_{cen}~^{(c)}$ & $\sigma_{min}~^{(d)}$ & $V^{max}_{rc}/\sigma_{cen}$ & $\Delta V^{mean}_c~^{(e)}$ & $R_{last}~^{(f)}$ \\
 & & $km~s^{-1}$ & $km~s^{-1}$ & $km~s^{-1}$ & & $km~s^{-1}$ & $kpc$  \\
\hline
\endhead

\hline
\endfoot

\hline
\multicolumn{8}{l}{(a): Beam smearing parameter: $B=\frac{D_{25}/2}{s}$, $s$ being the seeing.}\\
\multicolumn{8}{l}{(b): Maximum velocity measured on the \rc\ along the major axis at $z=1.7$.}\\
\multicolumn{8}{l}{(c): Central velocity dispersion from the uncorrected \vdm.}\\
\multicolumn{8}{l}{(d): Mean velocity dispersion from the 20\%\ smallest values.}\\
\multicolumn{8}{l}{(e): Mean difference of the \rc\ measured along the major axis at $z=1.7$ with the}\\
\multicolumn{8}{l}{actual \rc\ at $z=0$.}\\
\multicolumn{8}{l}{(f): Radius of the last point (maps were cut at $D_{25}/2$).}\\
\endlastfoot
UGC 89 & $  4.6$ & $  340$ & $  100$ & $   28$ & $  3.4$ & $   66$ & $  9.8$ \\
UGC 94 & $  4.4$ & $  205$ & $   62$ & $   20$ & $  3.3$ & $   32$ & $ 11.7$ \\
UGC 508 & $  6.5$ & $  514$ & $   54$ & $   23$ & $  9.6$ & $   60$ & $ 21.7$ \\
UGC 528 & $  0.8$ & $   25$ & $   36$ & $   24$ & $  0.7$ & $   60$ & $  1.9$ \\
UGC 763 & $  1.6$ & $   89$ & $   47$ & $   24$ & $  1.9$ & $   23$ & $  7.1$ \\
NGC 542 & $  2.5$ & $-$ & $   43$ & $   23$ & $-$ & $-$ & $  6.6$ \\
UGC 1249 & $  1.6$ & $-$ & $   31$ & $   21$ & $-$ & $-$ & $  4.7$ \\
UGC 1256 & $  1.7$ & $   64$ & $   40$ & $   25$ & $  1.6$ & $   29$ & $  7.2$ \\
\end{longtable}
\twocolumn

\section{Maps}
\label{maps}
\begin{figure*}
\begin{minipage}{180mm}
\begin{center}
\includegraphics[width=16cm]{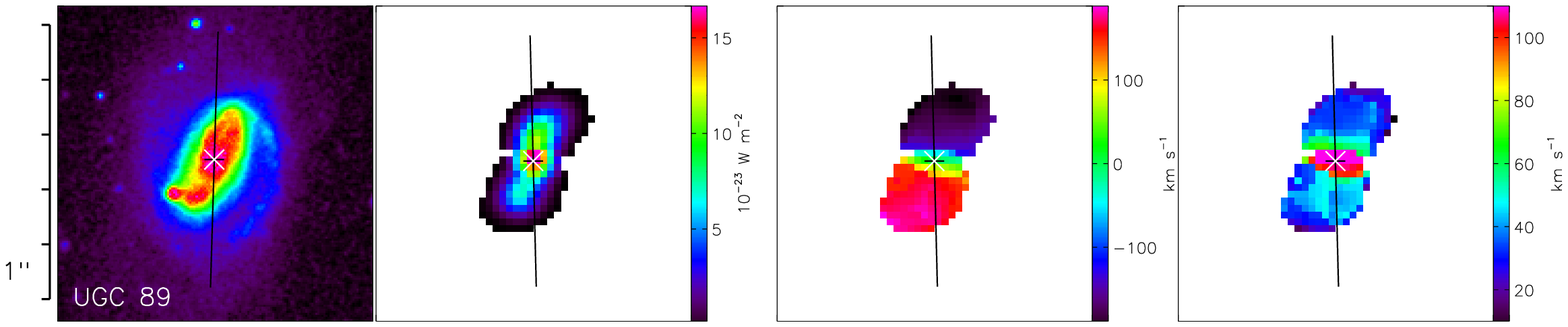}
\includegraphics[width=16cm]{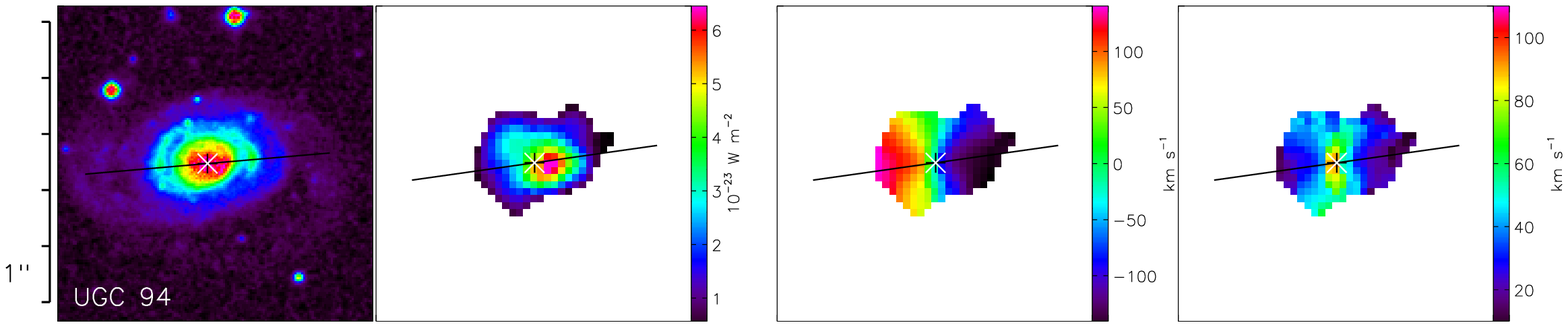}
\includegraphics[width=16cm]{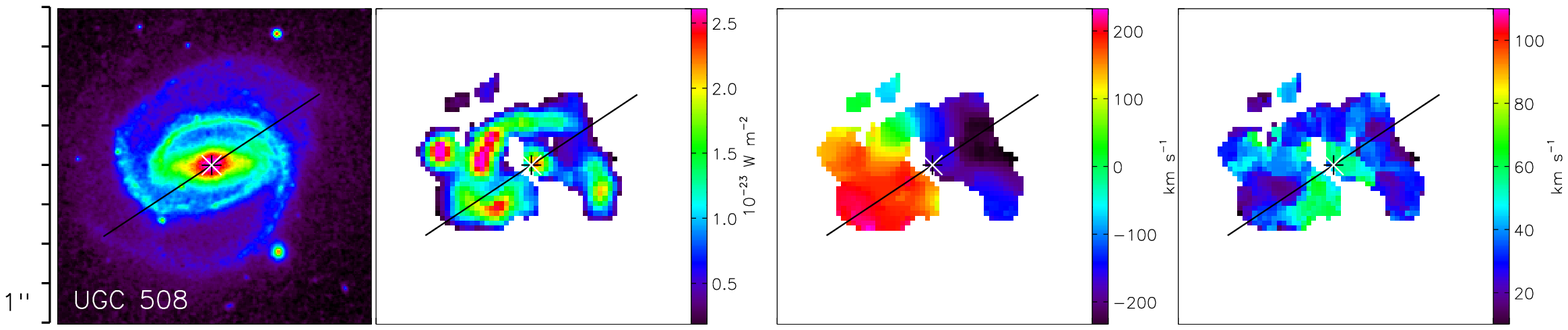}
\includegraphics[width=16cm]{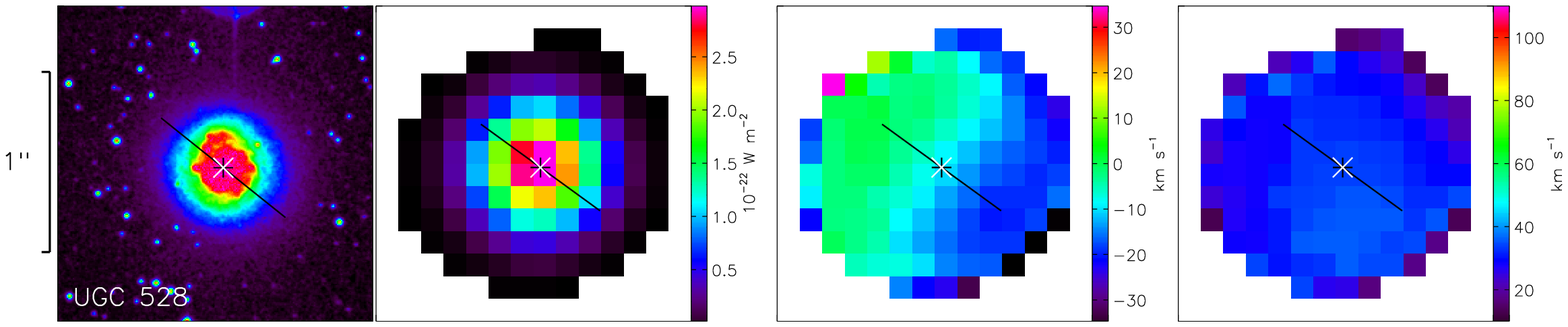}
\includegraphics[width=16cm]{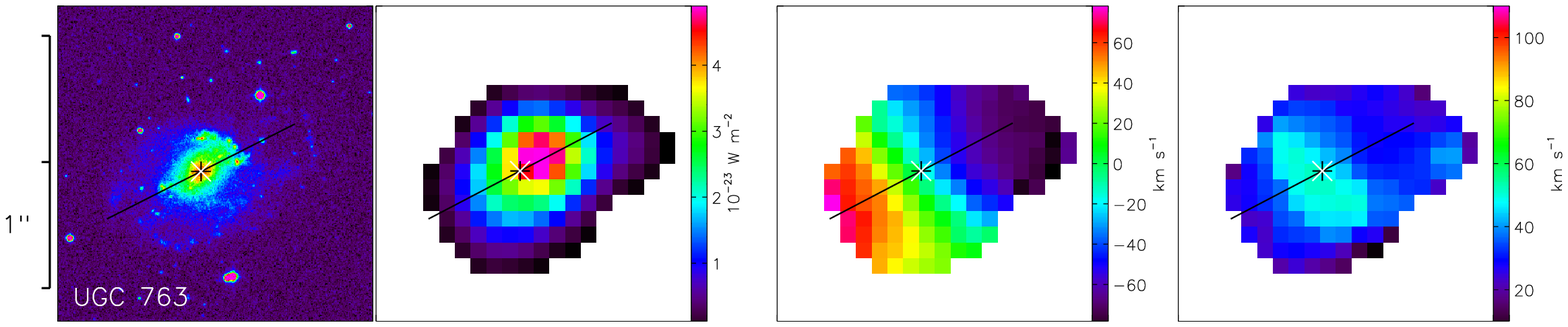}
\includegraphics[width=16cm]{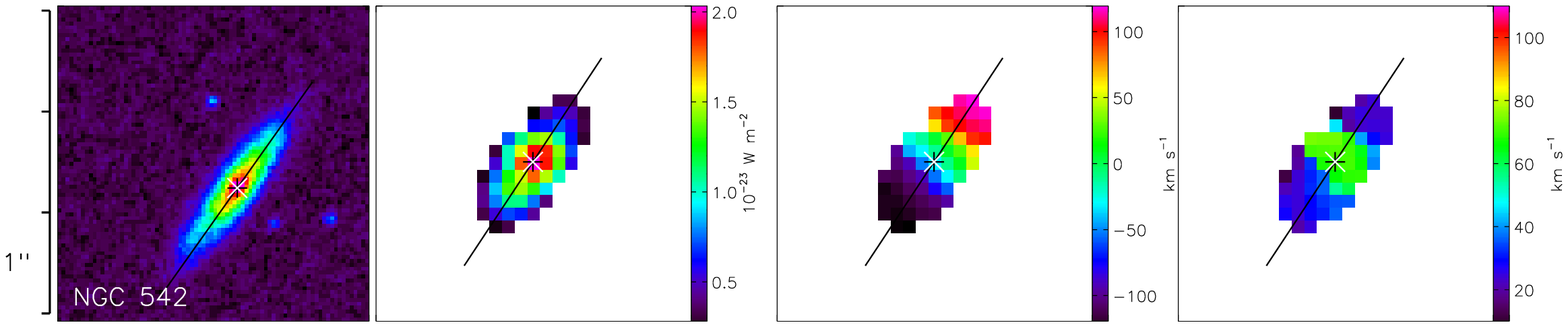}
\includegraphics[width=16cm]{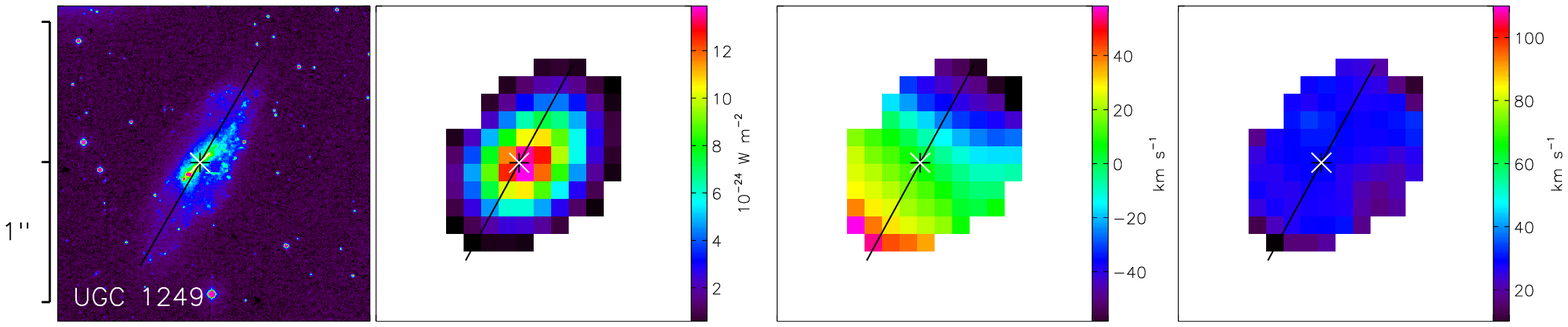}
\end{center}
\caption{From left to rigth: XDSS image, \Ha~monochromatic image, \vf, \vdm.
The white \& black crosses mark the kinematical center.
The black line is the major axis, its length represents the $D_{25}$. These maps are not truncated.}
\end{minipage}
\end{figure*}
\clearpage

\section{Rotation curves}
\label{rcz}
\begin{figure*}
\begin{minipage}{180mm}
\begin{center}
\includegraphics[width=5.5cm]{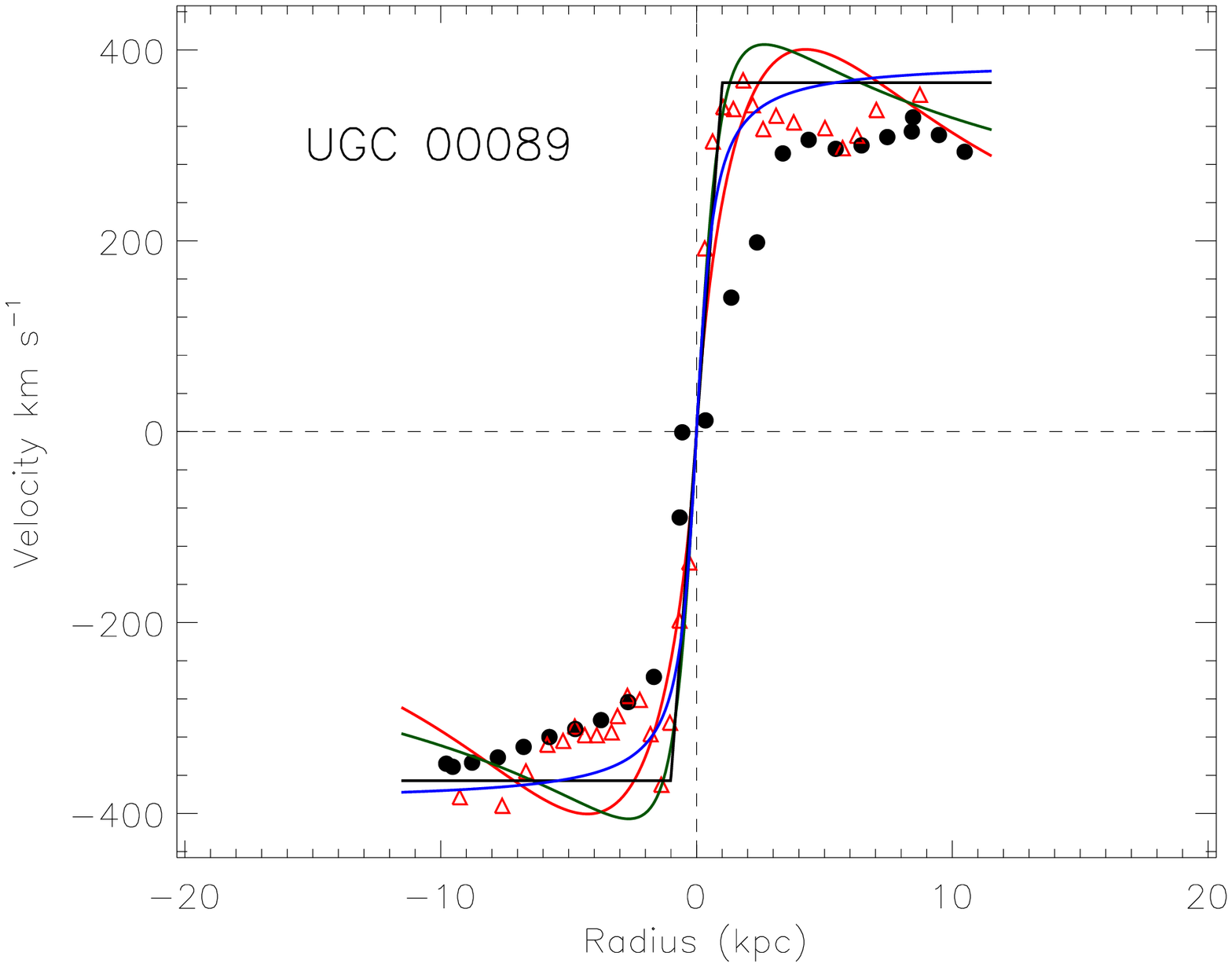}
\includegraphics[width=5.5cm]{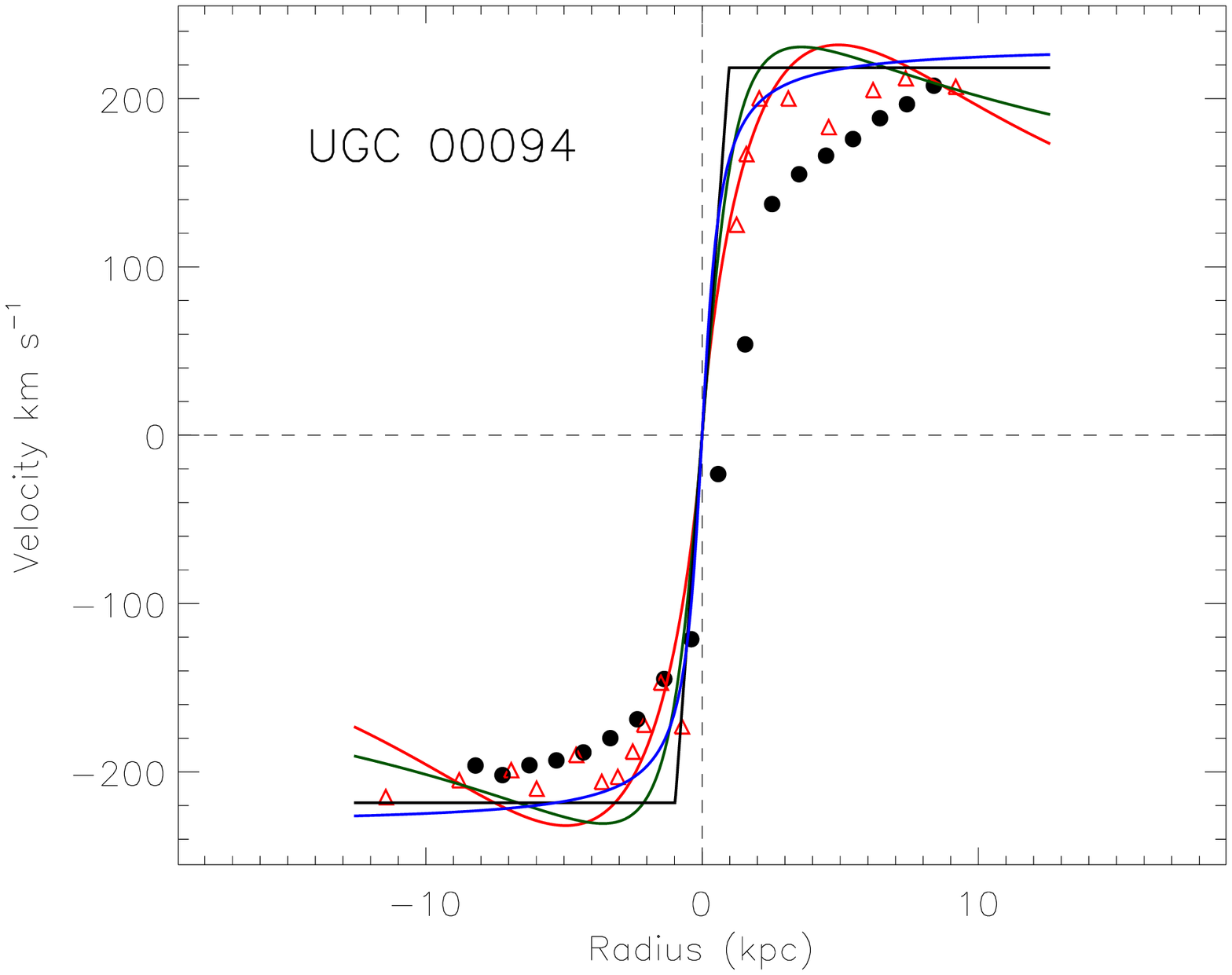}
\includegraphics[width=5.5cm]{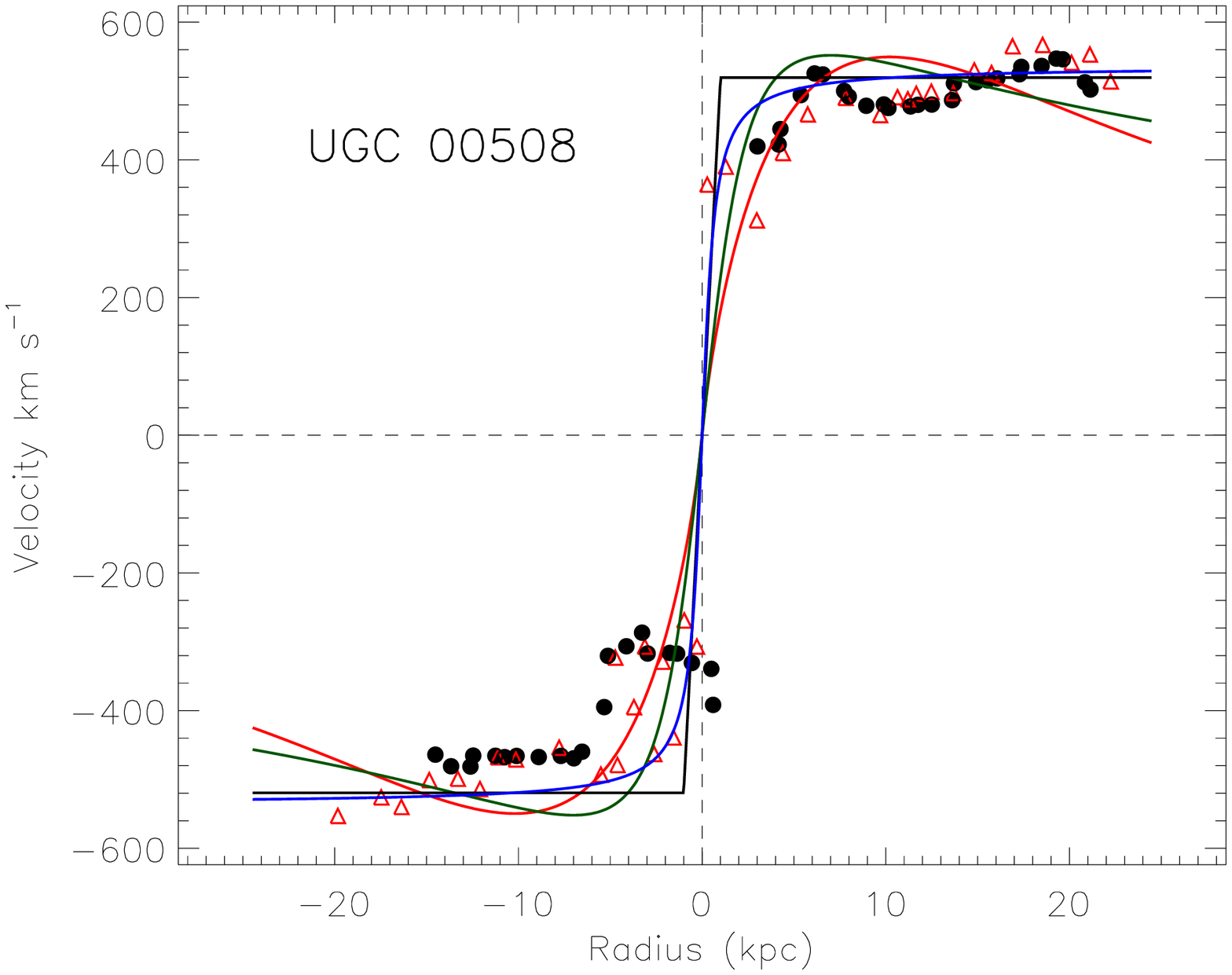}
\includegraphics[width=5.5cm]{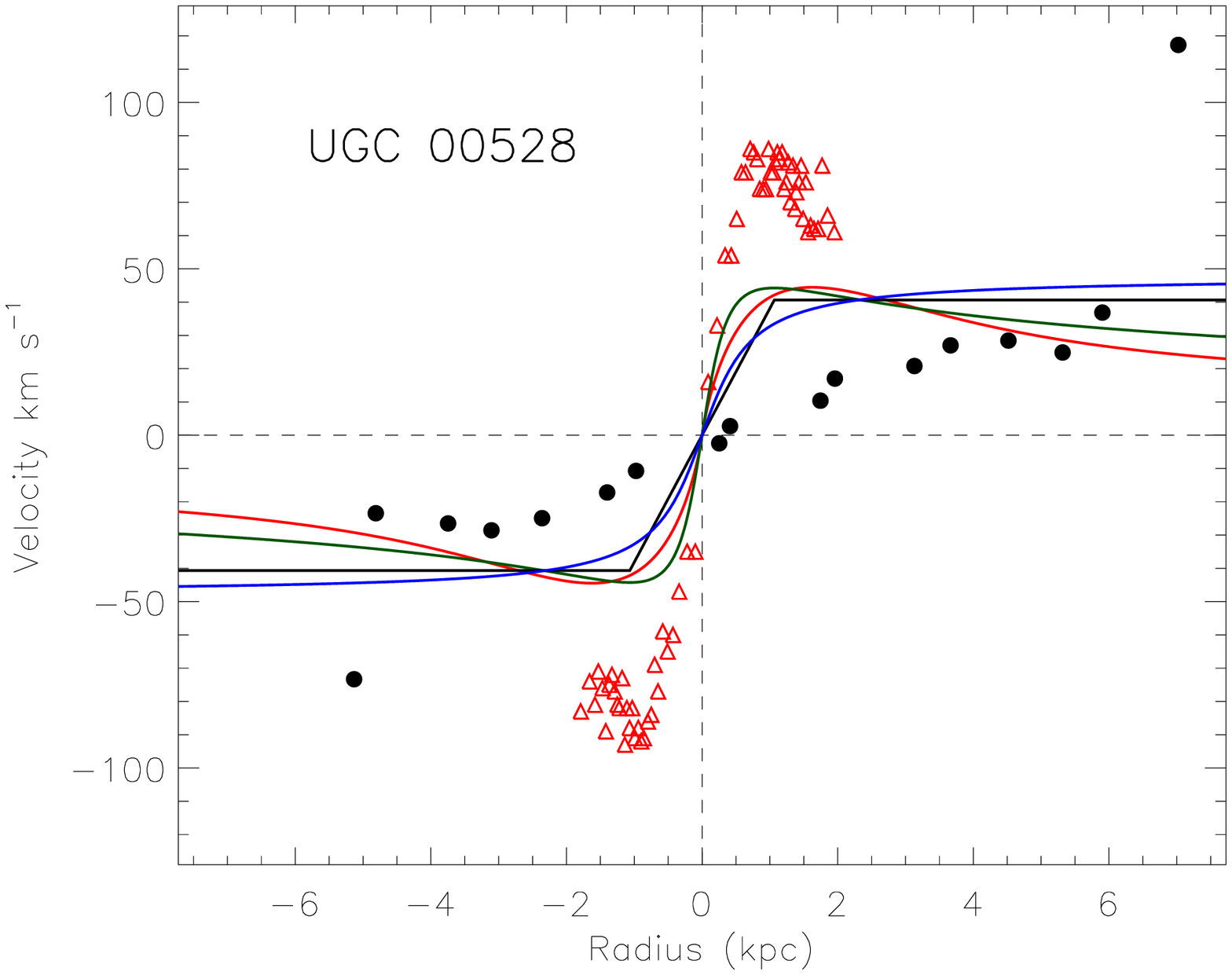}
\includegraphics[width=5.5cm]{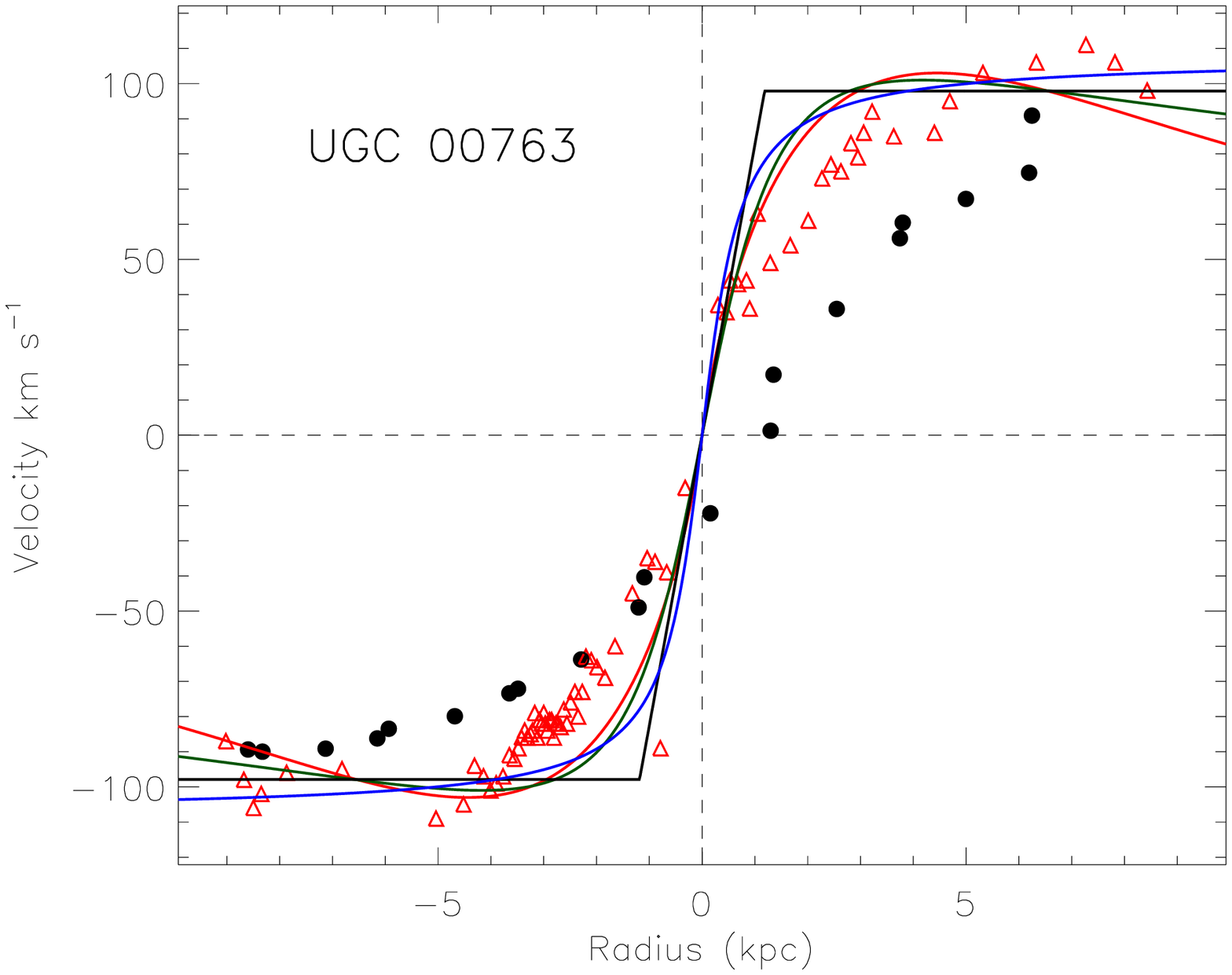}
\includegraphics[width=5.5cm]{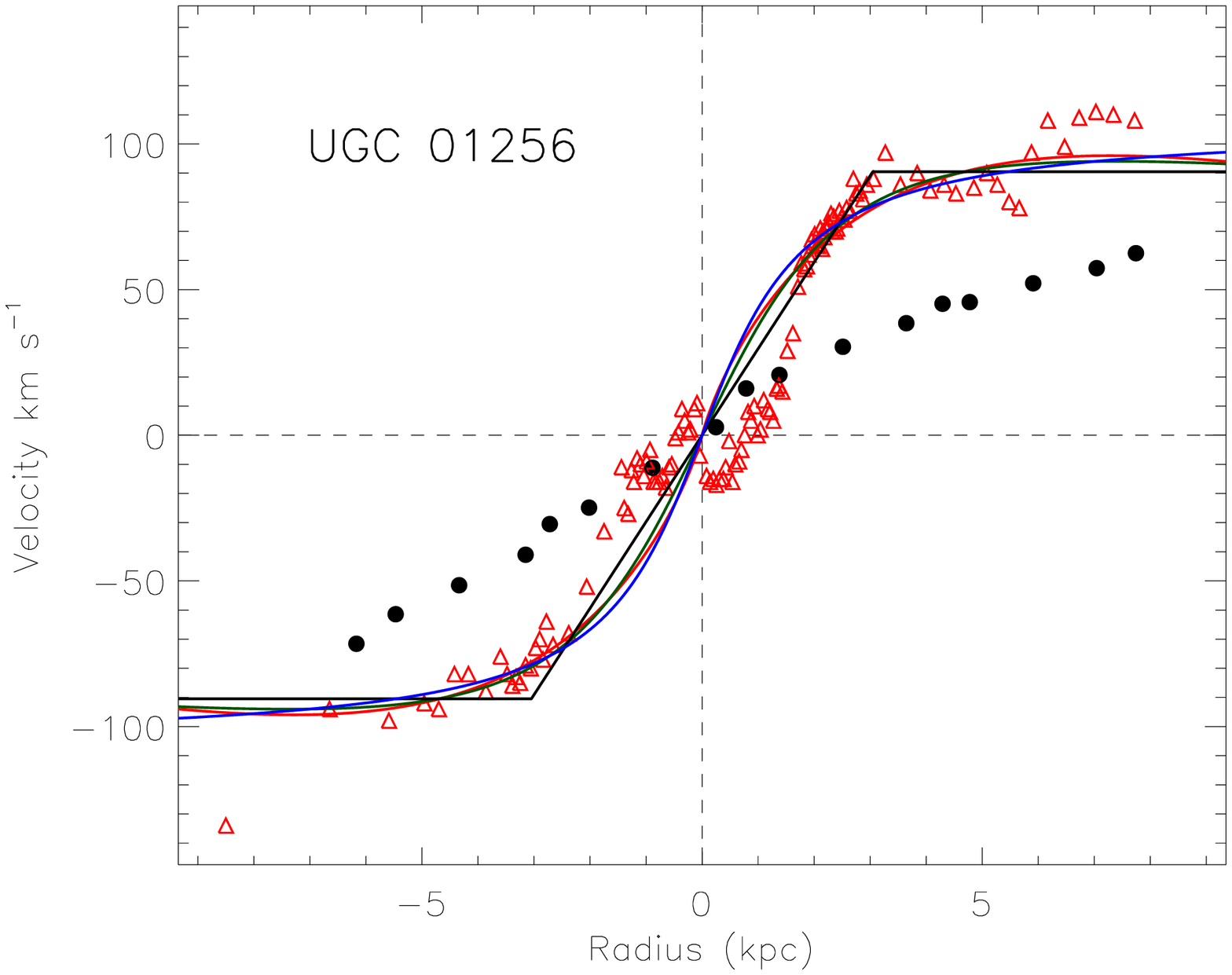}
\includegraphics[width=5.5cm]{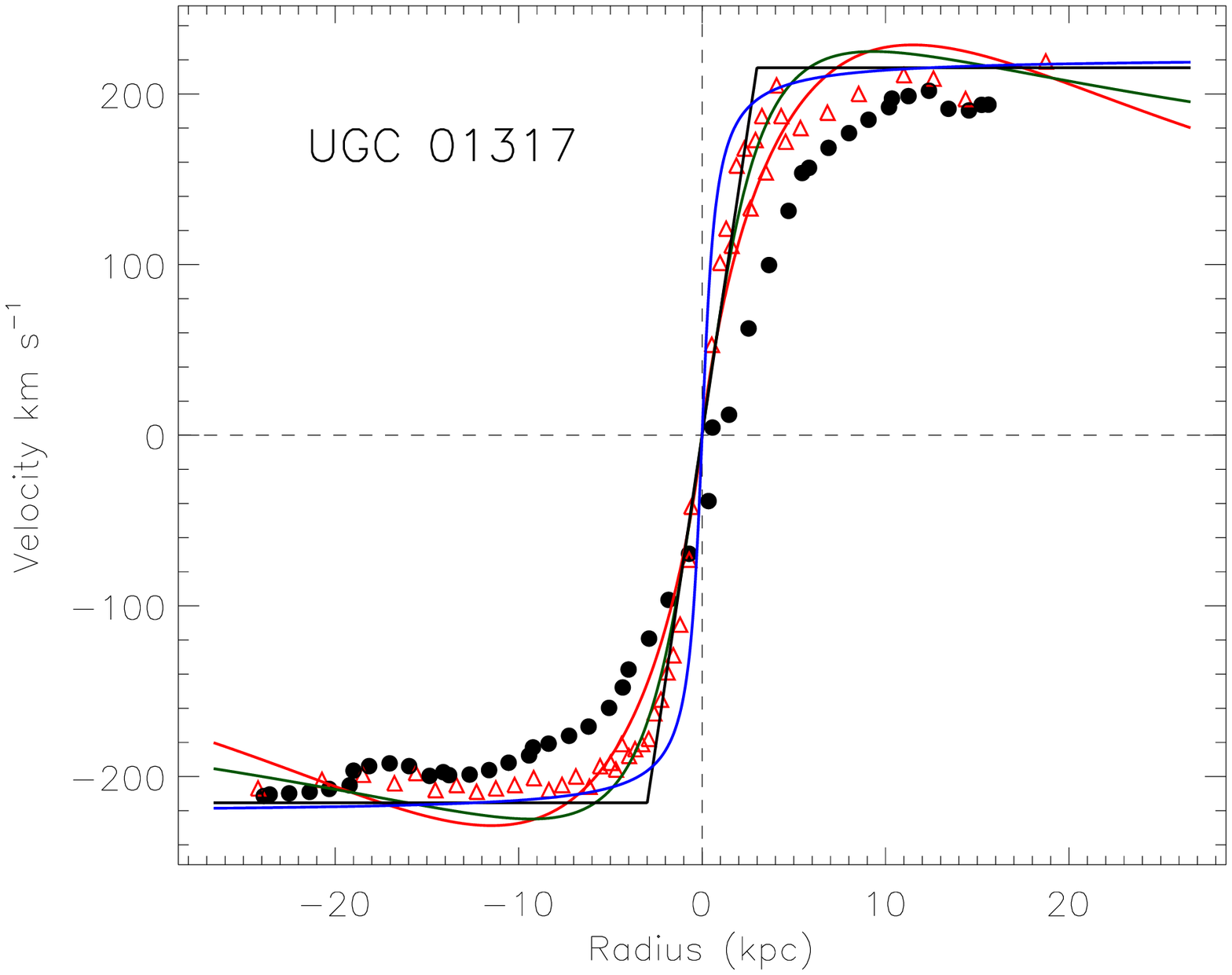}
\includegraphics[width=5.5cm]{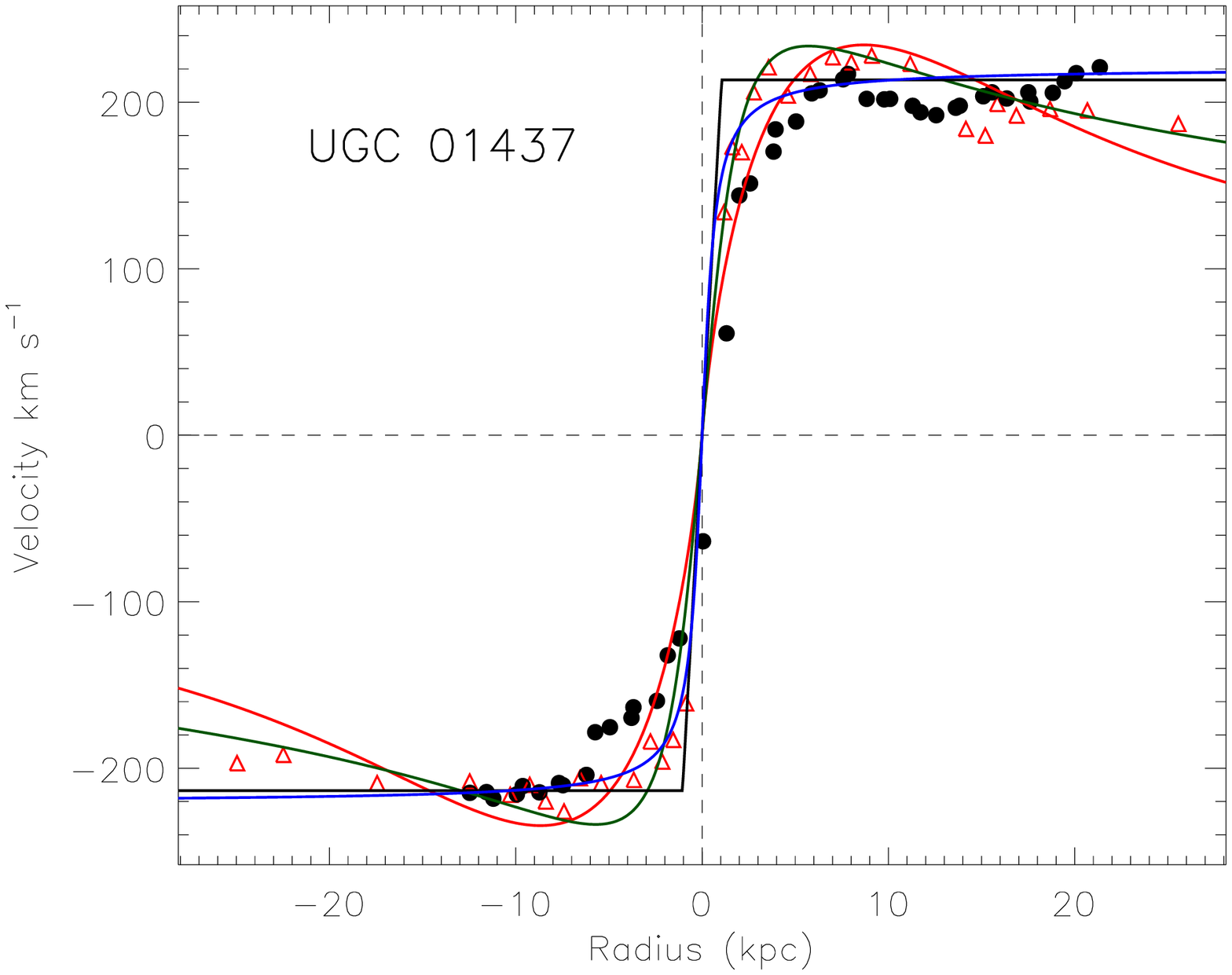}
\includegraphics[width=5.5cm]{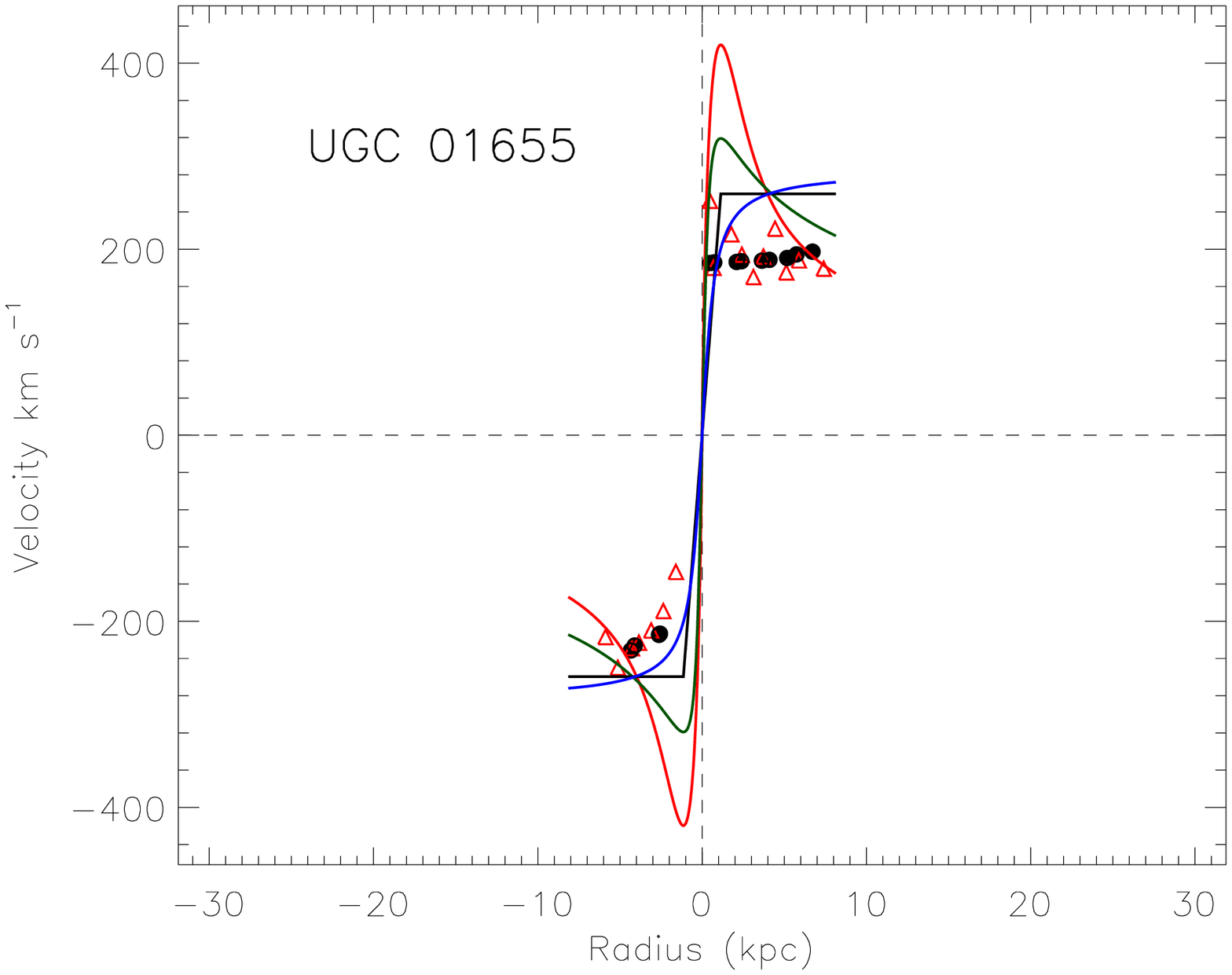}
\includegraphics[width=5.5cm]{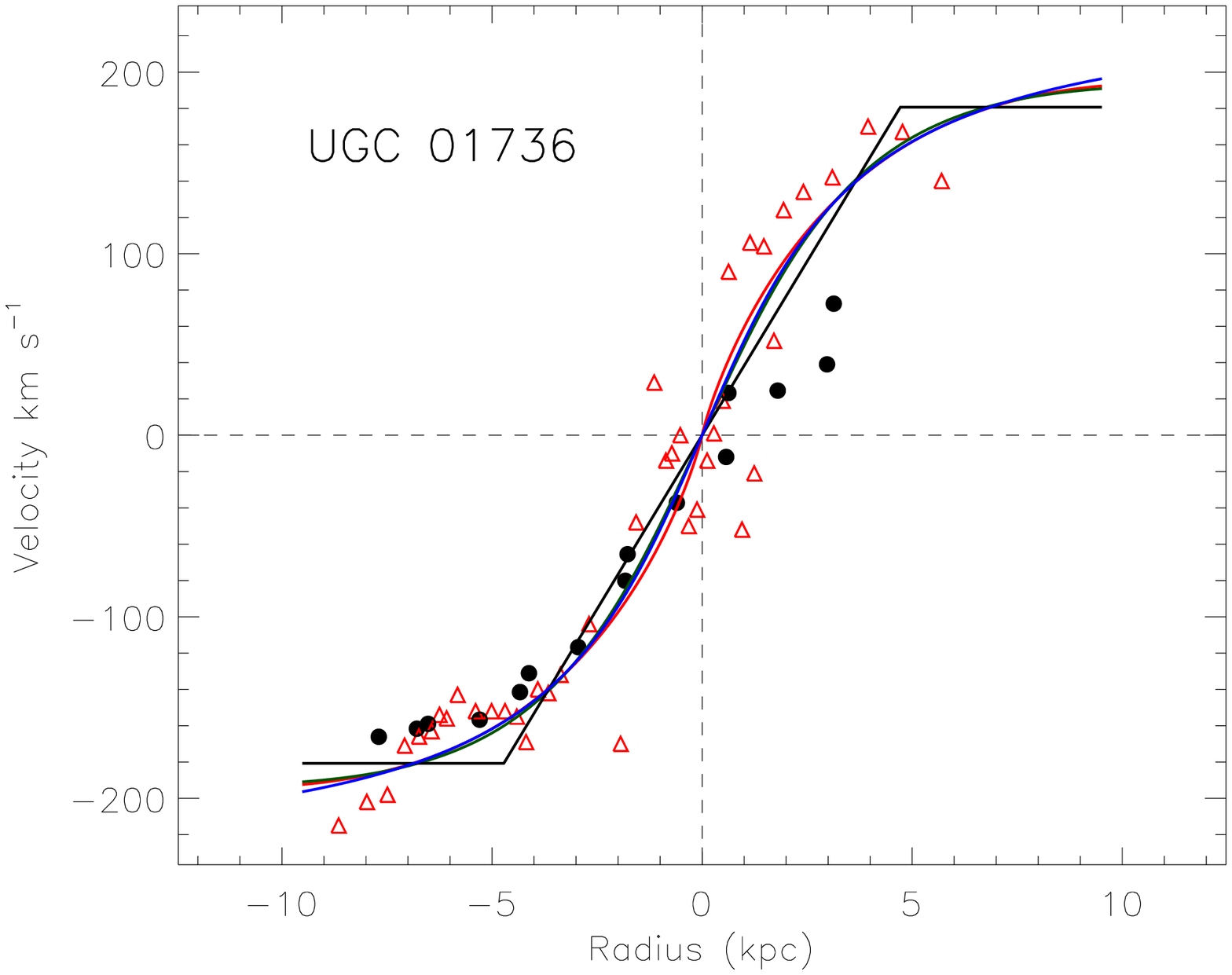}
\includegraphics[width=5.5cm]{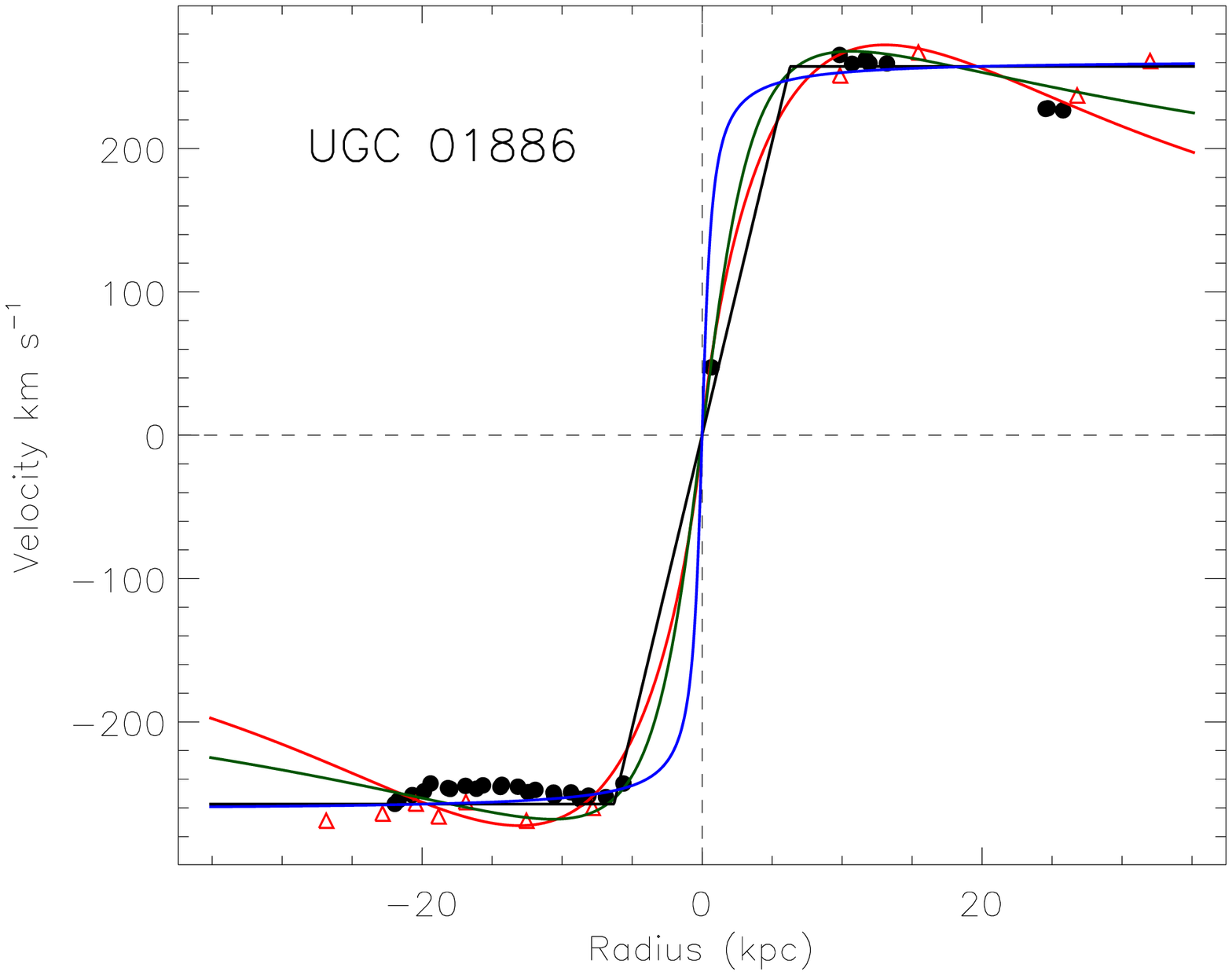}
\includegraphics[width=5.5cm]{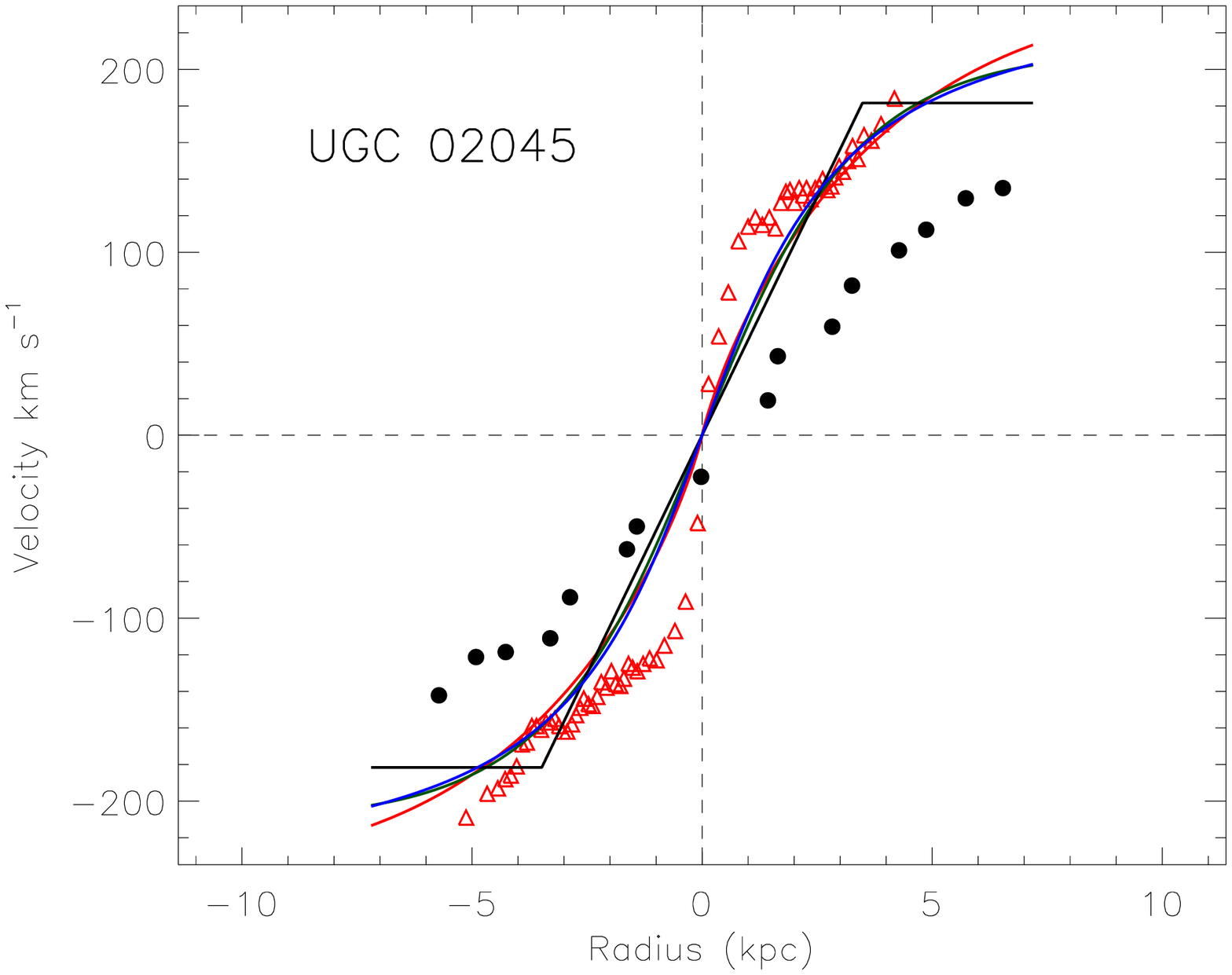}
\includegraphics[width=5.5cm]{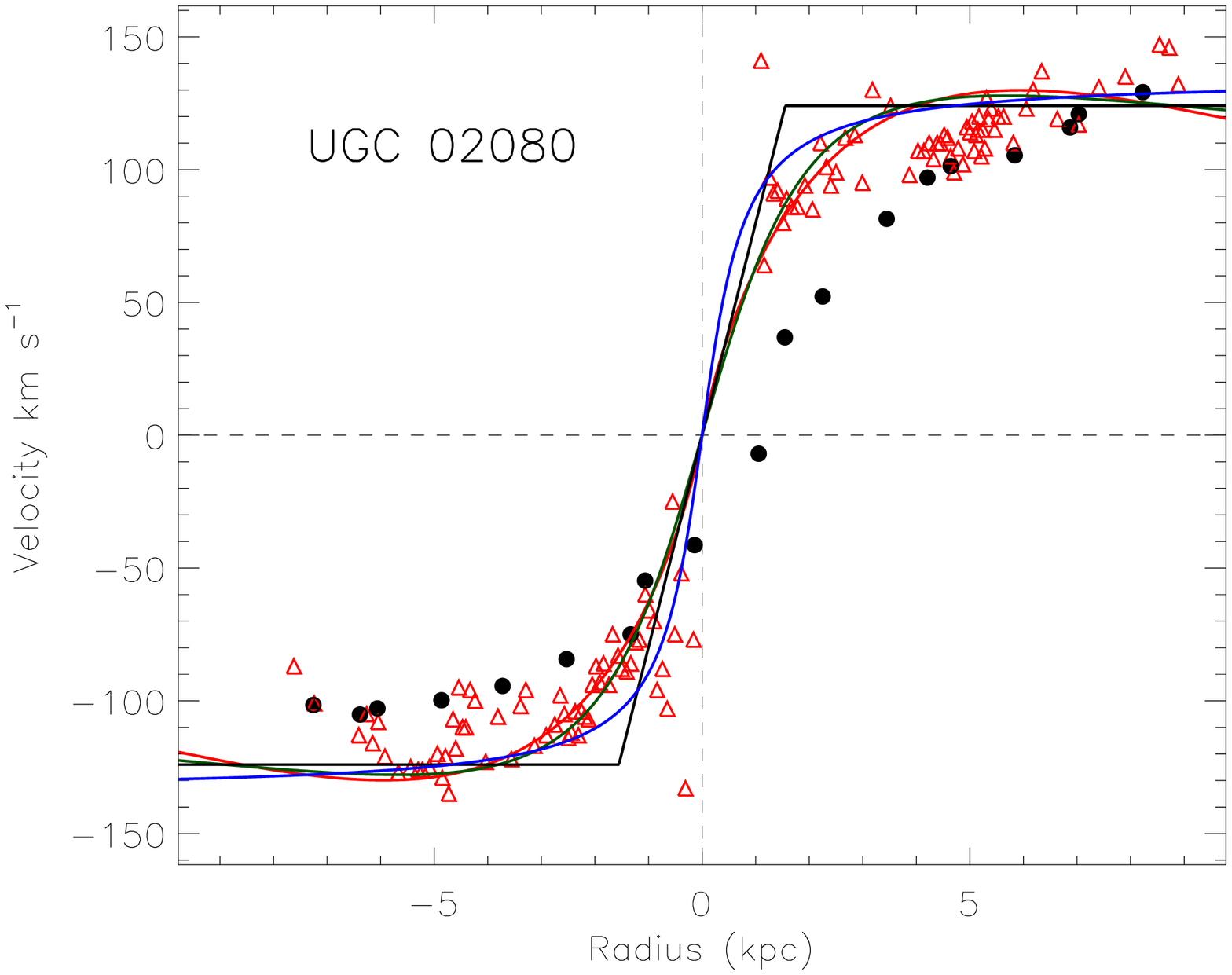}
\includegraphics[width=5.5cm]{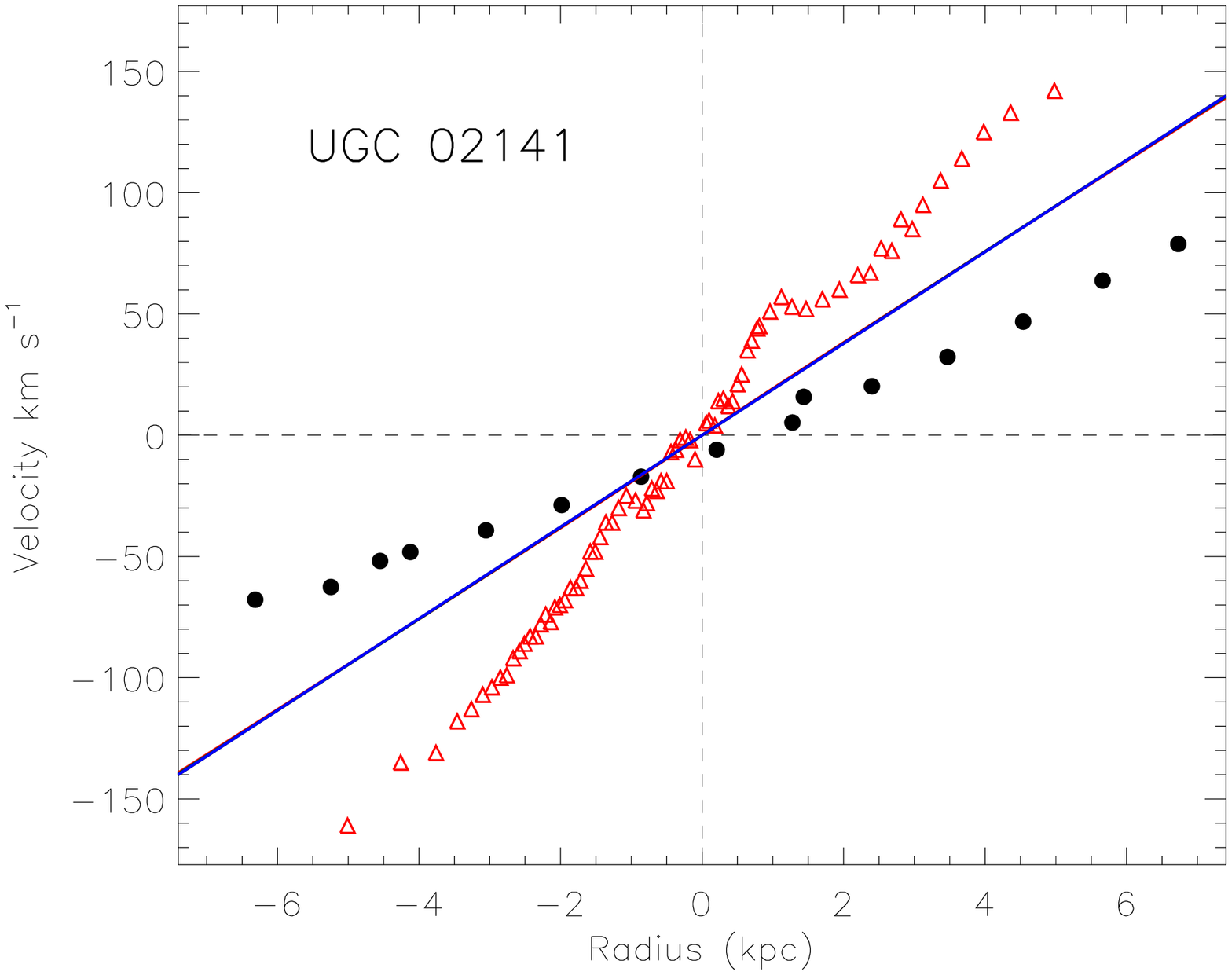}
\includegraphics[width=5.5cm]{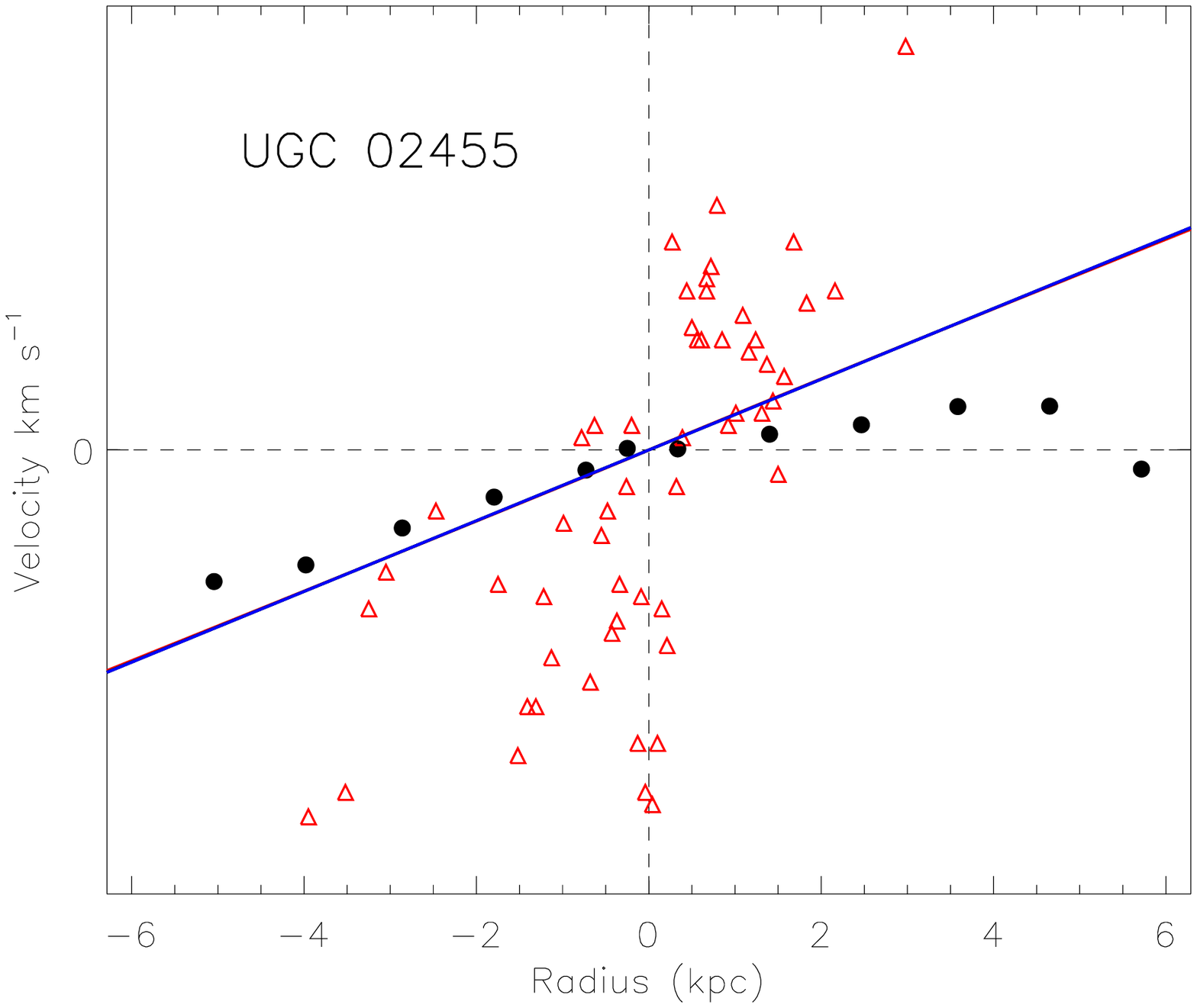}
\end{center}
\caption{High redshift rotation curves along the major axis (black dots), 
actual rotation curves at redshift zero (red-open triangles) and high resolution rotation curve models (red line: exponential disk; green line: isothermal sphere; black line: ``flat model''; blue line: arc-tangent function).}
\end{minipage}
\end{figure*}
\clearpage

\end{document}